\documentclass[a4paper,12pt]{article}
\pdfoutput=1 
\usepackage{a4wide}
\usepackage[bookmarks=false,colorlinks]{hyperref}
\usepackage[pdftex]{graphicx}
\usepackage{amsmath,amsfonts,amssymb,mathtools,bm}
\usepackage{rotating,amsmath,bm}
\usepackage{tabularx}
\usepackage{multicol,multirow}
\usepackage{latexsym}
\usepackage{relsize}
\usepackage[center]{subfigure}
\usepackage{float}
\usepackage{caption}
\usepackage{slashed}
\usepackage{cite,mcite}
\usepackage{appendix}
\usepackage{url}
\usepackage{placeins}
\usepackage[utf8]{inputenc}
\hypersetup{urlcolor=blue, citecolor=blue}

\normalsize

\newcommand{\HEPfit}{\texttt{HEPfit}}

\def\mev{{\rm MeV}}
\def\Heff{\mathcal{H}_{\rm eff}}
\def\AFB{\mathcal{A}_{\rm FB}}

\def\A{\mathcal{A}}
\def\B{\mathcal{B}}

\def\M{\mathcal{M}}

\def\Re{\mathcal{R}e}
\def\Im{\mathcal{I}m}

\def\Bbar{\overline{B}}

\def\cbar{\overline{c}}

\def\lbar{\overline{\ell}}
\def\nubar{\overline{\nu}}

\def\thl{{\theta_\ell}}
\def\thD{{\theta_D}}
\def\p{{\hat{\bm p}}}
\def\q{{\bm q}}

\def\Dst{{D^\ast}}
\def\DDst{{D^{(\ast)}}}

\def\htil{{\widetilde{h}}}
\def\Htil{{\widetilde{H}}}

\def\rlsq{{m_\ell^2 \over q^2}}
\newcommand{\bea}{\begin{eqnarray}}
\newcommand{\eea}{\end{eqnarray}}

\usepackage{relsize}
\def\Babar{{\mbox{\slshape B\kern-0.1em{\smaller A}\kern-0.1em B\kern-0.1em{\smaller A\kern-0.2em R}}}}

\begin{document}

\thispagestyle{empty} 
\begin{flushright}
DO-TH 19/10\\
LPT-Orsay-19-26\\
QFET-2019-06\\
TTP19-022
\end{flushright}
\vspace*{1cm}\par
\begin{center}	
{\par\centering \textbf{\LARGE  
\Large \bf Lepton Flavor Universality tests through angular}} \\
\vskip .25cm\par
{\par\centering \textbf{\LARGE
\Large  observables of $\bm{\Bbar\to\DDst\ell\nubar}$ decay modes}} \\
\vskip 1.25cm\par
{\scalebox{.88}{\par\centering \large  
\sc Damir Be\v cirevi\'c$^a$, Marco Fedele$^b$, Ivan Ni\v{s}and\v{z}i\'c$^c$, Andrey Tayduganov$^d$}} \\
{\par\centering \vskip 0.6 cm\par}
{\sl \small
$^a$~Laboratoire de Physique Th\'eorique,  B\^at.~210 (UMR 8627) \\
Universit\'e Paris Sud, Universit\'e Paris-Saclay, 91405 Orsay cedex, France.} \\
{\par\centering \vskip 0.3 cm\par}
{\sl \small
$^b$~Departament de F\'isica Qu\`antica i Astrof\'isica (FQA),
Institut de Ci\`encies del Cosmos (ICCUB), Universitat de Barcelona, Mart\'i i Franqu\`es 1, E08028 Barcelona, Spain.} \\
{\par\centering \vskip 0.3 cm\par}
{\sl \small
$^c$~Institut f\"{u}r Theoretische Teilchenphysik, \\
Karlsruher Institut f\"{u}r Technologie, Engesser-Str.7, D-76128 Karlsruhe, Germany.} \\
{\par\centering \vskip 0.3 cm\par}
{\sl \small
$^d$~Fakult\"{a}t Physik, TU Dortmund, Otto-Hahn-Str.4, D-44221 Dortmund, Germany.} \\

{\vskip 1.cm\par}
\end{center}

\begin{abstract}
We discuss the possibility of using the observables deduced from the angular distribution of the $B\to D^{(\ast)} \ell\bar{\nu}$ decays to test the effects of lepton flavor universality violation (LFUV). 
We show that the measurement of even a subset of these observables could be very helpful in distinguishing the Lorentz structure of the New Physics contributions to these decays. To do so 
we use the low energy effective theory in which besides the Standard Model contribution we add all possible Lorentz structures with the couplings (Wilson coefficients) that are determined by matching 
theory with the measured ratios $R{(D^{(\ast)})}^\mathrm{exp}$. 
We argue that even in the situation in which the measured $R{(D^{(\ast)})}^\mathrm{exp}$ becomes fully compatible with the Standard Model, one can still have significant New Physics contributions the size of which could  be probed by measuring the observables discussed in this paper and comparing them with their Standard Model predictions. 
\end{abstract}
\newpage
\setcounter{page}{1}
\setcounter{footnote}{0}
\setcounter{equation}{0} 
\noindent
\allowdisplaybreaks[1]
\renewcommand{\thefootnote}{\arabic{footnote}}
\setcounter{footnote}{0}


\section{Introduction}

Ever since the BaBar Collaboration reported on the first experimental indication of the lepton flavor universality violation (LFUV) in $B$-meson decays, after measuring the ratios~\cite{Aubert:2007dsa,Lees:2012xj}
\begin{equation}
R{(D^{(\ast)})} = \left. \dfrac{\mathcal{B}(B\to D^{(\ast)} \tau\bar{\nu})}{\mathcal{B}(B\to D^{(\ast)} l \bar{\nu})}\right|_{l\in \{e,\mu\}},
\label{eq:RD_definition}
\end{equation}
we witnessed an intense activity in trying to understand why $R{(D^{(\ast)})}^\mathrm{SM}<R{(D^{(\ast)})}^\mathrm{exp}$, where SM stands for the theoretically established value in the Standard Model. Since then, the experimentalists at Belle and LHCb corroborated the tendency that indeed $R{(D^{(\ast)})}^\mathrm{SM}<R{(D^{(\ast)})}^\mathrm{exp}$~\cite{Huschle:2015rga,Sato:2016svk,Aaij:2015yra,Aaij:2017uff}, while the theorists worked on scrutinizing the theoretical uncertainties within the SM~\cite{Fajfer:2012vx,Bernlochner:2017jka,Jaiswal:2017rve,Gambino:2019sif} and proposed various models of New Physics (NP) that could accommodate the observed discrepancies~\cite{Shi:2019gxi,Blanke:2019qrx,Murgui:2019czp,Popov:2019tyc,Cornella:2019hct,Alok:2019uqc,Blanke:2018yud,Iguro:2018vqb,Aebischer:2018iyb,Hu:2018veh,Angelescu:2018tyl,Heeck:2018ntp,Faber:2018qon,Huang:2018nnq,Feruglio:2018fxo,Becirevic:2018afm,Bordone:2018nbg,Azatov:2018knx,Alok:2018uft,Marzocca:2018wcf,Blanke:2018sro,Jung:2018lfu,Alok:2017qsi,Buttazzo:2017ixm,Altmannshofer:2017poe,Crivellin:2017zlb,Celis:2016azn,Alonso:2016gym,Freytsis:2015qca,Abada:2013aba,Tanaka:2012nw,Trifinopoulos:2019lyo}. The current values, including the most recent Belle result~\cite{Abdesselam:2019dgh}, are~\cite{Amhis:2016xyh}:
\bea\label{eq:HFLAV}
&&R(D)^\mathrm{exp} = 0.33(3)\,,\qquad R(D)^\mathrm{SM} = 0.299(3)\,,\\
&&R{(D^\ast)}^\mathrm{exp} = 0.30(2)\,,\qquad R{(D^\ast)}^\mathrm{SM} = 0.258(5)\,, 
\eea
which amounts to a combined $3.1\sigma$ disagreement between the measurements and the SM predictions. Measuring the ratios of similar decay rates is convenient because they are independent on the Cabibbo--Kobayashi-Maskawa (CKM) matrix element $|V_{cb}|$, and because of the cancellation of a significant amount of hadronic uncertainties. One should, however, be careful in assessing the uncertainties regarding the remaining non-perturbative QCD effects, especially in the case of $B\to D^\ast \ell\bar \nu_\ell$ $(\ell = e,\mu,\tau )$ for which the information concerning the shapes of hadronic form factors has never been deduced from the results based on numerical simulations of QCD on the lattice.~\footnote{Very recently the MILC collaboration presented preliminary results for the shapes of the $B\to D^\ast$ form factors~\cite{Aviles-Casco:2019vin,Vaquero:2019ary}. Numerical results, that could be used for phenomenology, are yet to be reported.} In the case of $B\to D \ell\bar \nu_\ell$, instead, both the vector and scalar form factors have been computed on the lattice at several different $q^2=(p_\ell+p_{\bar \nu_\ell})^2$ values~\cite{Lattice:2015rga,Harrison:2017fmw,Atoui:2013zza}. Moreover, the theoretical studies reported in Ref.~\cite{deBoer:2018ipi,Becirevic:2009fy} suggest that the soft photon radiation $B\to D  \ell\bar \nu_\ell\gamma$ could be an important source of uncertainty in $R(D)^\mathrm{SM}$, unaccounted for thus far. Further research in this direction is necessary and the associated theoretical uncertainty should be included in the overall error budget before the $5\sigma$ (or larger) discrepancy between theory (SM) and experiment is claimed.

Notice also that the LHCb Collaboration recently made another LFUV test based on $b\to c\ell\bar \nu_\ell$. They reported~\cite{Aaij:2017tyk} 
\bea
R(J/\psi) =   \dfrac{\mathcal{B}(B_c\to J/\psi \tau\bar{\nu})}{\mathcal{B}(B_c\to J/\psi \mu \bar{\nu})} = 0.71\pm 0.25\,,
\eea
again about $2\sigma$ larger than predicted in the SM, $R(J/\psi)^\mathrm{SM}<R(J/\psi)^\mathrm{exp}$.

The above-mentioned experimental indications of LFUV have motivated many theorists to propose a scenario beyond the SM which could accommodate the measured values of $R(D^{(\ast)})$ and $R(J/\psi)$. In terms of a general low energy effective theory, which will be described in the next Section, the NP effects can show up at low energies through an enhancement of the contribution arising from either the (axial-)vector current-current operators, the (pseudo-)scalar operators, the tensor one, or from a combination of those~\cite{Colangelo:2018cnj,Biancofiore:2013ki,Faller:2011nj,Becirevic:2016hea,Bhattacharya:2015ida,Ivanov:2015tru,Sakaki:2014sea,Sakaki:2013bfa,Cohen:2018vhw}. Another important aspect in describing the LFUV effects in most of the models proposed so far is the assumption that the NP couplings involving $\tau$ are the source of LFUV, whereas those related to $l=e,\mu$ are much smaller and can be neglected. A rationale for that assumption is that the LFUV effects have not been observed in other semileptonic decay modes (which involve lighter mesons and light leptons). In addition to that, the results of a detailed angular analysis of $B\to D^{(\ast)} l\bar \nu_l$ [$l\in (e,\mu)$] by BaBar and by Belle were found to be fully consistent with the SM predictions. 
We will follow the same practice in this paper and assume that the deviations from the lepton flavor universality arise from the NP couplings to the third generation of leptons. We will formulate a general effective theory scenario for $b\to c\ell\bar\nu_\ell$ decays by adding all operators allowed by the $CP$ and the Lorentz symmetries. In doing so we will not account for a possibility of the light right-handed neutrinos. Such models have been already proposed, but the problem they often encounter is that the NP amplitude does not interfere with the (dominant) SM one, and therefore getting $R(D^{(\ast)}) > R(D^{(\ast)})^\mathrm{SM}$ compatible with $R(D^{(\ast)})^\mathrm{exp}$ requires large NP couplings which could be in conflict with the bounds on these couplings that could be deduced from direct searches at the LHC.

Starting from the general effective theory we will provide the expressions for the full angular distribution of both decay processes, and then combine the coefficients involving the NP couplings in a number of observables. Some of these observables could be studied at the LHC, but most of them could be tested at Belle~II. It is therefore reasonable to explore the possibilities of testing the effects of LFUV not only via the ratios of branching fractions [such as $R(D^{(\ast)})$], but also by using the ratios of these newly defined observables. Since various observables are sensitive to different NP operators, we will argue that the experimental analysis of the ratios we propose to study could indeed help disentangling the basic features of NP.   
We will then also provide a phenomenological analysis to illustrate our claim.

Before we embark on the details of this work we should emphasize that for the phenomenological analysis we need a full set of form factors (including the scalar, pseudoscalar and the tensor ones) computed by means of lattice QCD, which is not the case right now. A dedicated lattice computation of all the form factors relevant to $B\to D^{(\ast )}\ell\bar \nu_\ell$ is of great importance for a reliable assessment of LFUV. Since we have to make (reasonable) assumptions about the form factors, our phenomenological results should be understood as a diagnostic tool to distinguish among various Lorentz structures of the NP contributions, while the accurate analysis will be possible only when the lattice QCD results become available.

The remainder of this paper is organized as follows: In Sec.~\ref{sec2} we provide the full angular distributions for both types of decays considered in this paper, and define all the observables which can be used to better test the effects of LFUV in these decays. In Sec.~\ref{sec3} we make the sensitivity study of the observables defined in Sec.~\ref{sec2} to the effects of LFUV, solely based on the deviations of the measured $R(D^{(\ast )})$ with respect to the SM predictions. We also made a short comment on the recently measured $F_L^{D^\ast}$. We finally summarize in Sec.~\ref{sec4}. Several definitions and technical details are collected in the Appendices.

\section{Full angular distributions of $\bm{\Bbar\to\DDst\ell\nubar}$\label{sec2}}

In this Section we define the effective Hamiltonian for a generic NP scenario and then derive the expressions for the full angular distribution of the differential decay rate of $B\to D \ell\bar \nu_\ell$, and of $B\to D^{\ast } (\to D\pi)\ell\bar \nu_\ell$. All angular coefficients will be expressed in terms of helicity amplitudes which are properly defined in terms of kinematical variables and hadronic form factors. For the reader's convenience the decomposition of all the matrix elements in terms of form factors is provided in Appendix~\ref{app:FF} of the present paper. With explicit expressions for the angular coefficients in hands, we will then proceed and define a set of observables that can be used to study the effects of LFUV. 

\subsection{Effective Hamiltonian}

Assuming the neutrinos to be left-handed, 
the most general effective Hamiltonian describing  the $b\to c\ell\nubar_\ell$ decays, containing all possible parity-conserving four-fermion dimension-6 operators, can be written as
\begin{equation}
\begin{split}
\Heff = \sqrt2 G_F V_{cb} & \left[ 
(1+g_V) (\cbar \gamma_\mu b) (\lbar_L \gamma^\mu \nu_L) + 
(-1+g_A) (\cbar \gamma_\mu\gamma_5 b) (\lbar_L \gamma^\mu \nu_L) \right. \\
& + g_S (\cbar b) (\lbar_R \nu_L)
  + g_P (\cbar \gamma_5 b) (\lbar_R \nu_L) \\
& + g_T (\cbar \sigma_{\mu\nu} b) (\lbar_R \sigma^{\mu\nu} \nu_L)
  + \left. g_{T5} (\cbar \sigma_{\mu\nu} \gamma_5 b) (\lbar_R \sigma^{\mu\nu} \nu_L) \right]  + \mathrm{h.c.}
\end{split}
\label{eq:Heff}
\end{equation}
Note that we use the convention such that  in the SM $g_i=0$, $\forall i\in \{S,P,V,A,T,T5\}$.

It is often more convenient to write the above Hamiltonian in the chiral basis of operators, 
\begin{equation}
\begin{split}
\Heff = {4G_F \over \sqrt2} V_{cb} & \left[ 
(1+g_{V_L}) (\cbar_L \gamma_\mu b_L) (\lbar_L \gamma^\mu \nu_L) + 
g_{V_R} (\cbar_R \gamma_\mu b_R) (\lbar_L \gamma^\mu \nu_L) \right. \\
& + g_{S_L} (\cbar_R b_L) (\lbar_R \nu_L)
  + g_{S_R} (\cbar_L b_R) (\lbar_R \nu_L) \\
& + \left. g_{T_L} (\cbar_R \sigma_{\mu\nu} b_L) (\lbar_R \sigma^{\mu\nu} \nu_L) \right] + \mathrm{h.c.}\,,
\end{split}
\label{eq:Heff_Jp}
\end{equation}
which is obviously equivalent to Eq.~\eqref{eq:Heff}, with the corresponding effective coefficients related as
\begin{equation}
g_{V,\,A} = g_{V_R} \pm g_{V_L} \,, \quad\qquad
g_{S,\,P} = g_{S_R} \pm g_{S_L} \,, \quad\qquad
g_T = -g_{T5} = g_{T_L} \,.
\label{eq:coeff_relations}
\end{equation}
The last relation is also an easy way to see that $(\cbar_L \sigma_{\mu\nu} b_R) (\lbar_R \sigma^{\mu\nu} \nu_L)$, the right-handed tensor operator,  cannot contribute to the decay amplitude.

\subsection{$\bm{\Bbar\to D\ell\nubar}$ decay\label{sec:Dlnu}}

We first focus on the decay to a pseudoscalar meson and write the differential decay rate as
\begin{equation}
{d^2\Gamma \over dq^2d\cos\thl} = {\sqrt{\lambda_{BD}(q^2)} \over 64(2\pi)^3m_B^3}\left(1-\rlsq\right) \sum_{\lambda_\ell}|\M^{\lambda_\ell}(\Bbar\to D\ell\nubar)|^2 \,,
\label{eq:dG2_D}
\end{equation}
where we made use of the definition $\lambda_{BD}(q^2) = m_B^4 + m_D^4 + q^4 - 2(m_B^2 m_D^2 + m_B^2 q^2 + m_D^2 q^2)$. 
It is convenient to separate the angular from the $q^2$ dependence and write the above expression as 
\begin{equation}
{d^2\Gamma \over dq^2d\cos\thl} = a_\thl(q^2) + b_\thl(q^2) \cos\thl + c_\thl(q^2) \cos^2\thl \,,
\label{eq:dG2_thl_D}
\end{equation}
where $\thl$ is the polar angle of the lepton momentum in the rest frame of the $\ell\nubar$-pair with respect to the $z$ axis which is defined by the $D$-momentum in the rest frame of $\Bbar$. The corresponding angular coefficients can be written as
\begin{subequations}
\begin{align}
\begin{split}
a_\thl(q^2) =& {G_F^2 |V_{cb}|^2 \over 256\pi^3 m_B^3} q^2 \sqrt{\lambda_{BD}(q^2)} \left(1-\rlsq\right)^2 \biggl[ |\htil_0^-|^2 + \rlsq |\htil_t|^2 \biggr] \,,
\end{split} \\
& \nonumber \\
\begin{split}
b_\thl(q^2) =& {G_F^2 |V_{cb}|^2 \over 128\pi^3 m_B^3} q^2 \sqrt{\lambda_{BD}(q^2)} \left(1-\rlsq\right)^2 \rlsq \Re[\htil_0^+ \htil_t^*] \,,
\end{split} \\
& \nonumber \\
\begin{split}
c_\thl(q^2) =& -{G_F^2 |V_{cb}|^2 \over 256\pi^3 m_B^3} q^2 \sqrt{\lambda_{BD}(q^2)} \left(1-\rlsq\right)^2 \biggl[ |\htil_0^-|^2 - \rlsq |\htil_0^+|^2 \biggr] \,,
\end{split}
\end{align}
\label{eq:abc_thl_D}
\end{subequations}
where $\htil_\lambda^{\lambda_\ell}$ are the linear combinations of the ``{\it standard}'' hadronic helicity amplitudes which are defined and computed in Appendix~\ref{app:helicity_amp} of the present paper.

Integration over the polar angle $\thl$ leads to the expression for the differential decay rate,
\begin{equation}
\begin{split}
{d\Gamma \over dq^2} & = \, 2a_\thl(q^2)+{2\over3}c_\thl(q^2) \\
& = {G_F^2 |V_{cb}|^2 \over 192\pi^3 m_B^3} q^2 \sqrt{\lambda_{BD}(q^2)} \left(1-\rlsq\right)^2 \times\biggl\{ \biggr. 
|\htil_0^-|^2 + {m_\ell^2 \over 2q^2}|\htil_0^+|^2 + {3\over2}{m_\ell^2 \over q^2}|\htil_t|^2 \biggl.\biggr\} \,.
\end{split}
\label{eq:dGdq2_D}
\end{equation}
We see that the linear dependence on $\cos\thl$ in Eq.~\eqref{eq:dG2_thl_D} is lost after integration in $\thl$, but it can be recovered by considering  the forward-backward asymmetry,
\begin{equation}
\AFB(q^2) = {\displaystyle { \int_0^1 {d^2\Gamma \over dq^2d\cos\thl} d\cos\thl - \int_{-1}^0 {d^2\Gamma \over dq^2d\cos\thl} d\cos\thl } \over d\Gamma/dq^2 } = { b_\thl(q^2) \over {d\Gamma/dq^2} } \,.
\label{eq:AFB_D}
\end{equation}

Another interesting observable for the study of the NP effects is the lepton polarization asymmetry defined from differential decay rates with definite lepton helicity:
\begin{subequations}
\begin{align}
\begin{split}
{d\Gamma^{\lambda_\ell=+1/2} \over dq^2} =& {G_F^2 |V_{cb}|^2 \over 192\pi^3 m_B^3} q^2 \sqrt{\lambda_{BD}(q^2)} \left(1-\rlsq\right)^2 {m_\ell^2 \over 2q^2} \biggl[ |\htil_0^+|^2 + 3|\htil_t|^2 \biggr] \,,
\end{split} \\
& \nonumber \\
\begin{split}
{d\Gamma^{\lambda_\ell=-1/2} \over dq^2} =& {G_F^2 |V_{cb}|^2 \over 192\pi^3 m_B^3} q^2 \sqrt{\lambda_{BD}(q^2)} \left(1-\rlsq\right)^2 |\htil_0^-|^2 \,.
\end{split}
\end{align}
\label{eq:dGdq2_lambdal_D}
\end{subequations}
The corresponding polarization asymmetry reads
\begin{equation}
\A_{\lambda_\ell}(q^2)  = { d\Gamma^{\lambda_\ell=-1/2}/dq^2 - d\Gamma^{\lambda_\ell=+1/2}/dq^2 \over d\Gamma/dq^2 } =  1-2{ d\Gamma^{\lambda_\ell=+1/2}/dq^2 \over d\Gamma/dq^2} \,.
\label{eq:A_lambdal_D}
\end{equation}

\subsection{$\bm{\Bbar\to\Dst(\to D\pi)\ell\nubar}$ decay}

\begin{figure}[t!]
\begin{center}
{\resizebox{11.4cm}{!}{\includegraphics{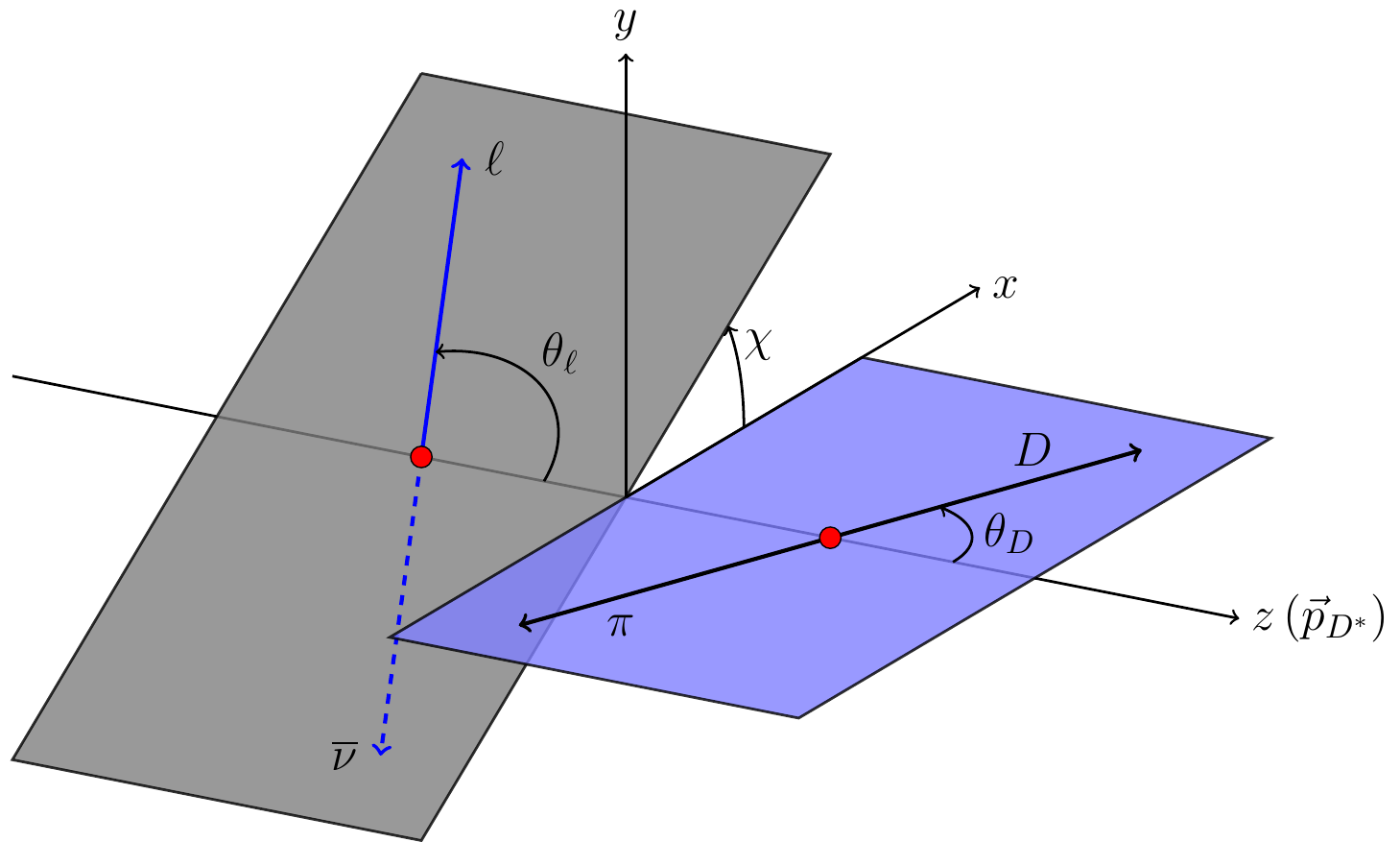}}} 
\caption{\footnotesize{\sl Kinematics of the $\Bbar\to\Dst(\to D\pi)\ell\nubar$ decay.}}
\label{fig:kinematics}
\end{center}
\end{figure}

We now discuss the case of the $\bar B$-meson decaying to the vector meson in the final state. In this work, for the computation of the helicity amplitudes and differential decay rates, we define the system of coordinates as depicted in Fig.~\ref{fig:kinematics}: the $z$-axis is set along the $\Dst$ momentum in the $\Bbar$ rest frame, and the $x$-axis is chosen in a way that the $D$ momentum in the $\Dst$ rest frame lies in the $x-z$ plane and has the positive $x$-component.

The full angular distribution reads,
\begin{equation}
{d^5\Gamma \over dq^2 dm_{D\pi}^2 d\cos\thD d\cos\thl d\chi} = {\sqrt{\lambda_{B\Dst}(q^2)} \over 256(2\pi)^6 m_B^3} \left(1-\rlsq\right) {|\p_D| \over m_{D\pi}} \sum_{\lambda_\ell} \biggl| \sum_{\lambda_\Dst} \M^{\lambda_\Dst,\,\lambda_\ell} (\Bbar\to D\pi\ell\nubar_\ell) \biggr|^2 \,,
\label{eq:dG5_Dst}
\end{equation}
where again $\lambda_{B\Dst}(q^2) = m_B^4 + m_\Dst^4 + q^4 - 2(m_B^2 m_\Dst^2 + m_B^2 q^2 + m_\Dst^2 q^2)$, and 
\bea
 |\p_D| = {\sqrt{\lambda(m_{D\pi}^2,m_D^2,m_\pi^2)} \over 2m_{D\pi}}\ .
\eea
After integrating over $m_{D\pi}^2$ around the ${D^\ast}$-resonance, the above distribution becomes
\begin{equation}
\begin{split}
{d^4\Gamma \over dq^2 d\cos\thD d\cos\thl d\chi} = {9 \over 32\pi} \biggl\{ \biggr. & I_{1c}\cos^2\thD + I_{1s}\sin^2\thD \\
&+ \bigl[ I_{2c}\cos^2\thD + I_{2s}\sin^2\thD \bigr] \cos2\thl \\
&+ \bigl[ I_{6c}\cos^2\thD + I_{6s}\sin^2\thD \bigr] \cos\thl \\
&+ \bigl[ I_3\cos2\chi + I_9\sin2\chi \bigr] \sin^2\thl \sin^2\thD \\
&+ \bigl[ I_4\cos\chi + I_8\sin\chi \bigr] \sin2\thl \sin2\thD \\
&+ \bigl[ I_5\cos\chi + I_7\sin\chi \bigr] \sin\thl \sin2\thD \biggl.\biggr\} \,,
\label{eq:fullangdist}
\end{split}
\end{equation}
where the angular coefficients $I_i\equiv I_i(q^2)$ are given by:
\begin{subequations}
\begin{align}
I_{1c} &= 2N \biggl[ |\Htil_0^-|^2 +  \rlsq |\Htil_0^+|^2 + 2\rlsq |\Htil_t|^2 \biggr] \,, \\
& \nonumber \\
I_{1s} &= {N\over2} \biggl[ 3\bigl( |\Htil_+^-|^2 + |\Htil_-^-|^2 \bigr) + \rlsq \bigl( |\Htil_+^+|^2 + |\Htil_-^+|^2 \bigr) \biggr] \,, \\
& \nonumber \\
I_{2c} &= 2N \biggl[ -|\Htil_0^-|^2 + \rlsq |\Htil_0^+|^2 \biggr] \,, \\
& \nonumber \\
I_{2s} &= {N\over2} \biggr[ |\Htil_+^-|^2 + |\Htil_-^-|^2 - \rlsq \bigl( |\Htil_+^+|^2 + |\Htil_-^+|^2 \bigr) \biggr] \,, \\
& \nonumber \\
I_3 &= -2N \, \Re\biggl[ \Htil_+^- \Htil_-^{-*} - \rlsq \Htil_+^+ \Htil_-^{+*} \biggr] = -2N \beta_\ell^2 \, \Re \bigl[ H_+ H_-^* - 4 H_{T,+} H_{T,-}^* \bigr] \,, \\
& \nonumber \\
\begin{split}
I_4 &= N \, \Re\biggl[ (\Htil_+^- + \Htil_-^-) \Htil_0^{-*} - \rlsq (\Htil_+^+ + \Htil_-^+) \Htil_0^{+*} \biggr] \\
&= N \beta_\ell^2 \, \Re\bigl[ (H_+ + H_-) H_0^{*} - 4(H_{T,+} + H_{T,-}) H_{T,0}^* \bigr]\,,
\end{split} \\
& \nonumber \\
I_5 &= 2N \, \Re\biggl[ (\Htil_+^- - \Htil_-^-) \Htil_0^{-*} - \rlsq (\Htil_+^+ + \Htil_-^+) \Htil_t^* \biggr] \,, \\
& \nonumber \\
I_{6c} &= 8N \rlsq \, \Re\bigl[ \Htil_0^+ \Htil_t^* \bigr] \,, \\
& \nonumber \\
I_{6s} &= 2N \bigl( |\Htil_+^-|^2 - |\Htil_-^-|^2 \bigr) \,, \\
& \nonumber \\
I_7 &= 2N \, \Im\biggl[ (\Htil_+^- + \Htil_-^-) \Htil_0^{-*} - \rlsq (\Htil_+^+ - \Htil_-^+) \Htil_t^* \biggr] \,, \\
& \nonumber \\
\begin{split}
I_8 &= N \, \Im\biggl[ (\Htil_+^- - \Htil_-^-) \Htil_0^{-*} - \rlsq (\Htil_+^+ - \Htil_-^+) \Htil_0^{+*} \biggr] \\
&= N \beta_\ell^2 \, \Im\bigl[ (H_+ - H_-) H_0^* - 4(H_{T,+} - H_{T,-}) H_{T,0}^* \bigr]\,,
\end{split} \\
& \nonumber \\
I_9 &= -2N \, \Im\biggl[ \Htil_+^- \Htil_-^{-*} - \rlsq \Htil_+^+ \Htil_-^{+*} \biggr] = -2N \beta_\ell^2 \, \Im \bigl[ H_+ H_-^* - 4 H_{T,+} H_{T,-}^* \bigr] \,,
\end{align}
\label{eq:Is}
\end{subequations}
with
\begin{equation}
N\equiv N(q^2) = \B(\Dst\to D\pi) {G_F^2 |V_{cb}|^2 \over 48(2\pi)^3 m_B^3} q^2 \sqrt{\lambda_{B\Dst}(q^2)} \left(1-\rlsq\right)^2 \,.
\end{equation}
The explicit expressions for the hadronic helicity amplitudes $H_\lambda$, as well as their linear combinations $\Htil_\lambda^{\lambda_\ell}$, are given in Appendix~\ref{app:helicity_amp}. 

It is now possible to write down three partially integrated decay distributions, integrating all but one 
angle at a time. 
\begin{itemize}
\item $\thl$ distribution :
\begin{equation}
\begin{split}
{d^2\Gamma \over dq^2 d\cos\thl} &= a_\thl(q^2) + b_\thl(q^2)\cos\thl + c_\thl(q^2)\cos^2\thl \,, \\
&a_\thl(q^2) = {3\over8} \left( I_{1c} + 2I_{1s} - I_{2c} - 2I_{2s} \right) \,, \\
&b_\thl(q^2)  = {3\over8} \left( I_{6c} + 2I_{6s} \right) \,, \\
&c_\thl(q^2)  = {3\over4} \left( I_{2c} + 2I_{2s} \right) \,.
\end{split}
\end{equation}
\item $\thD$ distribution :
\begin{equation}
\begin{split}
{d^2\Gamma \over dq^2 d\cos\thD} &= a_\thD(q^2) + c_\thD(q^2)\cos^2\thD \,, \\
&a_\thD(q^2) = {3\over8} \left( 3I_{1s} - I_{2s} \right) \,, \\
&c_\thD(q^2) = {3\over8} \left(3I_{1c} - 3I_{1s} - I_{2c} + I_{2s} \right) \,.
\end{split}
\end{equation}
\item $\chi$ distribution :
\begin{equation}
\begin{split}
{d^2\Gamma \over dq^2 d\chi} &= a_\chi(q^2) + c_\chi^c(q^2)\cos2\chi + c_\chi^s(q^2)\sin2\chi \,, \\
&a_\chi(q^2) = {1\over8\pi} \left( 3I_{1c} + 6I_{1s} - I_{2c} - 2I_{2s} \right)  \\
&c_\chi^c(q^2) = {1\over2\pi} I_3 \,, \\
&c_\chi^s(q^2) = {1\over2\pi} I_9 \,.
\end{split}
\end{equation}
\end{itemize}
Illustration of the $q^2$-dependence of the above observables, for fixed values of $g_i$'s, is provided in Fig.~\ref{fig:angular_plots_Dst}.

It is interesting to point out that the terms proportional to $\cos\thD$, $\cos\chi$ and $\sin\chi$ are absent in the distributions written above. Such terms can arise if a non-zero interference with the $S$-wave 
contribution, $\Bbar\to(D\pi)_S\ell\nubar$, is included in the distribution, as discussed in Ref.~\cite{Becirevic:2016hea}.

\subsubsection{Observables}

From the full angular distribution~\eqref{eq:fullangdist} we isolate various coefficients and combine them into various quantities normalized to the differential decay rate. In the following we define $12$ such quantities with a goal to use them in order to scrutinize the effects of LFUV that 
we propose to study. When necessary, before defining an observable, we will indicate how the corresponding coefficients can be isolated from the full angular distribution.

\begin{itemize}
\item \underline{Differential decay rate}
\begin{equation}
{d\Gamma \over dq^2} = {1\over4} \left( 3I_{1c} + 6I_{1s} - I_{2c} - 2I_{2s} \right) \,,
\label{eq:dGdq2_Dst}
\end{equation}

\item \underline{Forward-backward asymmetry}
\begin{equation}
\AFB(q^2) = { b_\thl(q^2) \over d\Gamma/dq^2 } = {3\over8} {\left( I_{6c}+2I_{6s} \right) \over d\Gamma/dq^2} \,, \\
\label{eq:AFB_Dst}
\end{equation}

\item \underline{Lepton polarization asymmetry}
\begin{equation}
\A_{\lambda_\ell}(q^2) = {d\Gamma^{\lambda_\ell=-1/2}/dq^2 - d\Gamma^{\lambda_\ell=+1/2}/dq^2 \over d\Gamma/dq^2} \,,
\label{eq:A_lambdal_Dst}
\end{equation}
with
\begin{equation}
\begin{split}
{d\Gamma^{\lambda_\ell=+1/2} \over dq^2} =& {1\over4} \left( I_{1c} + 2I_{1s} + I_{2c} - 6I_{2s} \right) + I_n \\
{d\Gamma^{\lambda_\ell=-1/2} \over dq^2} =& {1\over2} \left( I_{1c} + 2I_{1s} - I_{2c} + 2I_{2s} \right) - I_n \,,
\end{split}
\end{equation}
where we introduced an additional coefficient
\begin{equation}
I_n = 2N \rlsq |\Htil_t|^2 \,,
\end{equation}
which is not present in Eq.~\eqref{eq:fullangdist} because in the full angular distribution we have summed over the lepton polarization states. 
\item \underline{$\Dst$ polarization fraction}
\begin{equation}
R_{L,T}(q^2) = {d\Gamma_L/ dq^2 \over d\Gamma_T/dq^2} \,,
\label{eq:RLT}
\end{equation}
where $\Gamma_L$ and $\Gamma_T$ represent the longitudinal and transverse $\Dst$ polarization decay rates,
\begin{equation}
\begin{split}
{d\Gamma_L \over dq^2} &= {2\over3} \left[ a_\thD (q^2) + c_\thD (q^2) \right] = {1\over4} \left( 3I_{1c} - I_{2c} \right) \,, \\
{d\Gamma_T \over dq^2} &= {4\over3} a_\thD (q^2) = {1\over2} \left( 3I_{1s} - I_{2s} \right) \,.
\end{split}
\end{equation}
Alternatively, one can define the quantity $F^{D^\ast}_L$ which is a measure of the longitudinally polarized $D^\ast$'s in the whole ensemble of ${\Bbar\to\Dst \ell\nubar}$ decays, which is related to  
$R_{L,T}(q^2)$ as:
\bea\label{eq:FLDstar}
F_L^{D^\ast}(q^2)= \frac{R_{L,T}(q^2)}{1+ R_{L,T}(q^2)} \,=\, \frac{1}{2}\,\frac{3 I_{1c}-I_{2c}}{3\left( I_{1c} + I_{1s}\right) - I_{2c}- I_{2s}}\,.
\eea
$F_L^{D^\ast}$ is often referred to as $F_L^{D^\ast}(q^2)$ integrated over the available phase space.
\item \underline{$R_{A,B}$}
\begin{equation}
R_{A,B}(q^2) = {d\Gamma_A/ dq^2 \over d\Gamma_B/ dq^2} \,,
\label{eq:RAB}
\end{equation}
\begin{equation}
\begin{split}
{d\Gamma_A \over dq^2} &= {2\over3} \left[ a_\thl (q^2) - c_\thl (q^2) \right] = {1\over4} \left( I_{1c} + 2I_{1s} - 3I_{2c} - 6I_{2s} \right) \,, \\
{d\Gamma_B \over dq^2} &= {4\over3} \left[ a_\thl (q^2) + c_\thl (q^2) \right] = {1\over2} \left( I_{1c} + 2I_{1s} + I_{2c} + 2 I_{2s} \right) = {d\Gamma \over dq^2} - {d\Gamma_A \over dq^2} \,.
\end{split}
\end{equation}

\item \underline{$A_3$ and $A_9$}
\begin{equation}
\begin{split}
A_3(q^2) &= {c_\chi^c(q^2) \over d\Gamma/dq^2} = {1\over2\pi} {I_3 \over d\Gamma/dq^2} \,, \\
A_9(q^2) &= {c_\chi^s(q^2) \over d\Gamma/dq^2} = {1\over2\pi} {I_9 \over d\Gamma/dq^2} \,.
\end{split}
\label{eq:A3A9}
\end{equation}

\item \underline{$A_4$ and $A_8$}\\
If, from the full angular distribution, we first define the auxiliary quantities:
\begin{equation}
\begin{split}
\Phi_{48}(q^2,\chi,\thl) &= \left[ \int_{-1}^0  - \int_{0}^1 \right] {d^4\Gamma \over dq^2d\chi d\cos\thl d\cos\thD} \ d\cos\thD \\
\widetilde\Phi_{48}(q^2,\chi) &= \left[ \int_{-1}^0  - \int_{0}^1 \right] \Phi_{48}(q^2,\chi,\thl) \ d\cos\thl \,,
\end{split}
\end{equation}
then we can extract another two observables as,
\begin{equation}
\begin{split}
A_4(q^2) &= {\displaystyle{\left[ \int_{\pi/2}^{3\pi/2}-\int_{0}^{\pi/2} -\int_{3\pi/2 }^{2\pi}  \right] \widetilde\Phi_{48}(q^2, \chi) \ d\chi} \over d\Gamma/dq^2} = -{2\over\pi} {I_4 \over d\Gamma/dq^2} \,. \\
A_8(q^2) &= {\displaystyle{\left[ \int_{0}^{\pi} - \int_{\pi}^{2\pi} \right] \widetilde\Phi_{48}(q^2, \chi) \ d\chi} \over d\Gamma/dq^2 } = {2\over\pi} {I_8 \over d\Gamma/dq^2} \,.
\end{split}
\label{eq:A4A8}
\end{equation}

\item \underline{$A_5$ and $A_7$}\\
Similarly, if one first defines the asymmetry of the full angular distribution with respect to $\theta_D$, namely, 
\begin{equation}
\Phi_{57}(q^2,\chi) = 
\left[ \int_{-1}^0 - \int_{0}^1 \right] {d^3\Gamma \over dq^2 d\chi d\cos\thD} \ d\cos\thD \,,
\end{equation}
one can define two additional observables as follows:
\begin{equation}
\begin{split}
A_5(q^2) &= -{\displaystyle{\left[ \int_{\pi/2}^{3\pi/2}-\int_{0}^{\pi/2} - \int_{3\pi/2 }^{2\pi} \right] \Phi_{57}(q^2,\chi) \ d\chi} \over d\Gamma/dq^2} = -{3\over4} {I_5 \over d\Gamma/dq^2} \,, \\
A_7(q^2) &= {\displaystyle{\left[ \int_{0}^{\pi} - \int_{\pi}^{2\pi} \right] \Phi_{57}(q^2,\chi) \ d\chi} \over d\Gamma/dq^2} = -{3\over4} {I_7 \over d\Gamma/dq^2} \,.
\end{split}
\label{eq:A5A7}
\end{equation}

\item \underline{$A_{6s}$}\\
Finally from the asymmetry with respect to $\theta_\ell$.
\begin{equation}
\Phi_6(q^2,\thD) = \left[ \int_{-1}^0 - \int_{0}^1 \right] {d^3\Gamma \over dq^2 d\cos\thD d\cos\thl} \ d\cos\thl \,,
\end{equation}
we can isolate a term proportional to $I_{6s}$ as
\begin{equation}
A_{6s}(q^2) = {\displaystyle{\left[ 7 \int_{-1/2}^{1/2}-\int_{1/2}^{1} - \int_{-1}^{-1/2}\right] \Phi_6(q^2,\thD) \ d\cos\thD} \over d\Gamma/dq^2} = -{27\over8} {I_{6s} \over d\Gamma/dq^2} \,.
\label{eq:A6s}
\end{equation}
\end{itemize}
In the above definitions we played with Eq.~\eqref{eq:fullangdist} to isolate each of the angular coefficients $I_i\equiv I_i(q^2)$. Alternatively, with a large enough sample one can fit the full data set to 
Eq.~\eqref{eq:fullangdist} and extract each of the coefficients with respect to the full (differential) decay rate. The SM values of all
\begin{equation}
\label{eq:defAVI}
\langle I_i \rangle_\ell \, = \, \frac{1}{\Gamma(B\to D^\ast \ell\bar \nu_\ell)} \times\int\displaylimits_{m_\ell^2}^{(m_B-m_{D^\ast})^2} I_i^\ell(q^2) \, dq^2 \,
\end{equation}
for each of the leptons in the final state, are given in Tab.~\ref{tab:Is}.
\begin{table}[htbp!]
\renewcommand{\arraystretch}{1.7}
\centering
\scalebox{0.73}{
\begin{tabular}{|c|cccccccccccc|}
\hline 
$\ell$ 
& $\langle I_{1c} \rangle_\ell $ 
& $\langle I_{1s} \rangle_\ell $
& $\langle I_{2c} \rangle_\ell $
& $\langle I_{2s} \rangle_\ell $
& $\langle I_{3} \rangle_\ell $
& $\langle I_{4} \rangle_\ell $
& $\langle I_{5} \rangle_\ell $
& $\langle I_{6c} \rangle_\ell $
& $\langle I_{6s} \rangle_\ell $
& $\langle I_{7} \rangle_\ell $
& $\langle I_{8} \rangle_\ell $
& $\langle I_{9} \rangle_\ell $ \\ \hline\hline
$e$ 
& 0.521(2)
& 0.359(2)
& -0.520(2)
& 0.120(1)
& -0.170(3)
& -0.304(1)
& 0.26(1)
& 0
& -0.32(1)
& 0
& 0
& 0
 \\[0.3em]
    \hline
$\mu$ 
& 0.524(2)
& 0.359(2)
& -0.510(2)
& 0.119(1)
& -0.170(3)
& -0.302(1)
& 0.26(1)
& 0.014(1)
& -0.32(1)
& 0
& 0
& 0
 \\[0.3em]
    \hline
$\tau$ 
& 0.56(1)
& 0.379(4)
& -0.166(2)
& 0.064(1)
& -0.105(1)
& -0.138(1)
& 0.299(5)
& 0.36(2)
& -0.26(1)
& 0
& 0
& 0
 \\[0.3em]
    \hline
\end{tabular} 
}
\caption{ \sl \small  Standard Model values of the coefficients appearing in the angular distribution~\eqref{eq:fullangdist}, integrated over the full available phase space, as indicated in Eq.~\eqref{eq:defAVI}.  The values are obtained by using the form factors that are in the text referred to as CLN+HQET.
}
\label{tab:Is}
\end{table}
 
\begin{figure}[p!]
\begin{center}
\includegraphics[width=0.32\linewidth]{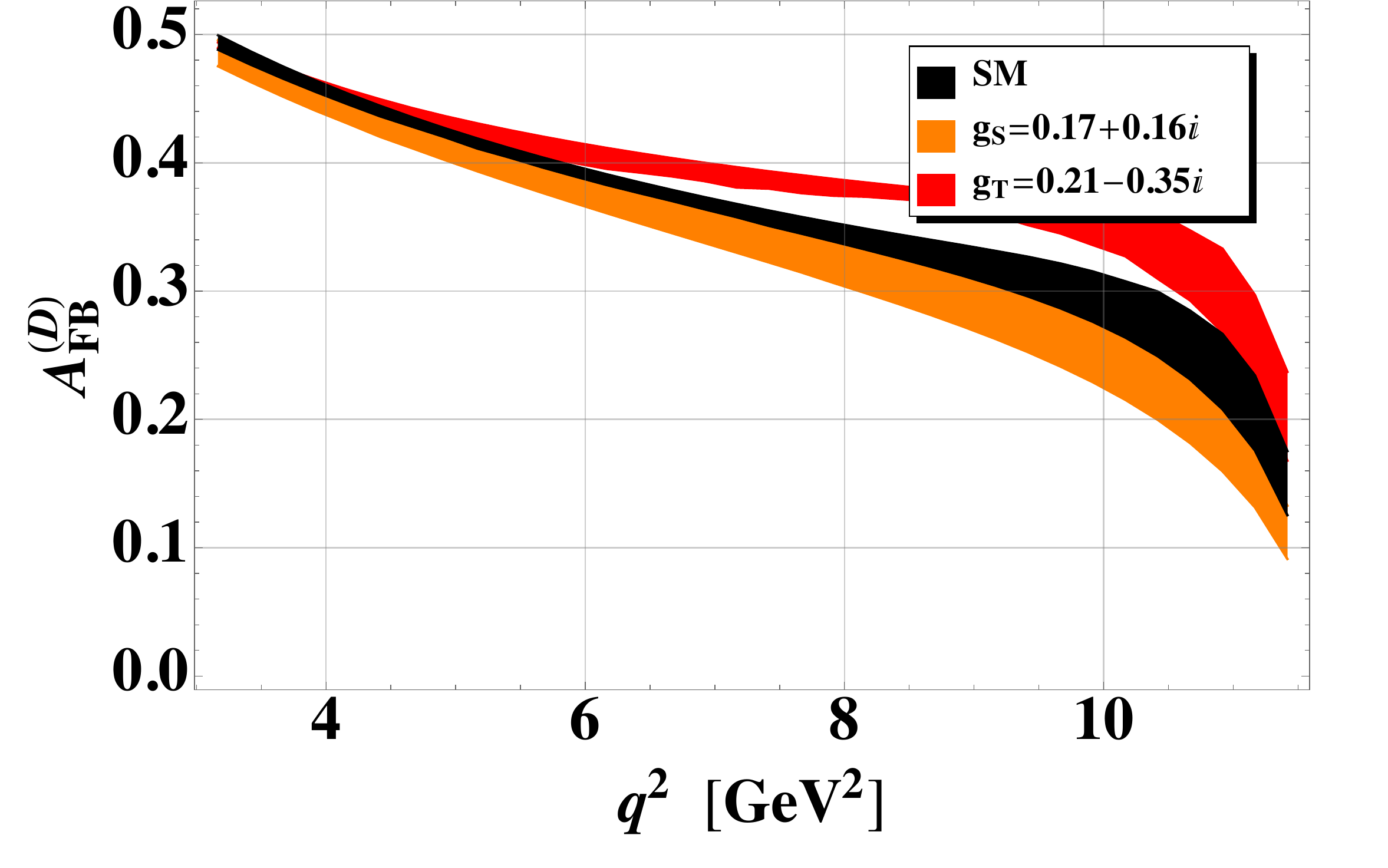}
\includegraphics[width=0.32\linewidth]{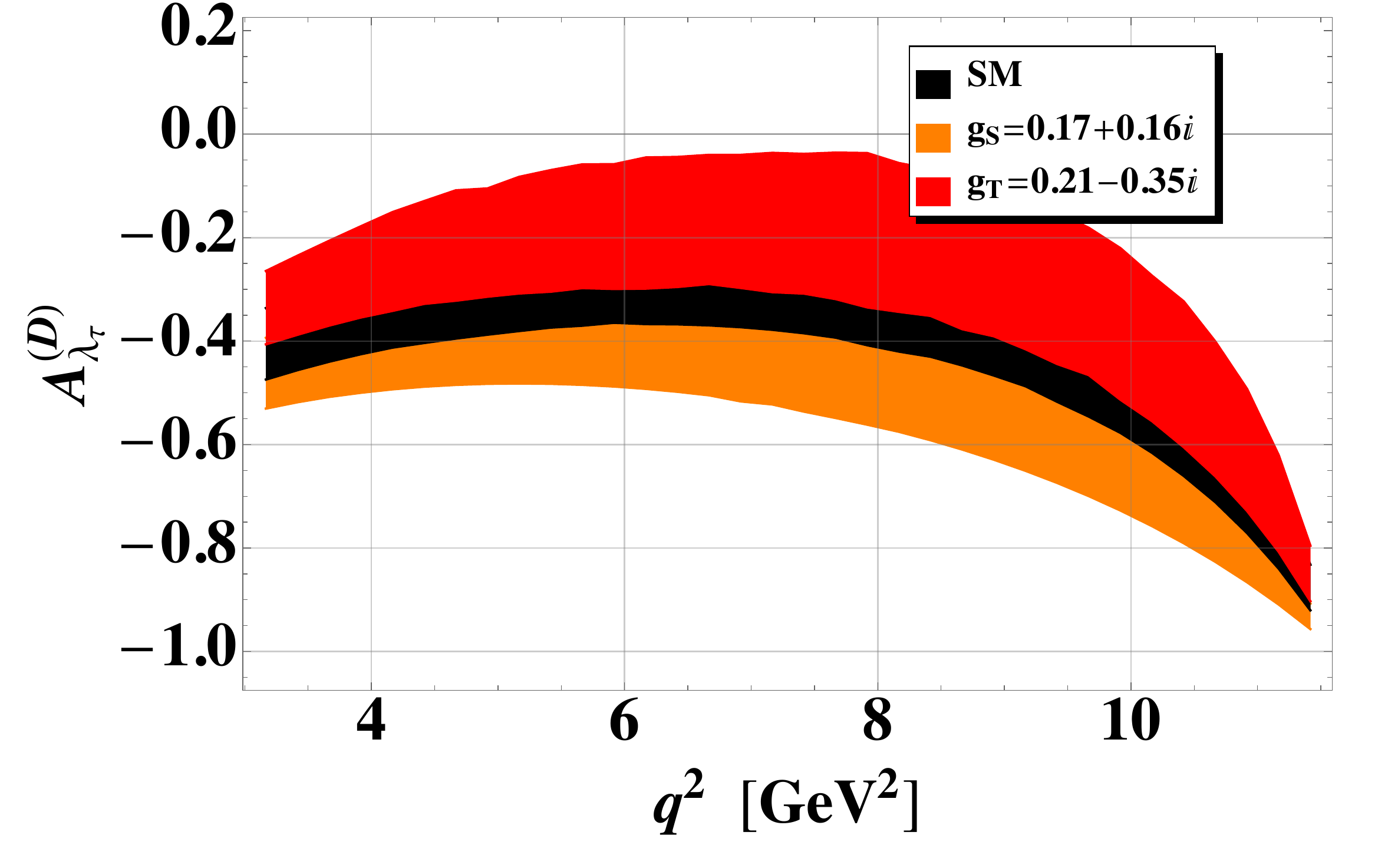}
\\ \vspace{5mm}
\includegraphics[width=0.32\linewidth]{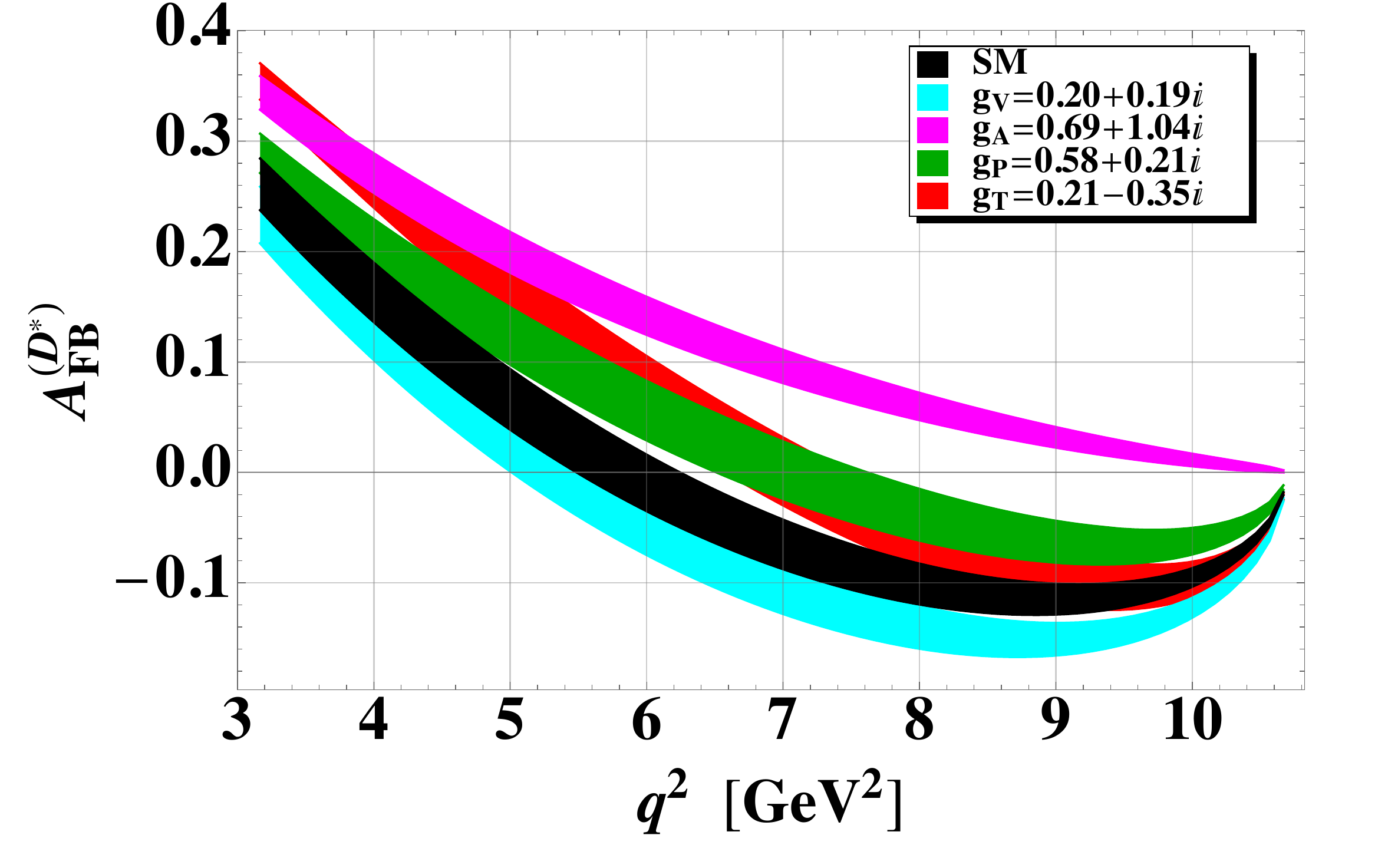}
\includegraphics[width=0.32\linewidth]{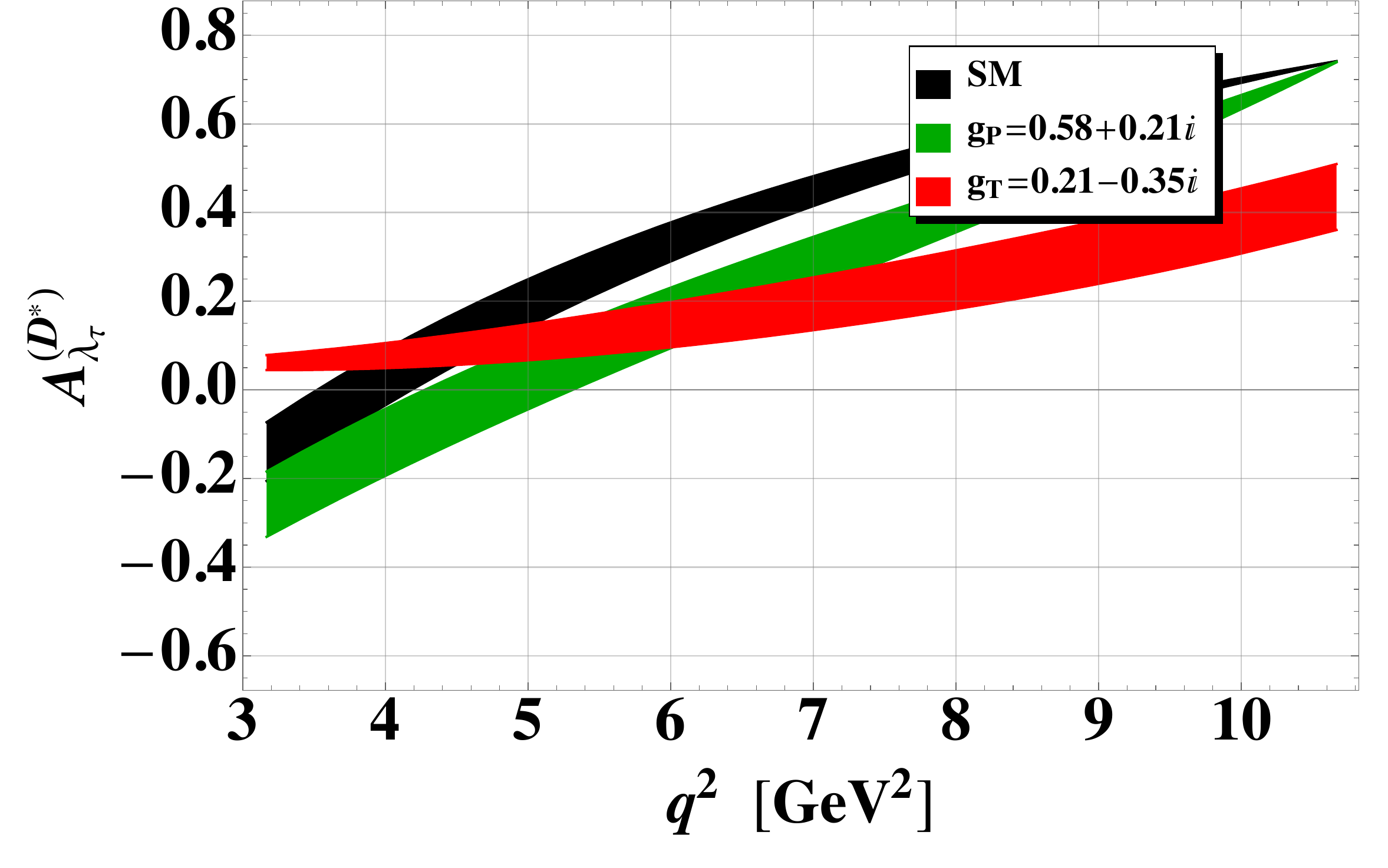}
\includegraphics[width=0.32\linewidth]{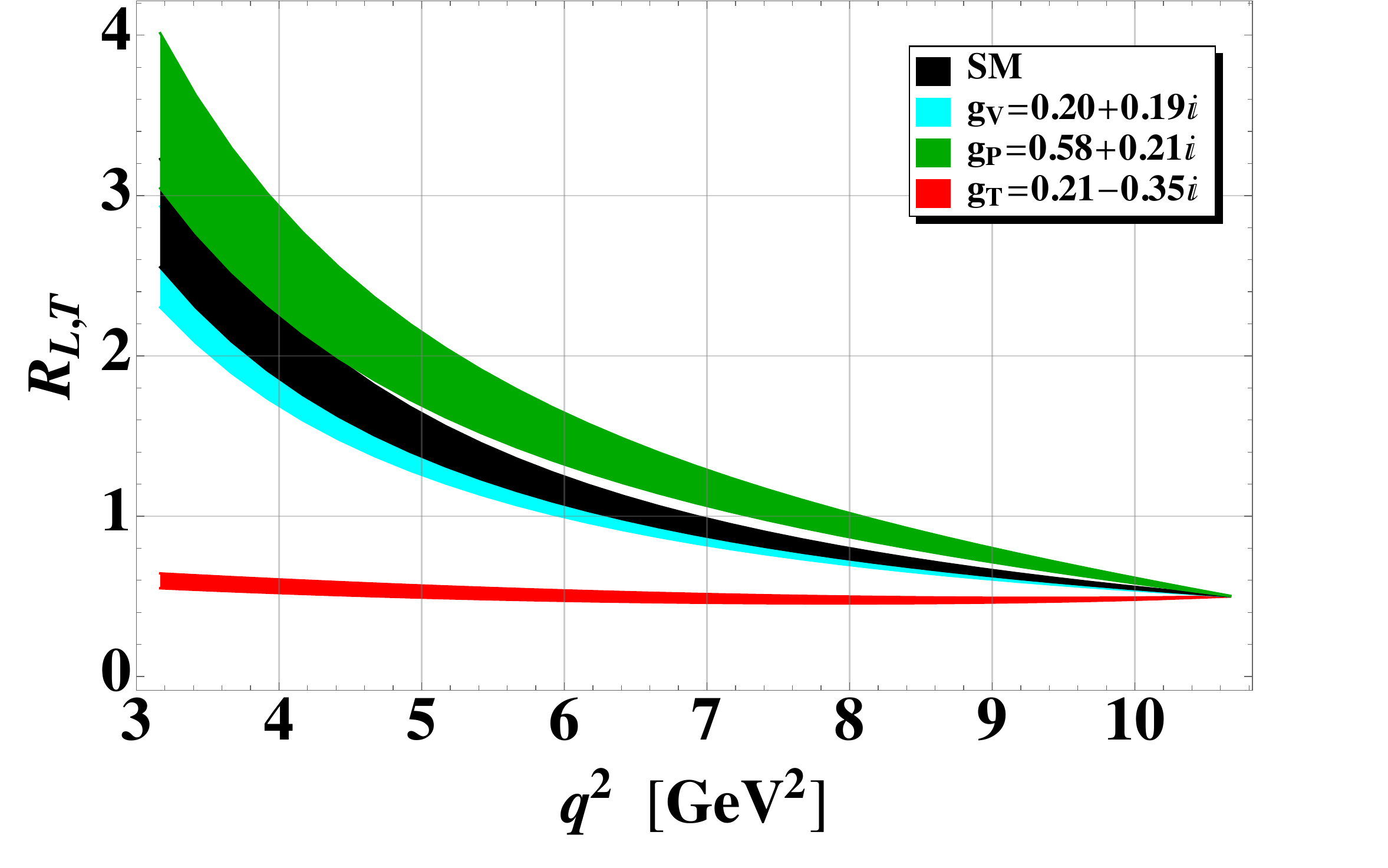}
\\ \vspace{5mm}
\includegraphics[width=0.32\linewidth]{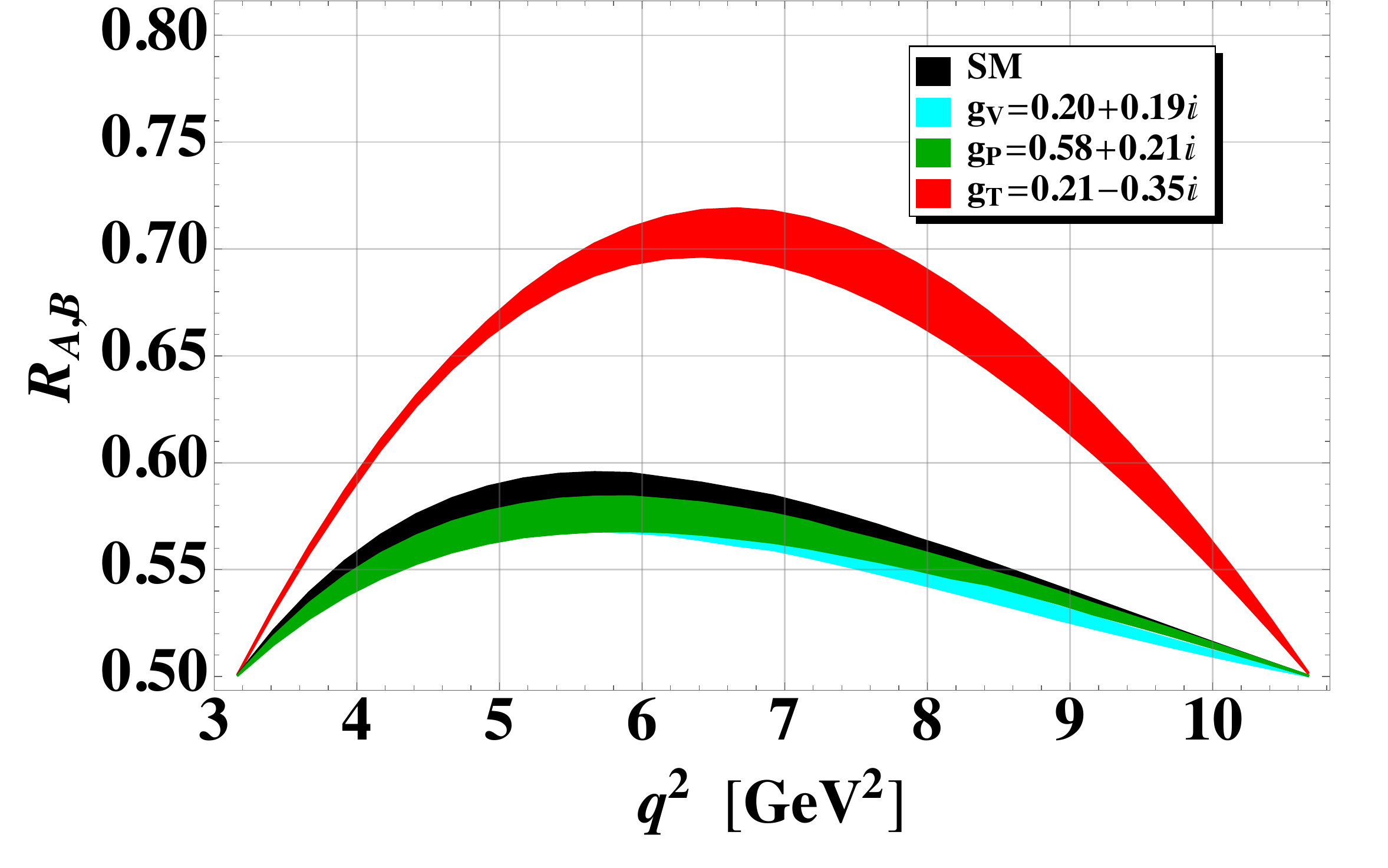}
\includegraphics[width=0.32\linewidth]{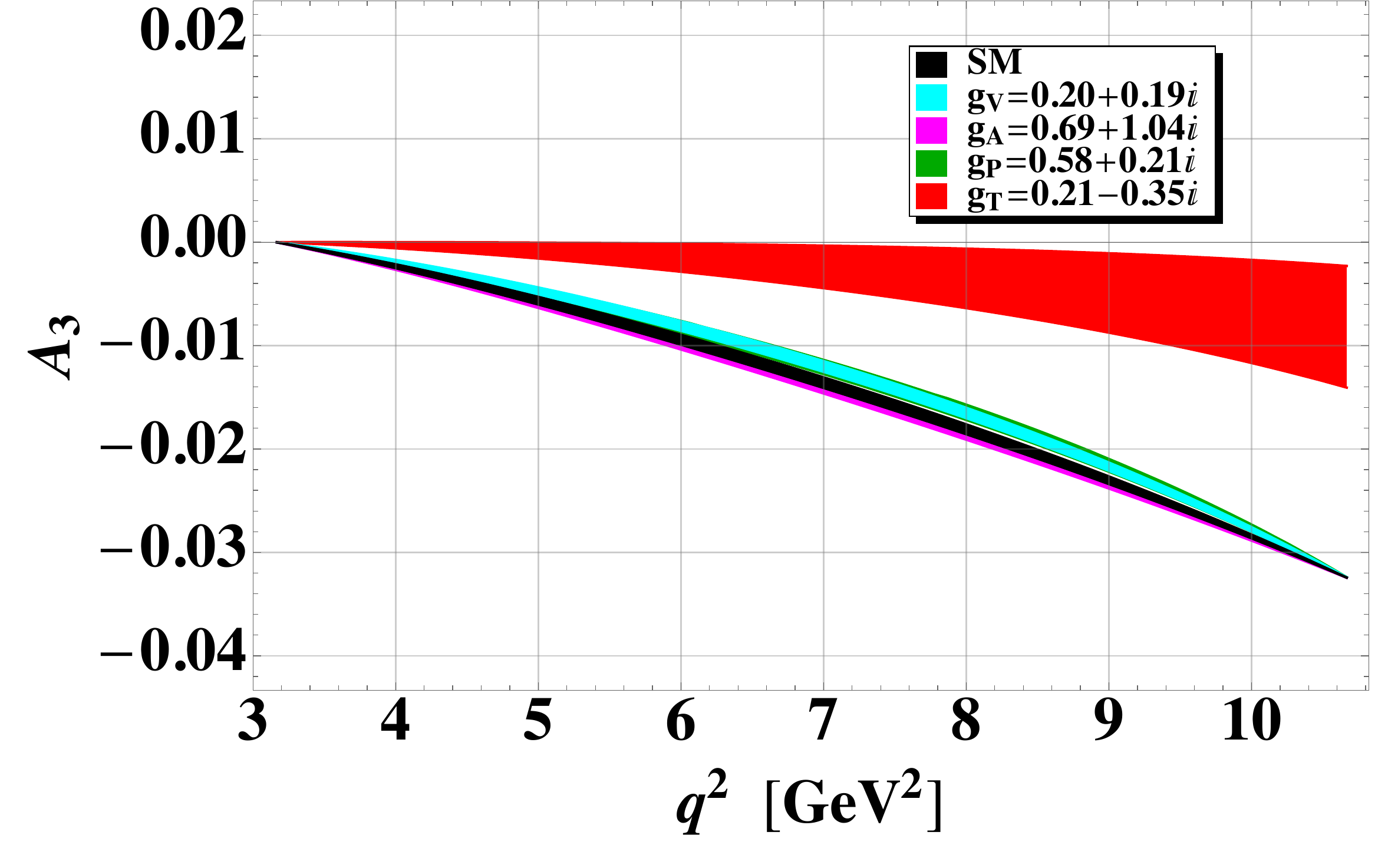}
\includegraphics[width=0.32\linewidth]{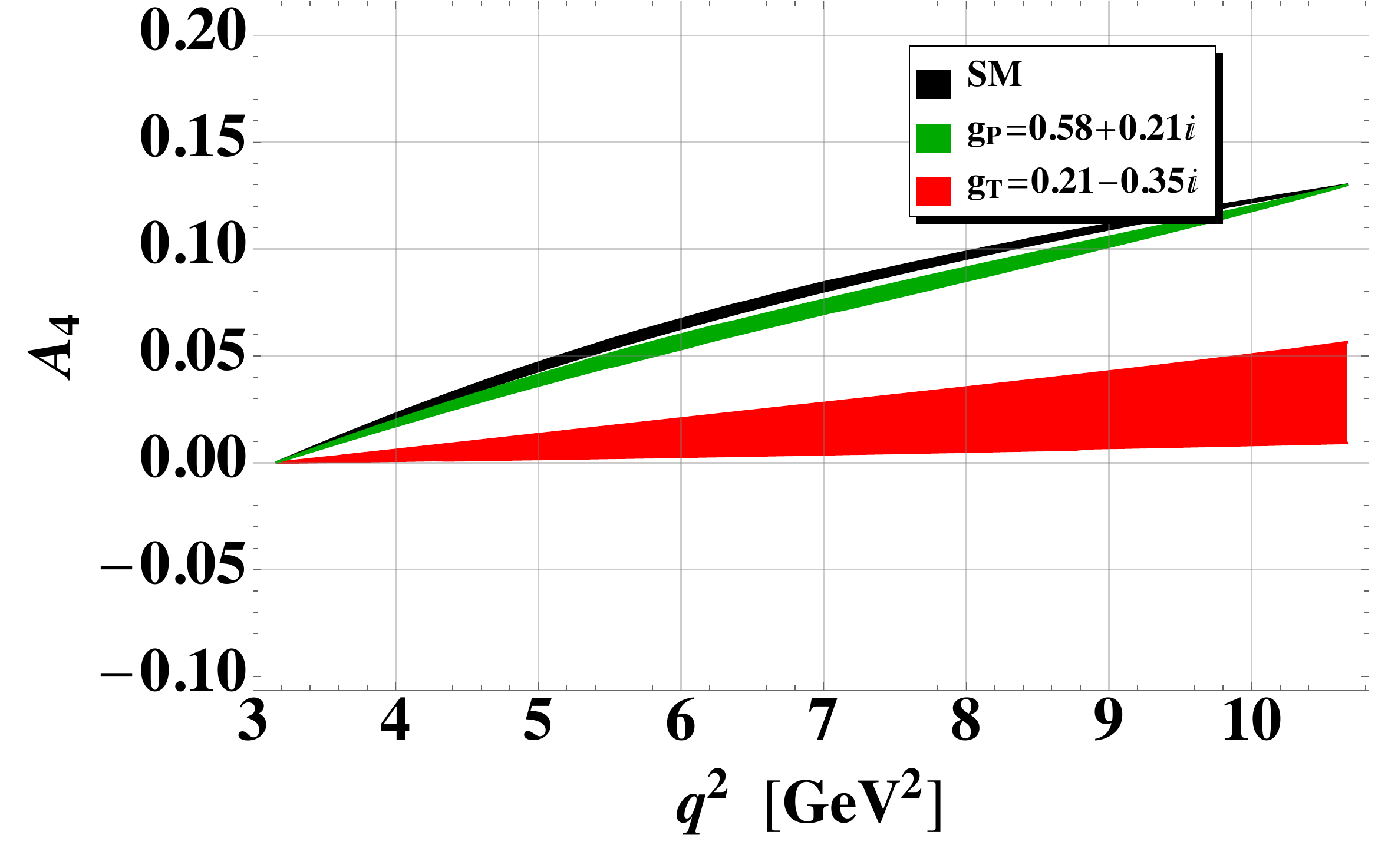}
\\ \vspace{5mm}
\includegraphics[width=0.32\linewidth]{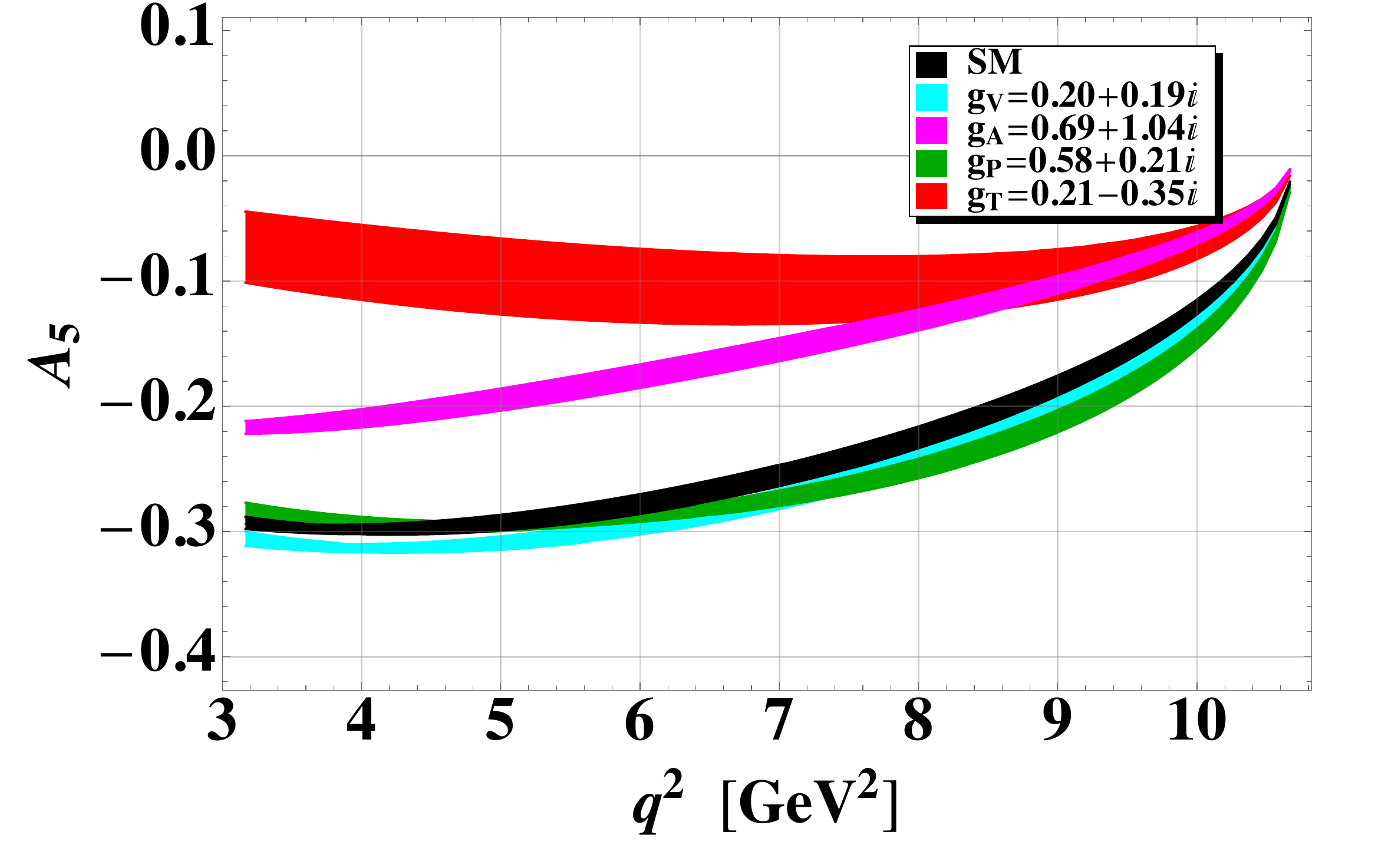}
\includegraphics[width=0.32\linewidth]{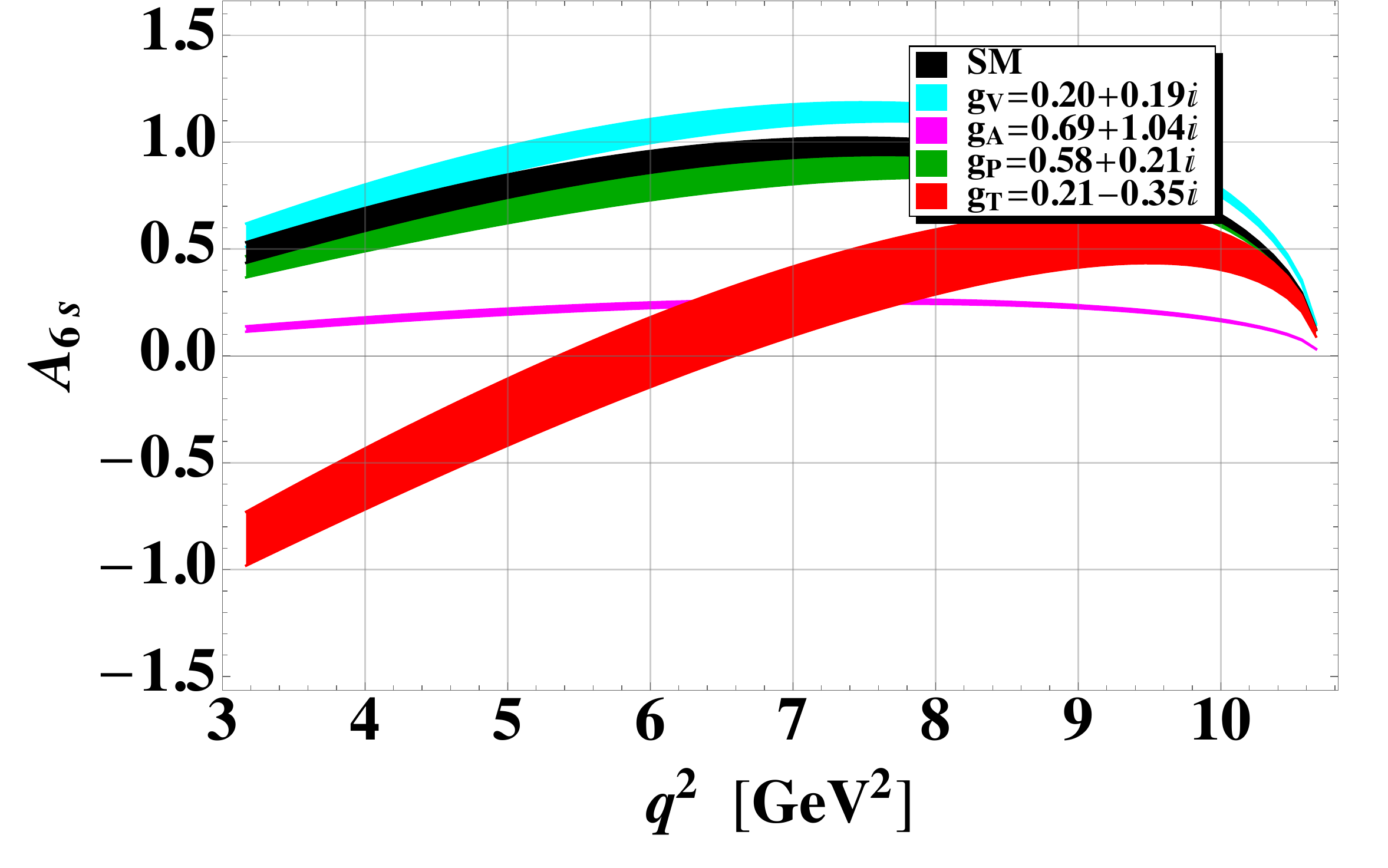}
\includegraphics[width=0.32\linewidth]{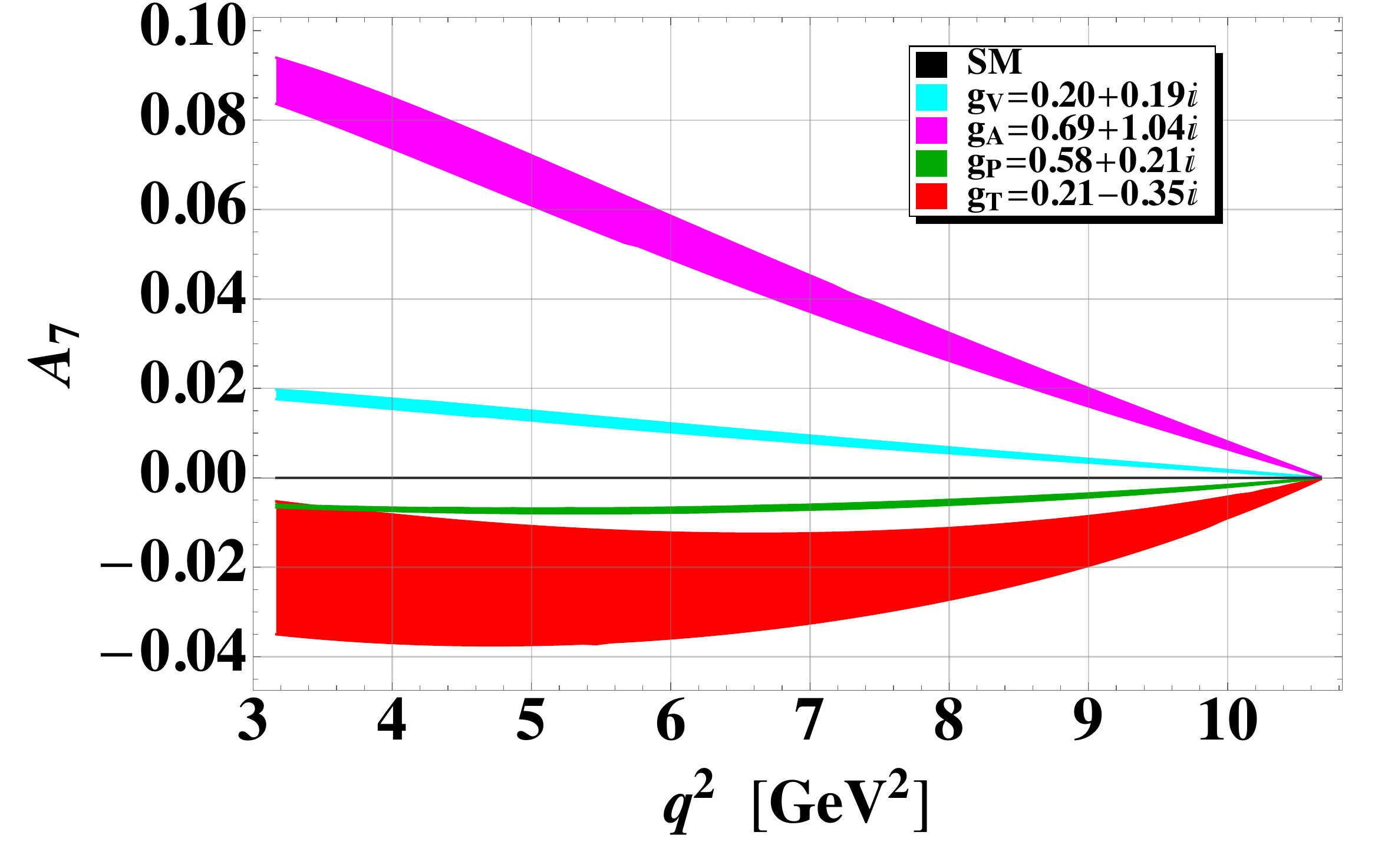}
\\ \vspace{5mm}
\includegraphics[width=0.32\linewidth]{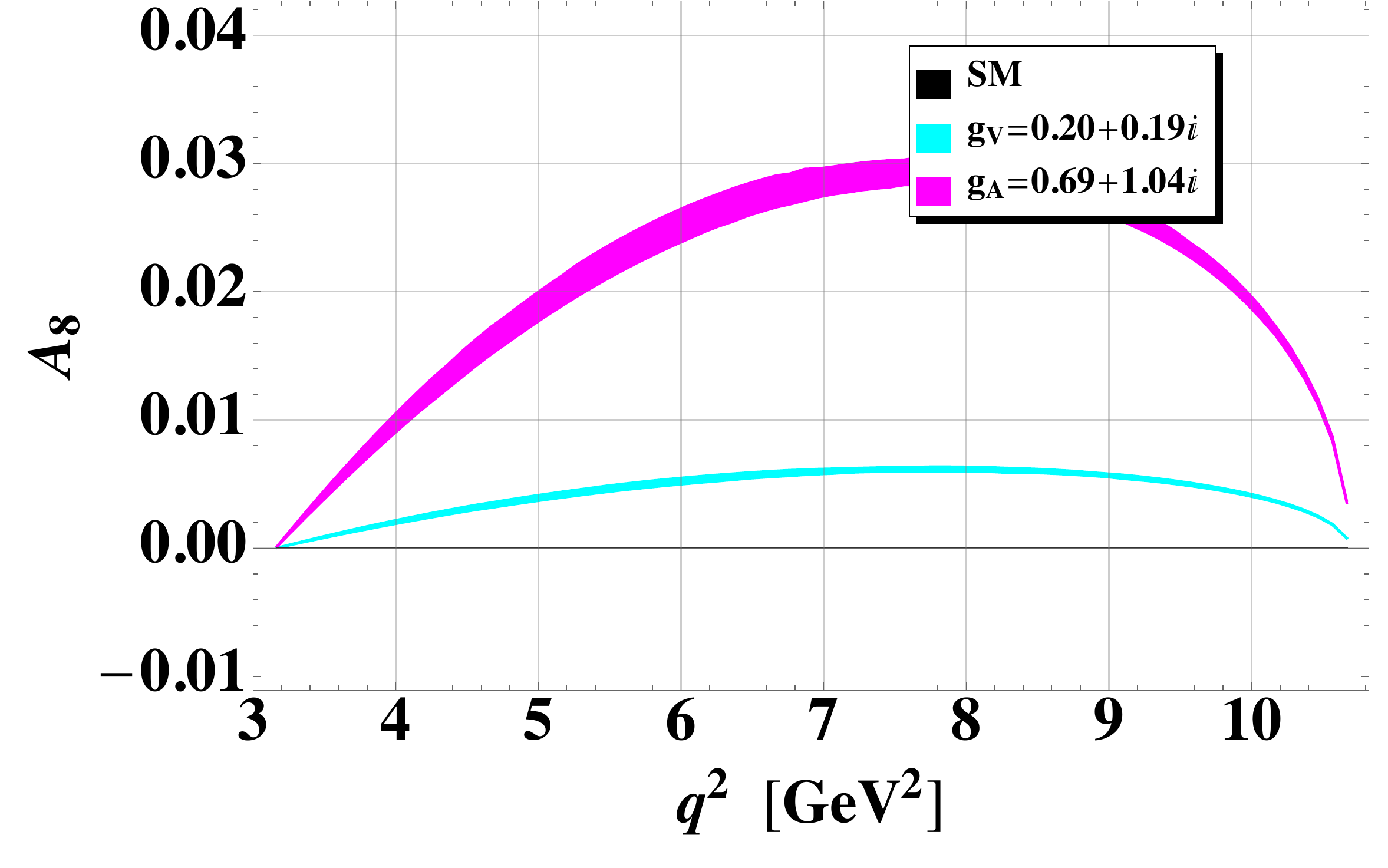}
\includegraphics[width=0.32\linewidth]{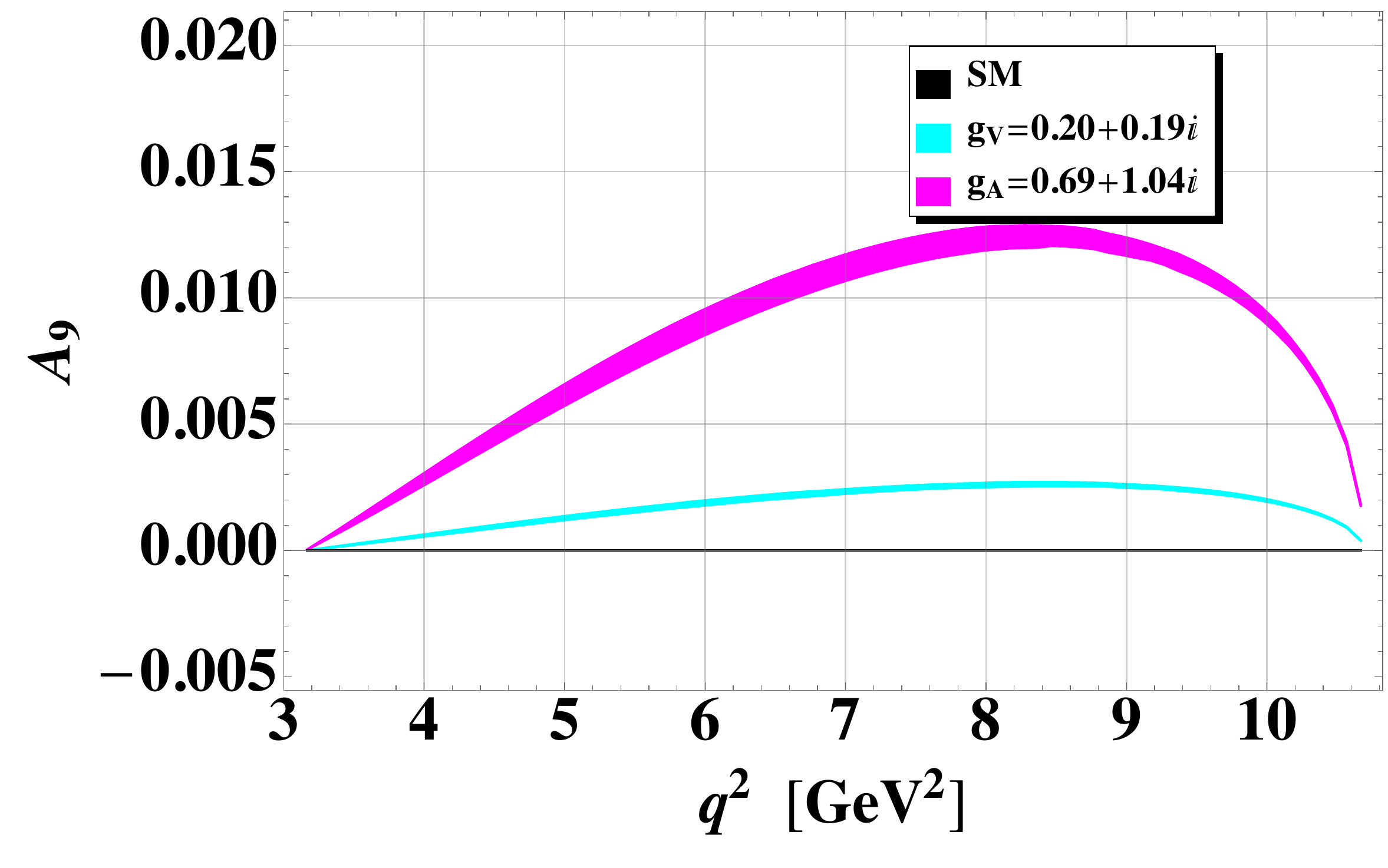}
\caption{\footnotesize{\sl $\A_{\rm FB}$, $\A_{\lambda_\tau}$ observables relevant to $\Bbar\to D\tau\nubar$ and 
$\A_{\rm FB}$, $\A_{\lambda_\tau}$, $R_{L,T}$, $R_{A,B}$ and $A_{3-9}$ relevant to $\Bbar\to \Dst\tau\nubar$ are displayed for various values of NP couplings and as functions of $q^2$. The width of each curve comes from the theoretical uncertainties in hadronic form factors and quark masses. The benchmark $g_i$'s are chosen to be the best fit values, as discussed in the text.}}
\label{fig:angular_plots_Dst}
\end{center}
\end{figure}

The example plots of $q^2$-dependent observables for different benchmark values of NP couplings are depicted in Fig.~\ref{fig:angular_plots_Dst}.

\subsection{Testing LFUV through angular observables}\label{sec:obs}

One can now make the ratios between the above observables with the $\tau$-lepton in the final state and the same observables extracted from the decay to $l$ [$l\in ( e,\mu )$]. 
Of course this can only be done for the quantities which are nonzero in the SM. For the observables proportional to $I_{7,8,9}$, which are zero in the SM, we consider the differences, namely
\begin{equation}
D(A_{7,8,9}) \equiv \langle A_{7,8,9}^\tau \rangle - {1\over2}\biggl( \langle A_{7,8,9}^e \rangle + \langle A_{7,8,9}^\mu \rangle \biggr) \,.
\end{equation}
For all the other observables, defined in Eqs.~\eqref{eq:AFB_D}, \eqref{eq:A_lambdal_D} and \eqref{eq:AFB_Dst}--\eqref{eq:A6s}, we define the  LFUV observables as:
\begin{equation}
R(O_i) \equiv {\langle O_i^\tau \rangle \over {1\over2}\left( \langle O_i^e \rangle + \langle O_i^\mu \rangle \right)} \,,
\end{equation}
where each $O_i^\ell$ is integrated over the available phase space. The only exception to this definition is made in the case of $\mathcal{A}_\mathrm{FB}^D$ which we divide by $m_\ell^2$ in order to make the effect of the presence of NP more pronounced.~\footnote{In other words, $R(\mathcal{A}_\mathrm{FB}^D) = 
{\langle \mathcal{A}_\mathrm{FB}^{D\, \tau}/m_\tau^2 \rangle/ \left( 1/2 \times\langle \mathcal{A}_\mathrm{FB}^{D\, \mu}/m_\mu^2 \rangle +1/2 \times  \langle \mathcal{A}_\mathrm{FB}^{D\, e}/m_e^2 \rangle \right)}$. } 
Since most of the observables are written in terms of the ratios, $O_i^\ell(q^2)=\mathcal{N}_i^\ell(q^2)/\mathcal{D}_i^\ell(q^2)$ with  $\mathcal{N}$ and $\mathcal{D}$ being generically a numerator and a denominator, the integrated quantities are then defined as
\begin{equation}
\label{eq:integration}
\langle O_i^\ell \rangle = {\displaystyle{\int_{m_\ell^2}^{q_{\rm max}^2} \mathcal{N}_i^\ell(q^2) \ dq^2} \over \displaystyle{\int_{m_\ell^2}^{q_{\rm max}^2} \mathcal{D}_i^\ell(q^2) \ dq^2}} \,.
\end{equation}

As it can be seen from the previous subsections, most of the observables are normalized to the differential decay rate which makes it difficult to monitor a dependence on $g_{V,A,V_L}\neq 0$, since the dependence on these couplings would cancel in these observables. Since in many specific models it is proposed to accommodate $R(D^{(\ast )})^\mathrm{exp}> R(D^{(\ast )})^\mathrm{SM}$ by switching on $g_{V_L}\neq 0$, the only departure from the SM value would indeed be in $R(D^{(\ast )})$, while all the other observables would remain compatible with the SM predictions.

\section{Sensitivity to New Physics \label{sec3}}

To study the sensitivity of the LFUV observables defined in the previous Section on the non-zero values of $g_i$'s, 
we determine possible values of $g_i$'s from the fit to the measured $R(D)$ and $R(D^*)$. To do so we use the publicly available 
code \HEPfit~\cite{HEPfit} in which the Bayesian statistical approach is adopted. 

Since the hadronic uncertainties are the main source of the theoretical error, we have to be careful regarding the dependence of observables on the choice of 
form factors. In that respect we first used the set of form factors obtained in the constituent quark model of Ref.~\cite{Melikhov:2000yu} which contains all of 
the form factors needed for this study. 
The uncertainties of the results obtained by the quark models are, however, unclear and attributing $10\%$ uncertainty to each of the form factors is just an educated guess derived from the comparison between the predicted and measured decay rates.  
For that reason the results obtained by using the quark model form factors could only be considered as qualitative and the departures from the SM predictions only as indication (diagnostic) of the presence of LFUV.

Another choice of form factors consists in relying on the experimental analyses of the angular distribution of $\Bbar\to \Dst l \nubar$ [$l\in (e,\mu$)] decays from which the ratios of form factors can be extracted if one assumes 
the so called CLN parametrization of the dominant form factor $A_1(q^2)$~\cite{Caprini:1997mu}.  The experimental averages of the corresponding parameters [$\rho^2$, $R_1(1)$, $R_2(1)$] and the information about their correlations is provided by HFLAV~\cite{Amhis:2016xyh} and we use it in this work.~\footnote{Notice that the HFLAV results are obtained by using the CLN parametrization which fit well the data. Recent studies~\cite{Gambino:2019sif} show that the so called BGL parametrization of Ref.~\cite{Boyd:1997kz} should be preferred because the slopes of the ratios $R_{1,2}(w)$ are not fixed, but left as free parameters. In Ref.~\cite{Bigi:2017njr} it was shown, however, that both parametrizations provide good fit with data. The resulting $|V_{cb}|$ values, as obtained from fitting the data to these two parametrizations, were different. Since we are not interested in assessing the value of $|V_{cb}|$, the choice of parametrization is immaterial. Notice, however, that the most recent study by Belle~\cite{Abdesselam:2018nnh} in which a larger sample of data has been used, showed that (a) the form factor shapes are fully consistent with the results reported by HFLAV~\cite{Amhis:2016xyh}, and (b) the values of $|V_{cb}|$ inferred from the fits to two parametrizations (CLN and BGL) are consistent with each other, both being lower than $|V_{cb}|$ extracted from the inclusive decays.} As for the remaining form factors we used the expressions derived in heavy quark effective theory (HQET)~\cite{Bernlochner:2017jka} in which the 
leading $\alpha_s$ and the leading power corrections have been included to compute $A_0/A_1$, $f_T/f_+$, $T_{1,2,3}/A_1$ to each of which we attribute $10\%$ of error (considerably larger than those quoted in Ref.~\cite{Bernlochner:2017jka}). Finally, for the  
form factors $f_+$ and $f_0$ we use the results obtained in numerical simulations of QCD on the lattice~\cite{Aoki:2019cca}. 
We have checked that our final results obtained by using the form factors computed in the constituent quark model of Ref.~\cite{Melikhov:2000yu} are practically indistinguishable from those obtained with the form factors chosen as explained in this paragraph to which we will refer as ``CLN+HQET+LATT". Since less assumptions are needed in the error estimate of the form factors in the latter case, 
in the following all our results will be obtained by choosing CLN+HQET+LATT form factors~\cite{Caprini:1997mu,Amhis:2016xyh,Bernlochner:2017jka,Aoki:2019cca}.

We now proceed and allow for one complex valued coefficient $g_i$ to be non-zero at a time, i.e. we add $2$ extra parameters (compared to the SM case) in every fit. 
We emphasize once again that we assume the 
NP to affect only the decay modes with $\tau$ in the final state, namely, $\overline B \to D^{(*)} \tau \bar \nu_\tau$. The joint \emph{probability distribution
functions} (\emph{p.d.f.}'s) for the coefficients $g_i$ 
are therefore obtained by using $R(D)^\mathrm{exp}$ and $R(D^*)^\mathrm{exp}$ as constraint, and then they are employed to 
predict all of the LFUV observables discussed in the previous Section. From the comparison with the SM results we can see which quantity 
is more sensitive to the considered $g_i\neq 0$. Notice again that the $B \to D \tau \nu_\tau$ observables will be affected by NP effects in $g_{V,S,T}$, while $g_{V,A,P,T}\neq 0$ will
modify the $B \to D^* \tau \nu_\tau$ ones.

\begin{figure}[t!]
\begin{center}
\includegraphics[width=0.3\textwidth]{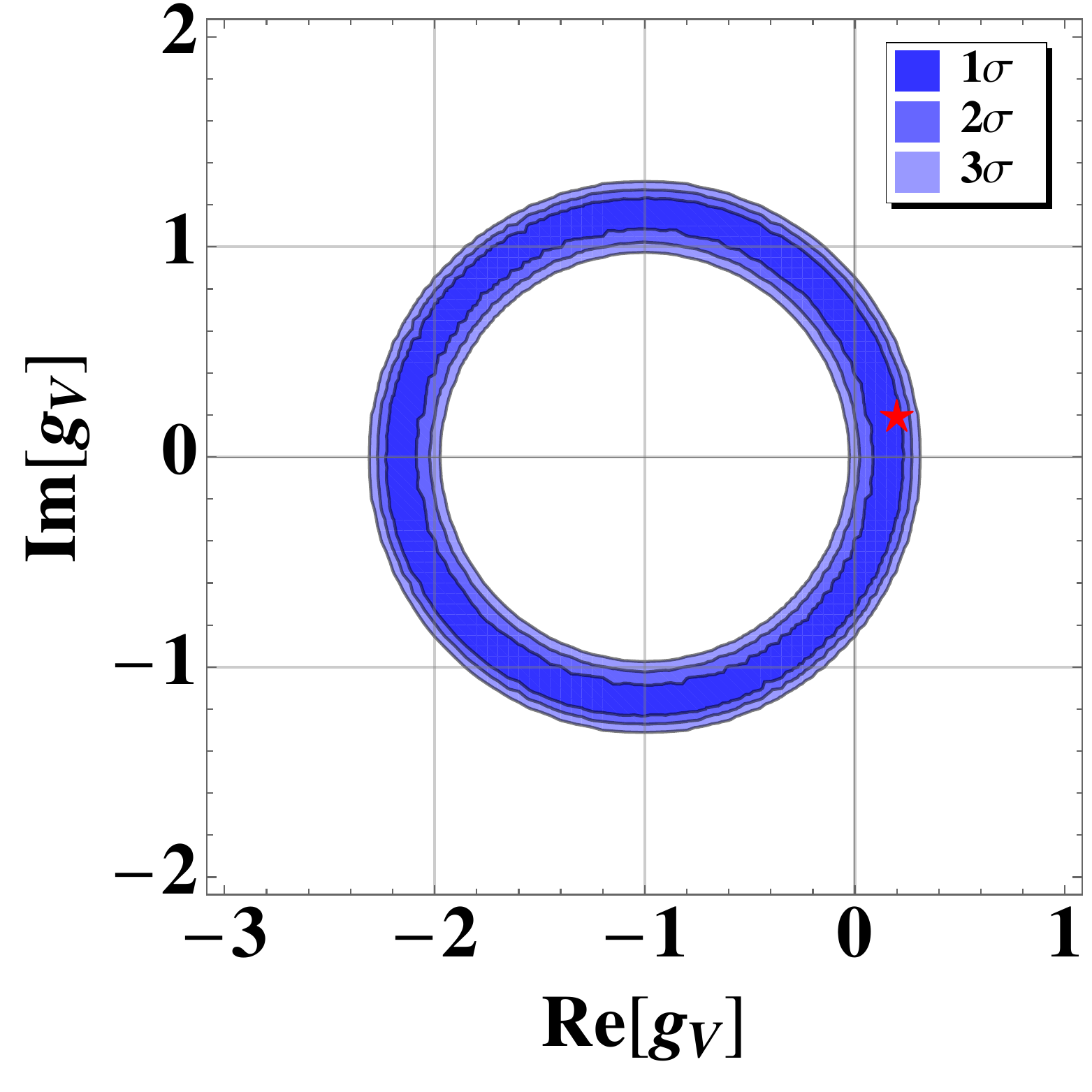}
\includegraphics[width=0.3\textwidth]{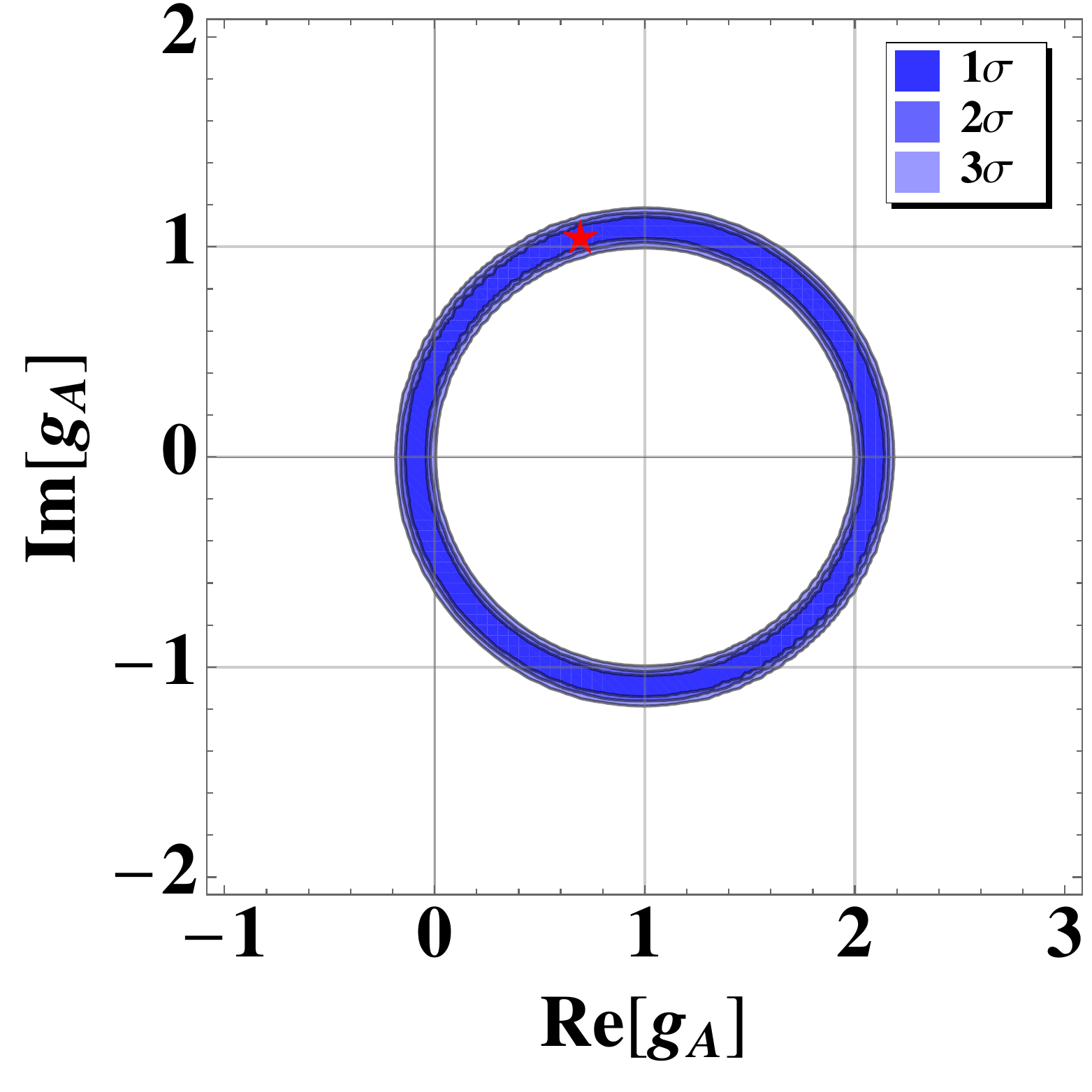}
\includegraphics[width=0.3\textwidth]{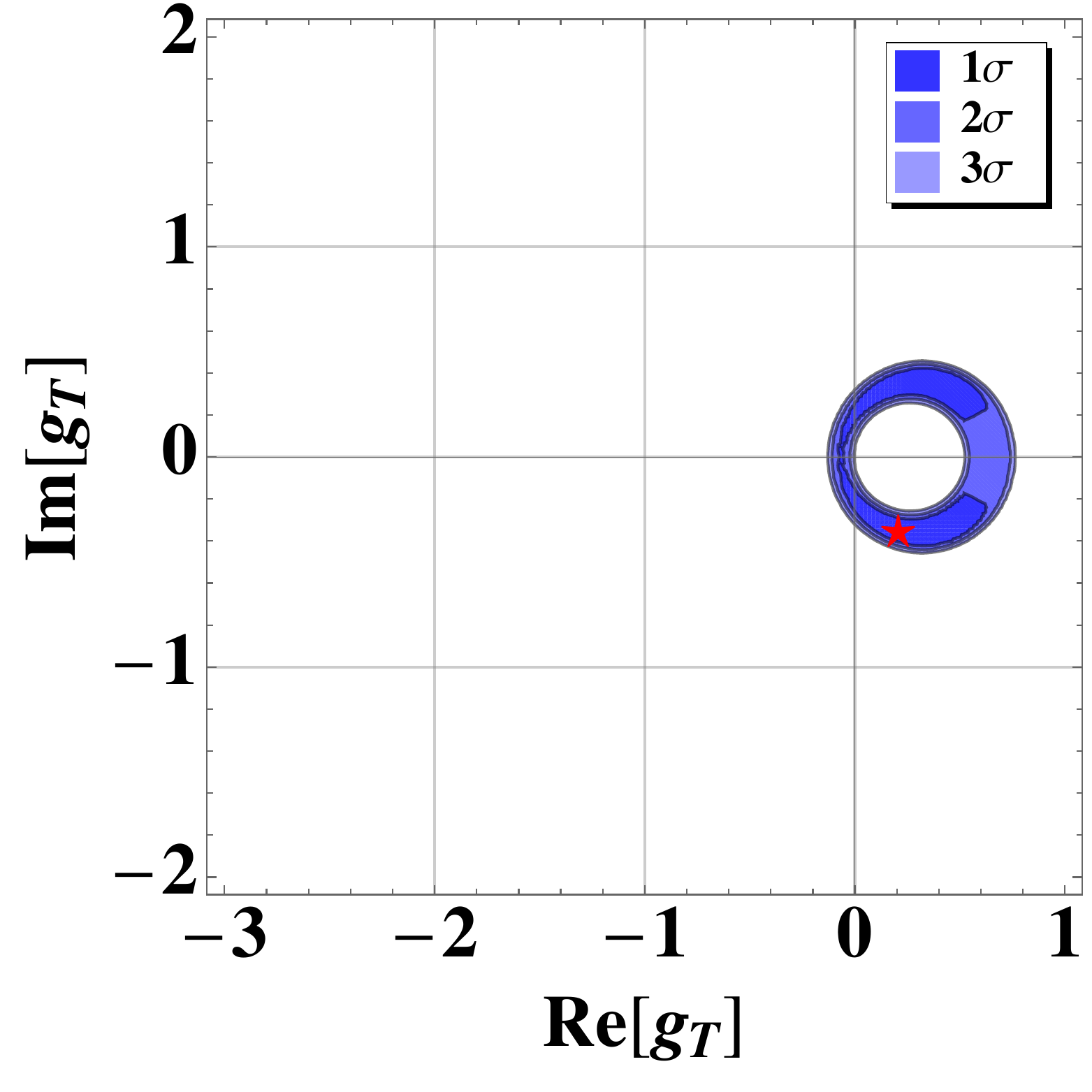}
\\ \vspace{3mm}
\includegraphics[width=0.3\textwidth]{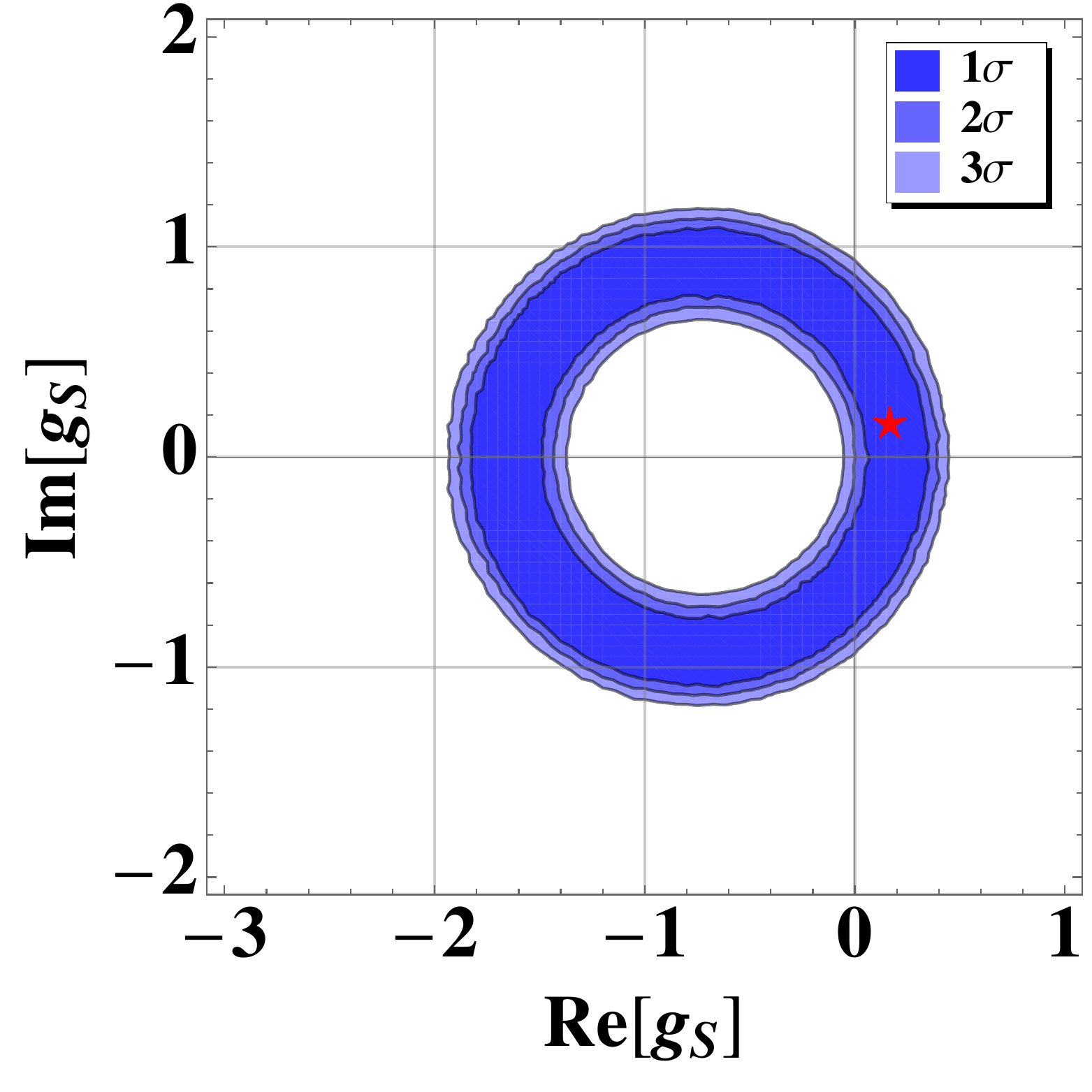}
\includegraphics[width=0.3\textwidth]{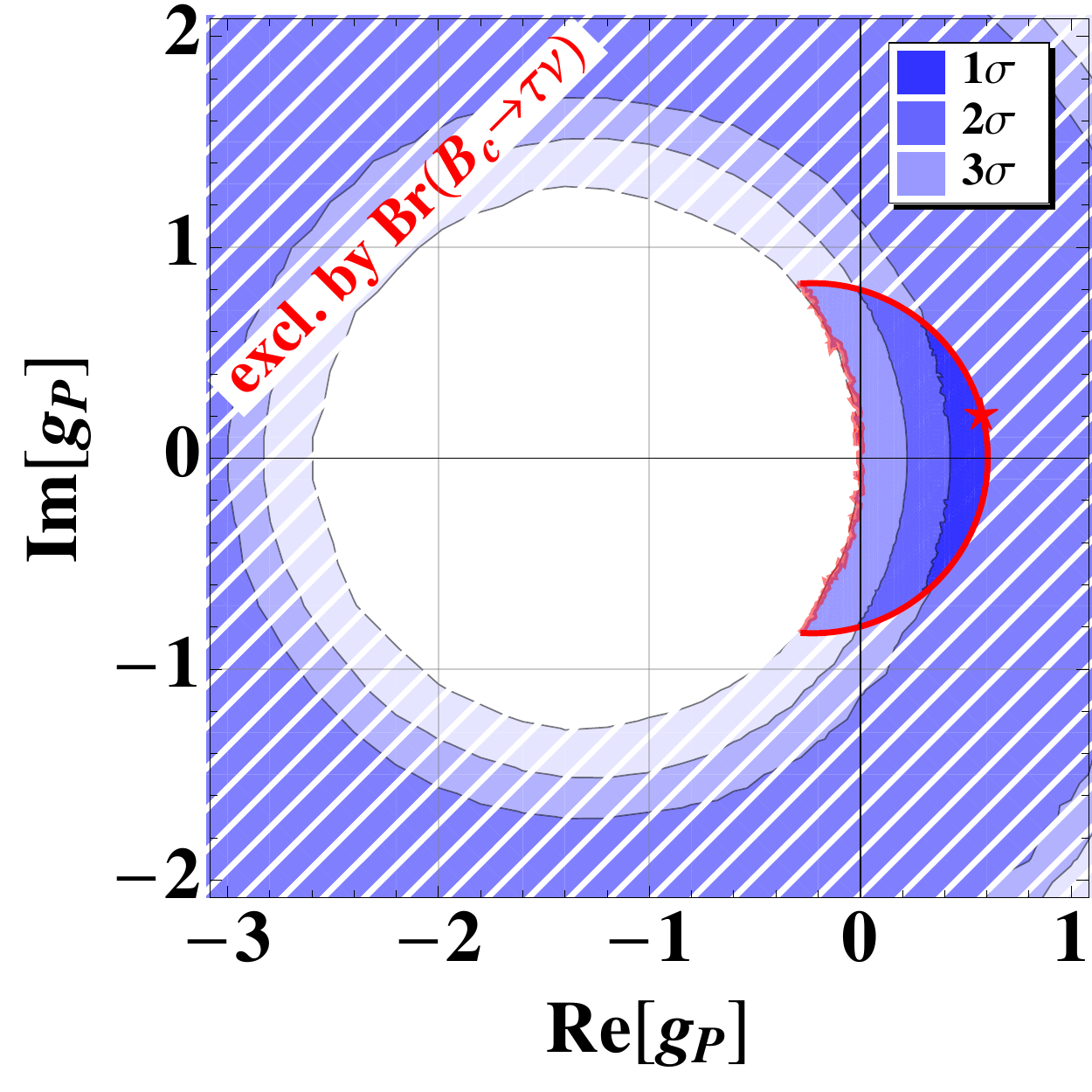}
\caption{\footnotesize{\sl The allowed values for the NP couplings in Eq.\eqref{eq:Heff} as obtained from the the fit with $R(D)^\mathrm{exp}$ and $R(D^*)^\mathrm{exp}$, and by switching one coupling $g_i$ at the time. Red stars denote the best fit values. Regarding the form factors we used those to which we refer in the text as CLN+HQET+LATT. }}
\label{fig:g_constraints}
\end{center}
\end{figure}

\subsection{Fit results}\label{sec:fitres}

\begin{figure}[t!]
\begin{center}
\includegraphics[width=0.3\textwidth]{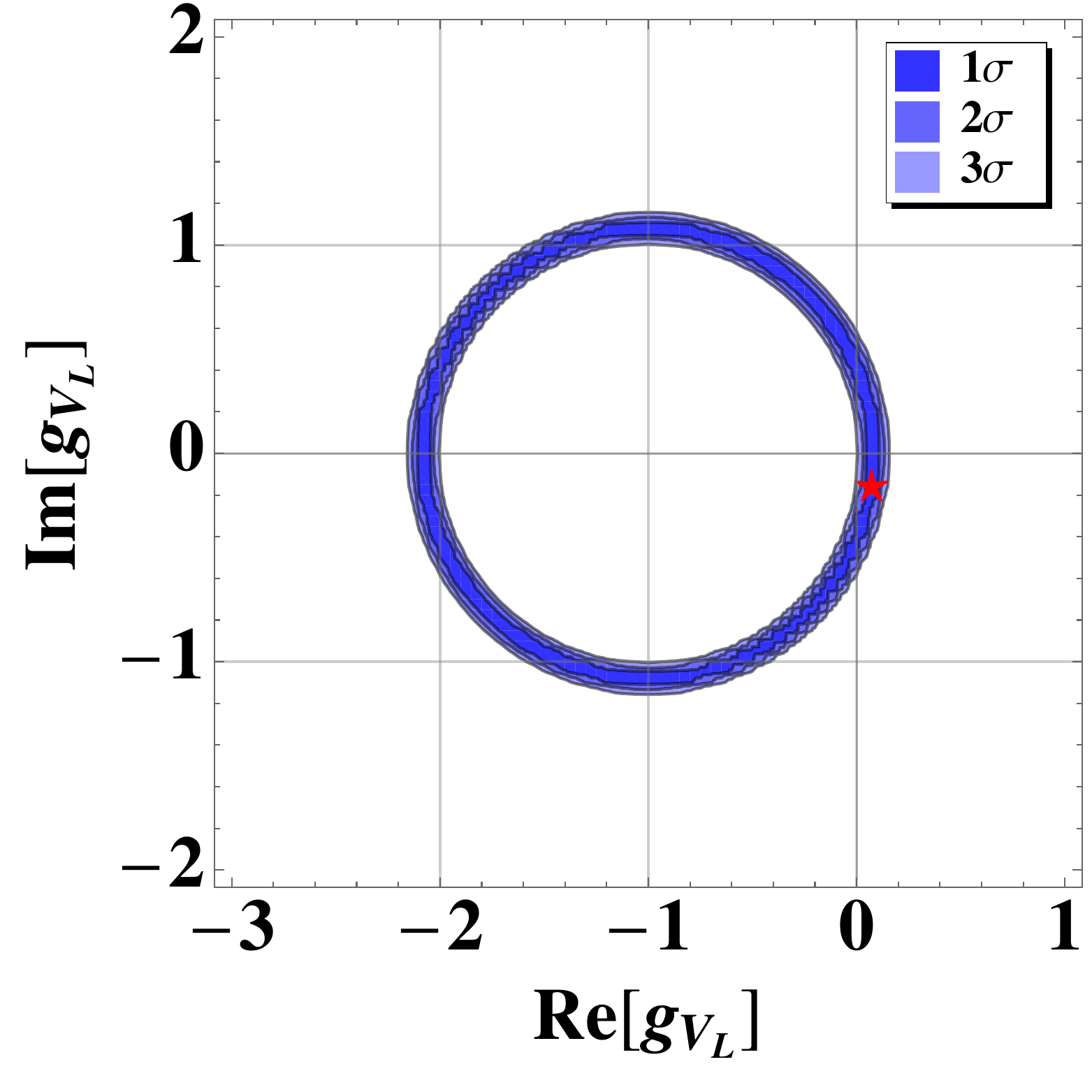}
\includegraphics[width=0.3\textwidth]{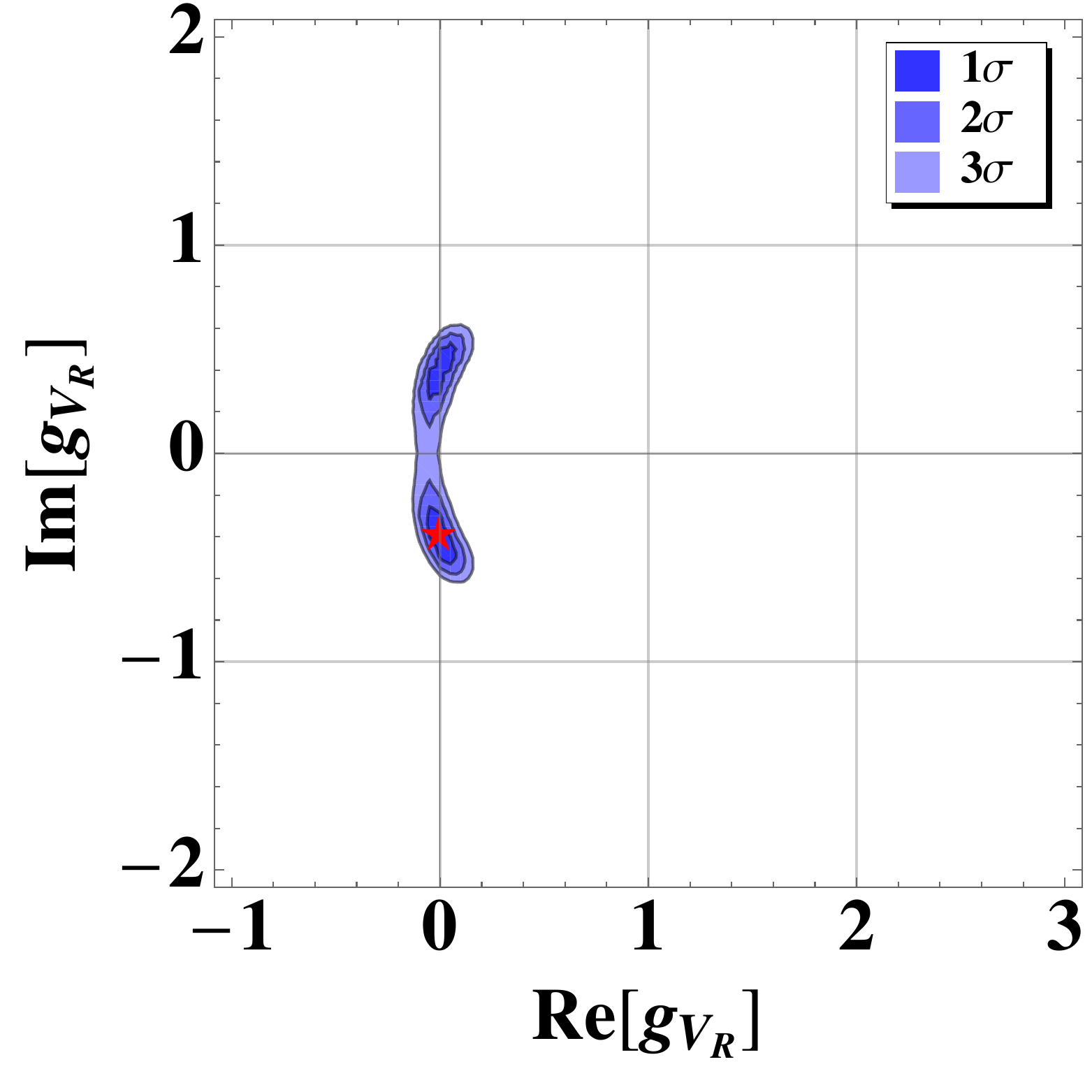}
\includegraphics[width=0.3\textwidth]{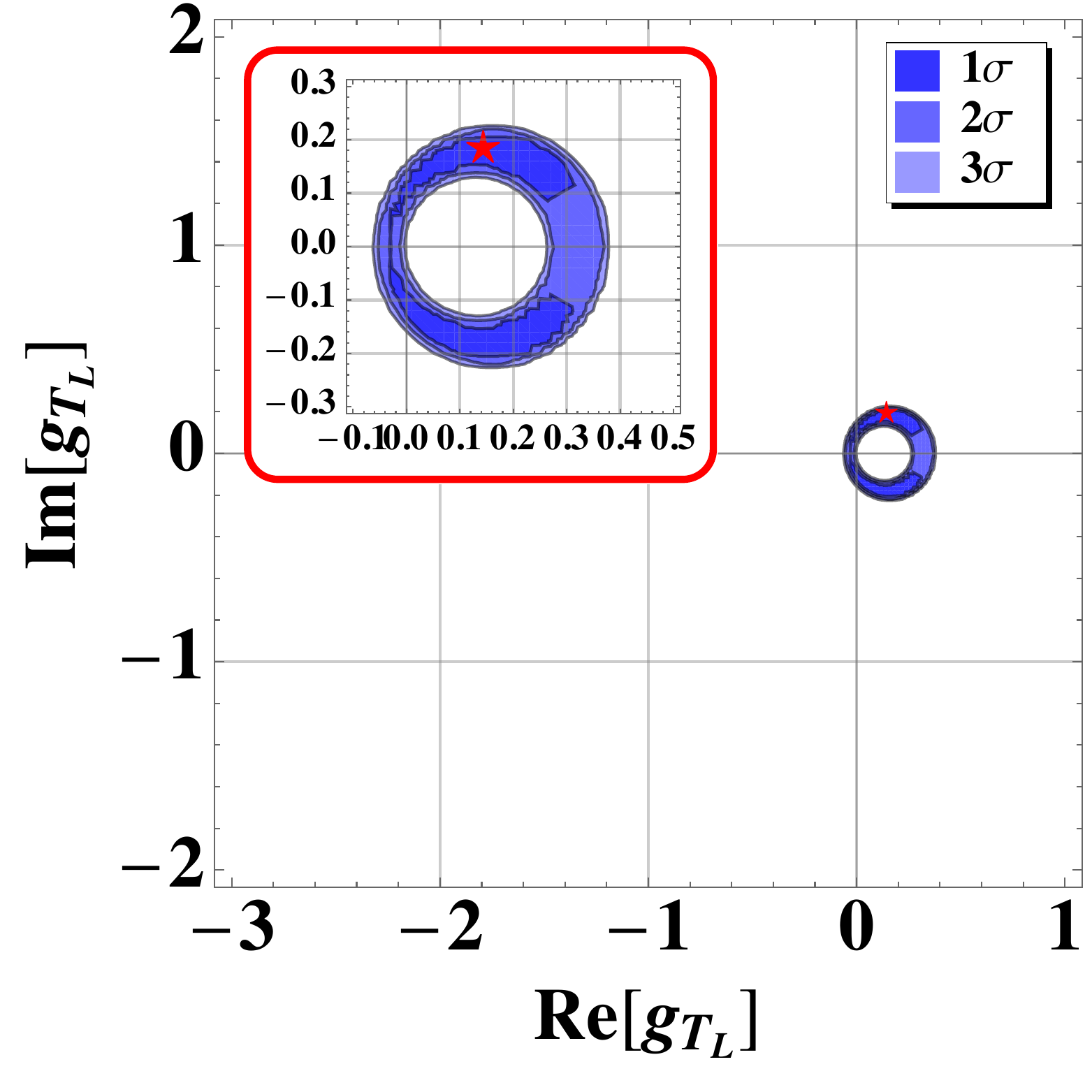}
\\ \vspace{3mm}
\includegraphics[width=0.3\textwidth]{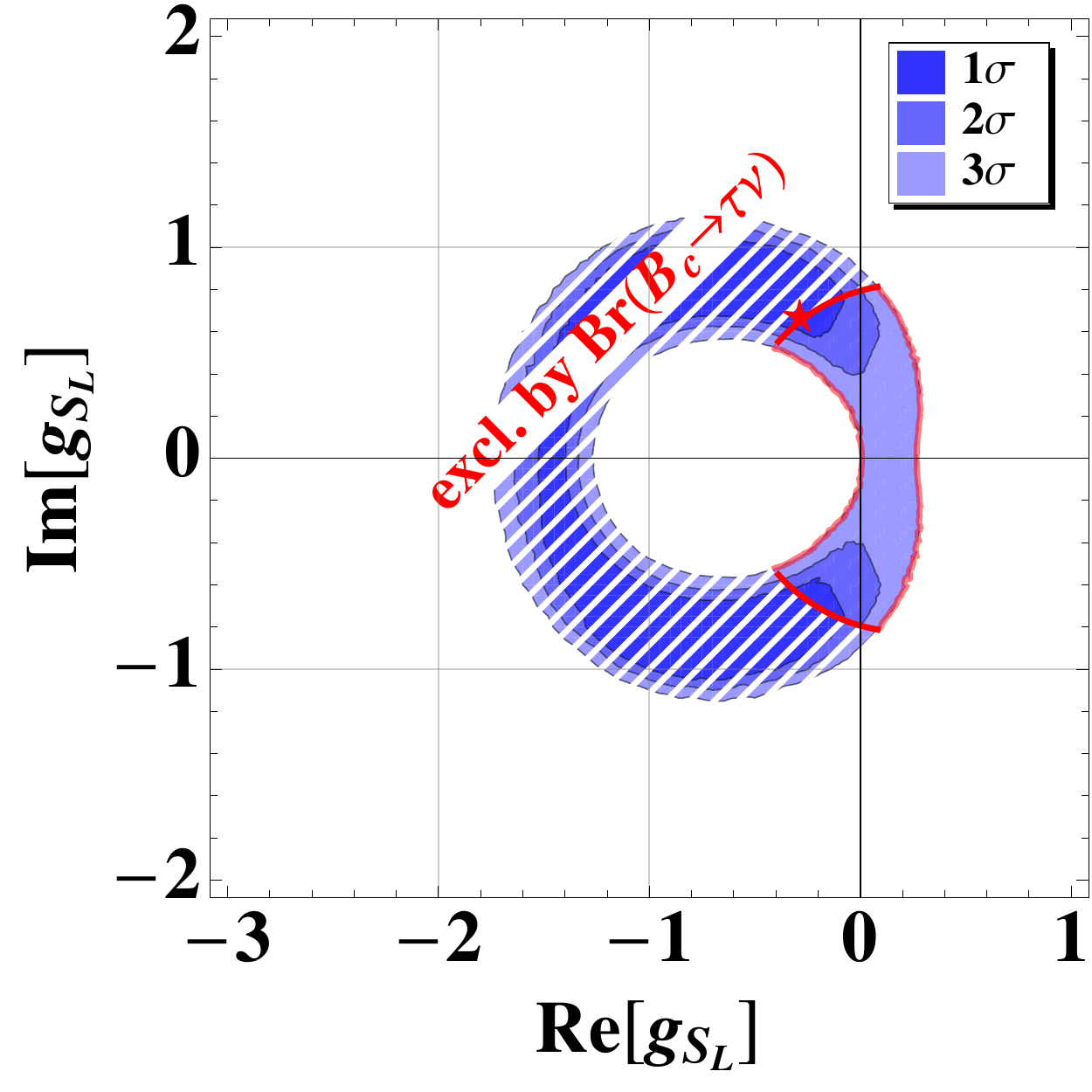}
\includegraphics[width=0.3\textwidth]{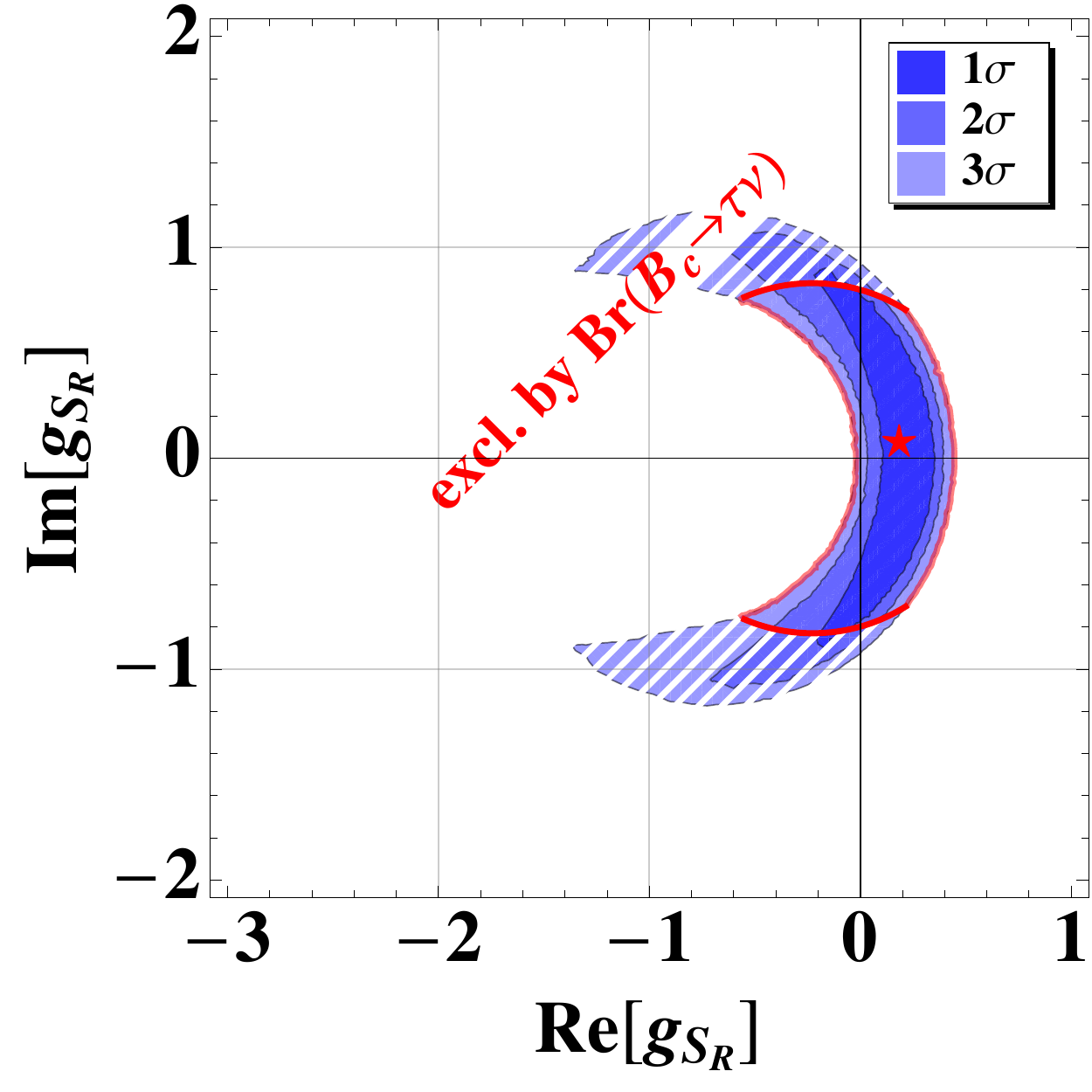}
\caption{\footnotesize{\sl Same as in Fig.~\ref{fig:g_constraints} but for the NP couplings appearing in Eq.\eqref{eq:Heff_Jp}.}}
\label{fig:g_constraints_LR}
\end{center}
\end{figure}

In Fig.~\ref{fig:g_constraints} and~\ref{fig:g_constraints_LR} we show the allowed regions for the NP couplings $g_i$ relevant to the basis 
\eqref{eq:Heff} and \eqref{eq:Heff_Jp}, respectively. We reiterate that the allowed regions for NP couplings are obtained by letting one (complex valued) coupling $g_i$ to be non-zero at a time. 
Any value of the $g_i$ derived in this way is plausible. 
We also include the limit derived from the $B_c$-meson lifetime as discussed in Refs.~\cite{Li:2016vvp,Alonso:2016oyd} which is particularly restrictive to the pseudoscalar NP contribution $g_P$, as well as in $g_{S_L,S_R}$. In this paper we take the conservative limit $\mathcal{B}(B_c\to \tau \bar{\nu}) \lesssim 30\%$ and use the expression
\begin{equation}
\label{eq:leptonicB}
\mathcal{B}(B_c\to \tau \bar{\nu}) = \tau_{B_c} \dfrac{m_{B_c} f_{B_c}^2 G_F^2 |V_{cb}|^2}{8 \pi} m_\tau^2 \left( 1- \dfrac{m_\tau^2}{m_{B_c}^2} \right)^2 \Bigg{|} 1+ g_{V_L} + \dfrac{ \left(g_{S_R}-g_{S_L}\right)m_{B_c}^2}{m_\tau (m_b+m_c)}\Bigg{|}^2\,, 
\end{equation}
where the $B_c$ decay constant $f_{B_c}=427(6)~\mev$ has been determined in the lattice QCD study of Ref.~\cite{McNeile:2012qf}.

The benchmark values, denoted by red stars in Figs.~\ref{fig:g_constraints}-\ref{fig:g_constraints_LR}, correspond to the best fit values. We get
\begin{align}
&g_V= 0.20 + i\ 0.19\,,&&  g_A= 0.69+ i\ 1.04\,,&&& \nonumber \\
&g_S= 0.17 + i\ 0.16\,,&&  g_P= 0.58 + i\ 0.21\,,& g_T-g_{T5}=0.21-i\ 0.35\,,&&
\label{eq:bestfits}
\end{align}
and in terms of couplings introduced in Eq.~\eqref{eq:Heff_Jp} we find
\begin{align}
&g_{V_L}= 0.07 - i\ 0.16\,,&&  g_{V_R}= -0.01 - i\ 0.39\,,&&& \nonumber \\
&g_{S_L}= -0.29 - i\ 0.67\,,&&  g_{S_R}= 0.19 + i\ 0.08\,,& g_{T_L}=0.11- i\ 0.18\,.&&
\label{eq:bestfits_LR}
\end{align}
In Fig.~\ref{fig:angular_plots_Dst} we show the $q^2$-dependence of each observable relevant to $\bar B\to D^\ast \tau\bar\nu$ discussed in the previous Section, both in the SM and with $g_i$ given in Eq.~\eqref{eq:bestfits}.

After integrating over $q^2$'s as described in Eq.~\eqref{eq:integration}, and by sweeping over the entire range of $g_i$ allowed by 
$R(D)^\mathrm{exp}$ and $R(D^*)^\mathrm{exp}$, we obtain the results listed in Tab.~\ref{tab:obsCLN} and Tab.~\ref{tab:obsCLN_LR}. 
The results are presented along with the SM ones in order to make a comparison simpler. Notice that the SM values we obtain are fully compatible with those 
quoted in Eq.~\eqref{eq:HFLAV}, obtained by HFLAV, even though the central values are slightly different owing to our choice of form factors. 
For each $g_i$ we then compute $R(D)$ and $R(D^*)$, which are now obviously compatible with experimental values, and the full list of the LFUV observables discussed in the previous 
Section. 
For the observables showing gaussian shapes we quote the mean values with the corresponding standard deviations. Some of the observables, however, are hardly gaussian for the reasons explained below. For such observables in Tabs.~\ref{tab:obsCLN} and \ref{tab:obsCLN_LR} we present the interval of values within $2\sigma$. 
The interested reader can also find in the tables of Appendix \ref{app:bestfit_tab} the predictions for all
the LFUV observables obtained assuming for the benchmark values for the NP couplings given in 
Eqs.~(\ref{eq:bestfits},\ref{eq:bestfits_LR}).

Let us now comment on the results we obtain, starting from the ones shown in Tab.~\ref{tab:obsCLN}.
\begin{itemize}
\item $\underline{g_V}$: Assuming NP affecting only the vector current, the constraints arising from
 $R(D)^\mathrm{exp}$ and $R(D^*)^\mathrm{exp}$ are such that either no or very small deviation of the LFUV observables 
 with respect to their SM values is predicted. The only exception is $ R(A_3)$ which, for allowed $g_V\neq 0$, 
 becomes lower than its SM counterpart. 
 
Notice that for several observables we give only the intervals due to their non-gaussian behaviour. They can be 
divided into two classes: those only sensitive to  $\Re (g_V)$, such as $ R(A_{FB}^{D^\ast})$, $ R(A_5)$ and $ R(A_6)$, 
and those only sensitive to $\Im (g_V)$, namely $ D(A_7)$, $ D(A_8)$ and $ D(A_9)$. 
Let us focus on one observable from the first class, say $R(A_5)$.  As stated above, its value would be indistinguishable from the SM 
for $g_V$ purely imaginary. From the plot shown in the first panel of 
Fig.~\ref{fig:g_constraints} it is apparent that there are $2$ distinct allowed solutions for $\Re (g_V)$ when $\Im (g_V)=0$, and therefore there are $2$ 
distinct predictions of $ R(A_5)$. After sweeping through various $\Im(g_V)$ we fill up the gap between the two distinct real solutions in the case of $\Im (g_V)=0$, hence producing the
anticipated non-gaussian prediction for $ R(A_5)$, shown in the left panel of Fig.~\ref{fig:RA5}.

\begin{figure}[t!]
\begin{center}
\includegraphics[width=0.45\textwidth]{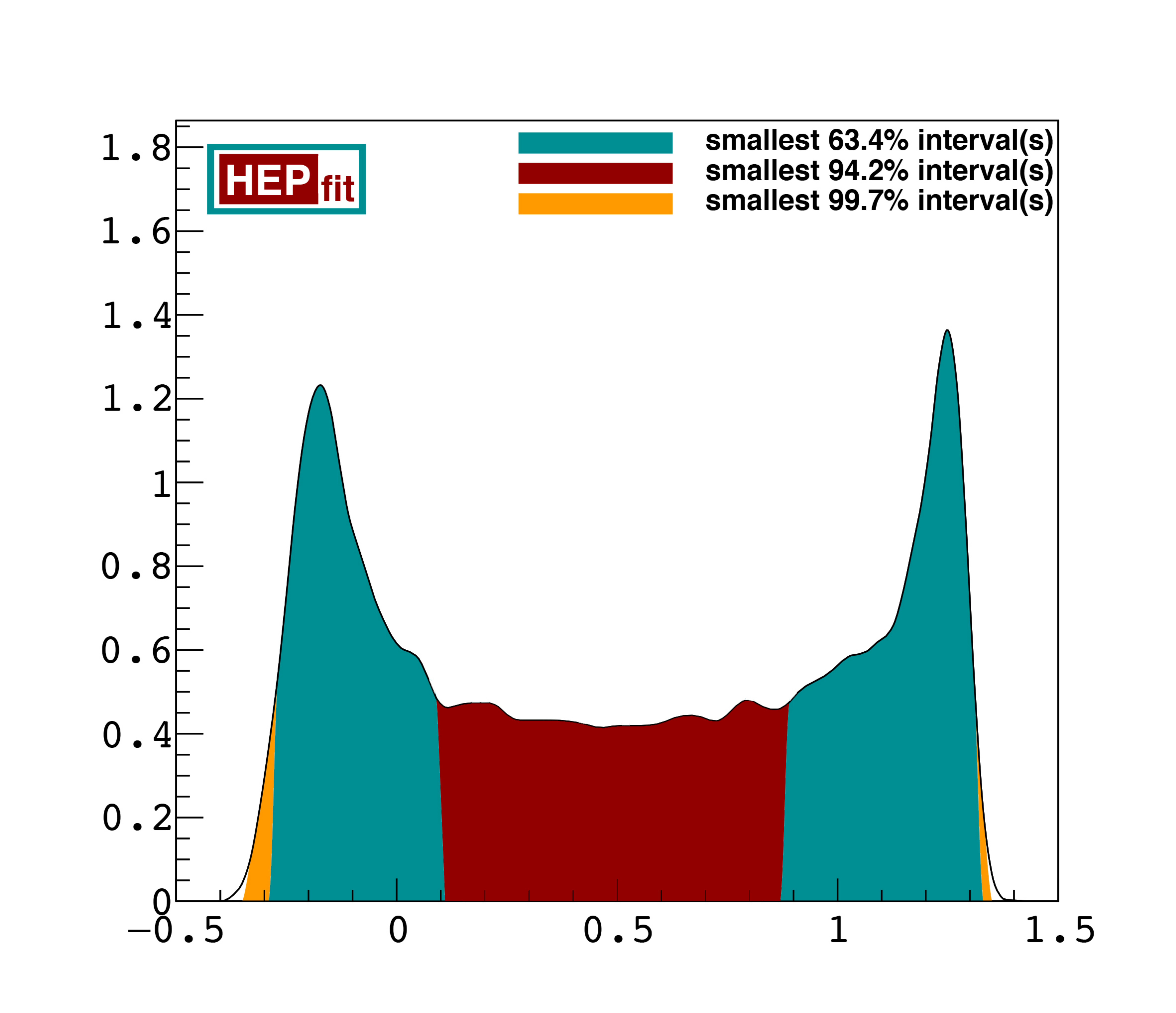}
\includegraphics[width=0.45\textwidth]{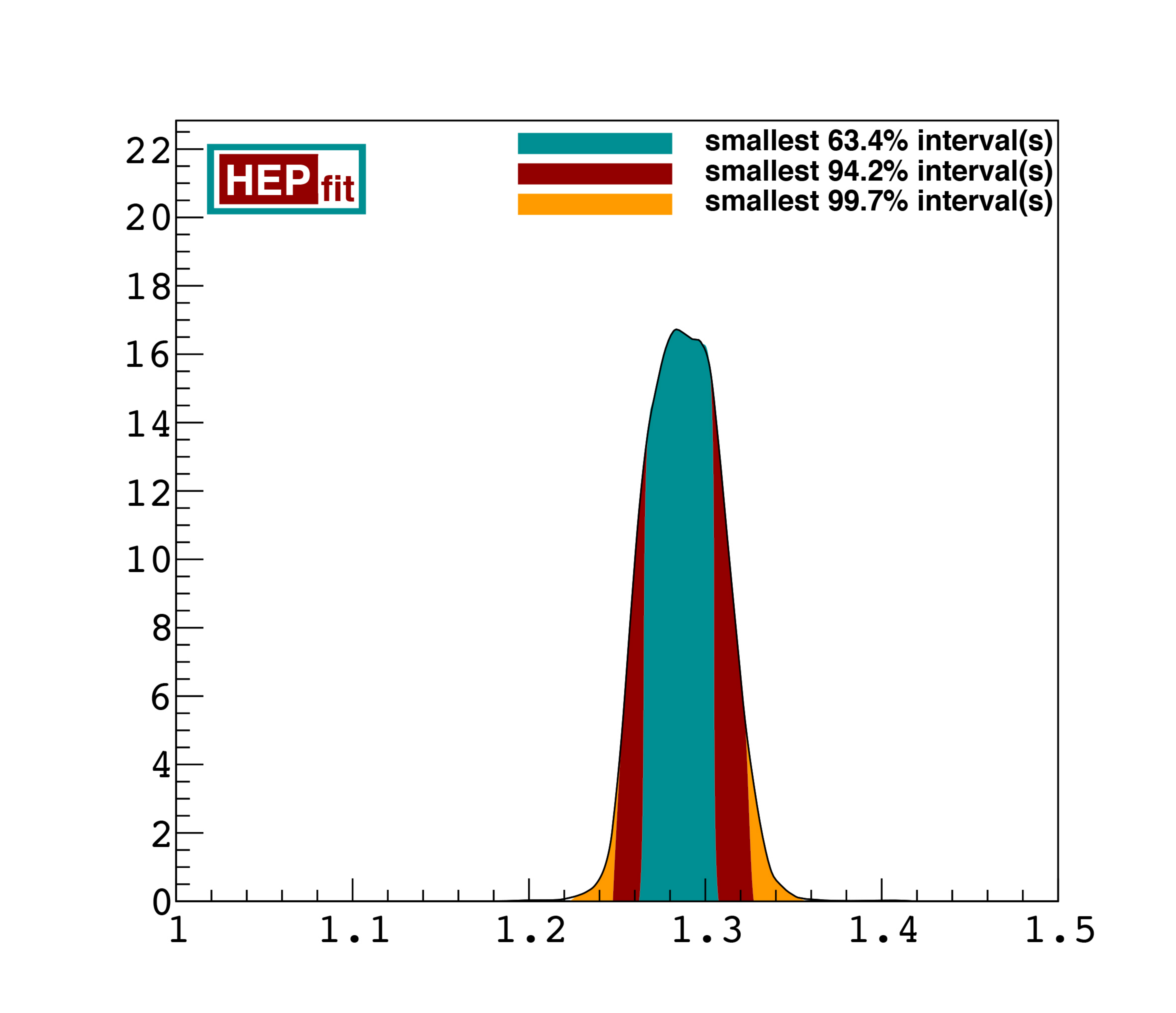}
\put(-320,-3){\footnotesize $R(A_5)$}
\put(-100,-3){\footnotesize $R(A_5)$}
\caption{\footnotesize{\sl Left panel: predicted p.d.f. for the observable $ R(A_5)$, assuming a complex
$g_V$. Right panel: predicted p.d.f. for the observable $ R(A_5)$, assuming that $g_V$ is equal
to the best fit value reported in Eq.~\eqref{eq:bestfits}.}}
\label{fig:RA5}
\end{center}
\end{figure}

At first sight, given the broad range of predicted values, this observable might not seem of particular use. However,
a careful reader will realize that a given value of $g_V$ will correspond to a point (and not a disk) in the plane depicted in the first panel of
Fig.~\ref{fig:g_constraints}. Therefore, for a given value of $Re(g_V)$, one gets a very sharp prediction for $R(A_5)$ as shown in the right panel of Fig.~\ref{fig:RA5}, where we assumed that $g_V$ is equal
to its best fit value reported in Eq.~\eqref{eq:bestfits}. Conversely, the measurement of $ R(A_5)$ would produce a sharp bound on the
real part of $g_V$, corresponding to a vertical stripe in the $Re(g_V)$-$Im(g_V)$ plane that intersected with the
disk produced by the measurements of $R(D)$ and $R(D^*)$, therefore severely reducing the allowed region for that coupling.

As stated above, a similar reasoning can be performed for $ R(A_{FB}^{D^\ast})$, $ R(A_6)$, 
$ D(A_7)$, $ D(A_8)$ and $ D(A_9)$, hence making all these observable extremely interesting. 

\item $\underline{g_A}$: If NP appears only in the axial current, only the $B \to D^* \tau \nu_\tau$ channel 
is affected. Similarly to the previous scenario the obtained bounds are such that no appreciable difference
is observed in the observables $ R(A_{\lambda_\ell}^{D^\ast})$, $ R(R_{L,T})$, $ R(R_{A,B})$, 
$ R(A_3)$ and $ R(A_4)$. On the other hand, analogously to the $g_V$ scenario, for 
$ R(A_{FB}^{D^\ast})$, $ R(A_5)$, $ R(A_6)$, $ D(A_7)$, $ D(A_8)$ and $ D(A_9)$ we get broad ranges of values, with the LFUV 
ratios affected by $Re(g_A)$ and the LFUV differences by $\Im(g_A)$.
Like in the vector scenario, the measurement of these observables would highly
constrain the allowed region for $g_A$. 

\item $\underline{g_S}$: If the NP effects come with the contribution arising solely from the scalar operator, then only $B \to D \tau \nu_\tau$ would be affected.
In particular, $ R(A_{\lambda_\ell}^D)$ is sensibly shifted compared to the SM prediction,
while $ R(A_{FB}^{D})$ displays once again a broad prediction, due to its sensitivity to $Re(g_S)$. Therefore,
its measurement would provide the strongest bound on $g_S$.

\item $\underline{g_P}$: In contrast to the previous case, the NP giving a nonzero contribution to a term proportional to the pseudoscalar current would 
result in the changes in the $B \to D^* \tau \nu_\tau$ 
channel only. The unaffected observables by this choice would be $ R(R_{A,B})$, $ D(A_8)$
and $ D(A_9)$. On the other hand, $ R(A_{\lambda_\ell}^{D^\ast})$, $ R(R_{L,T})$, $ R(A_3)$,
$ R(A_4)$ and $ R(A_6)$ are all predicted with good precision, and sensibly different
from the values predicted in the SM. Moreover, we again obtain broad ranges for $ R(A_{FB}^{D^\ast})$, $ R(A_5)$ and $ D(A_7)$, with $\Re(g_P)$ affecting the LFUV ratios, while $\Im(g_P)\neq 0$ would particularly affect the LFUV differences.

\item $\underline{g_T}$: If the NP effects give a nonzero contribution to the term proportional to the tensor current then both channels are 
affected. Furthermore, all the LFUV ratios are predicted with fairly good precision and
with sensible discrepancies when compared with the SM predictions. Regarding the LFUV differences, 
$ D(A_7)$ is the only one mildly affected by $g_T\neq 0$, even if perfectly compatible with
the SM prediction, while the remaining ones are both unaffected by this kind of NP.
\end{itemize}

Similar observations can be obtained analyzing the results from Tab.~\ref{tab:obsCLN_LR}, where we have not
listed the results relative to $g_{T_L}$ since they are the same as the ones obtained for $g_T$, owing to Eq.\eqref{eq:coeff_relations}.
\begin{itemize}
\item $\underline{g_{V_L}}$: If the effects come with the left-handed vector current alone then the present
experimental bounds are such that no sensible deviation can be appreciated in any of the LFUV quantities
defined above.

\item $\underline{g_{V_R}}$: In contrast to the previous case, if the coupling to the right-handed vector current is preferred then some LFUV observables in the $B \to D^* \tau \nu_\tau$ channel 
exhibit a sensitivity to its effects, namely $ R(A_{FB}^{D^\ast})$, 
$ R(A_5)$, $ R(A_6)$, $ D(A_7)$, $ D(A_8)$ and $ D(A_9)$. Moreover, given that the LFUV differences
depend on $\Im(g_{V_R})$ and that, as can be observed from the second panel of
Fig.~\ref{fig:g_constraints_LR}, the fit allows for two distinct solutions for $Im(g_{V_R})$, we obtain 
two separate predicted regions for each of these observables. Notice that the range of allowed $g_{V_R}$ is far more restricted by the data which is expected on the basis of the SM gauge invariance. 

\item $\underline{g_{S_L}}$ and  $\underline{g_{S_R}}$: Assuming NP effects in the scalar current, we obtain
both in the left-handed and in the right-handed cases a similar outcome: three observables are unaffected,
namely $ R(R_{A,B})$, $ D(A_8)$ and $ D(A_9)$, while $ R(A_{\lambda_\ell}^D)$ is found to be sensibly different with respect to its SM value. 
All the remaining observables display broad prediction ranges,
once again with the real (imaginary) part affecting the LFUV ratios (differences).

\end{itemize}

To further explore the possibilities of using the angular observables as tests for LFUV we performed an extra test, motivated
by the following observation: as can be seen from Eq.~\eqref{eq:dGdq2_Dst}, only a small subset of the angular
coefficients defined at Eq.~\eqref{eq:Is} appear in the definition of the decay rate and hence of $R(D^*)$, i.e. 
$I_{1c,1s}$ and $I_{2c,2s}$. Therefore, the measurement of the branching fraction can constrain only a part of the
coefficients appearing in the full angular distribution defined in Eq.~\eqref{eq:fullangdist}, with the remaining ones
still potentially affected by the NP effects. In order to test such effects we performed a new set of fits where we assumed
that $R(D)$ and $R(D^*)$ would be measured with a central value equal to the SM predictions with a $10\%$
error. The allowed regions for the couplings are found to be similar to the ones shown in
Figs.~\ref{fig:g_constraints} and \ref{fig:g_constraints_LR}, even if reduced in size and thickness. Therefore,
the LFUV observables that were previously showing a broad prediction due to their non-trivial interplay with the
real and the imaginary parts of the NP couplings will continue to display such a behavior.  In other words,
even if $R(D)$ and $R(D^*)$ are measured to agree with the SM predictions, there are
$7$ angular observables that might still display a behavior unambiguously related to NP effects: four LFUV 
ratios [$ R(A_{FB}^{D, D^\ast})$, $ R(A_{5,6})$], and three LFUV differences [$ D(A_{7,8,9})$].

All of the above observations show the impact that the measurement of even a small subset of the observables presented
in Sec.~\ref{sec:obs} would have on our understanding of the NP contribution to the $b \to c$ transitions. In particular, 
the measurement of any of the observables between $ R(A_{FB}^{D, D^\ast})$, $ R(A_{5,6})$ and $ D(A_{7,8,9})$ would 
be of great interest, given their twofold power: they would help deciphering the Lorentz structure of the NP contributions, 
and they would severely constrain the presently allowed region for the NP couplings. We emphasize once again that a measurement 
of these observables would be of great interest
even in the case of $R(D)$ and $R(D^*)$ being fully compatible with their SM values.

\begin{sidewaystable}[!t]
\fontsize{10.45}{13}\selectfont
\centering
\begin{tabular}{|c|c|c|c|c|c|c|c|}
\hline
&&&&&&&\\[-1mm]
\textbf{Obs.} & Exp. & SM & $ g_V $ & $ g_A $ & $ g_S $ & $ g_P $ & $ g_T $ \\[1mm]
&&&&&&&\\[-3mm]
\hline
&&&&&&&\\[-3mm]
$ R_D $
& $0.334 \pm 0.029$ 
& $0.30 \pm 0.02$ 
& $0.37 \pm 0.03$ 
& $-$ 
& $0.37 \pm 0.03$ 
& $-$ 
& $0.34 \pm 0.03$ 
\\
&&&&&&&\\[-3mm]
$ R_{D^*} $
& $0.297 \pm 0.015$ 
& $0.258 \pm 0.003$ 
& $0.261 \pm 0.003$ 
& $0.30 \pm 0.01$ 
& $-$ 
& $0.278 \pm 0.007$ 
& $0.30 \pm 0.01$ 
\\
&&&&&&&\\[-3mm]
\hline
&&&&&&&\\[-3mm]
$ R(A_{FB}^{D})$ 
& $ - $ 
& $0.077 \pm 0.004$ 
& $0.074 \pm 0.003$ 
& $-$ 
& $[-0.058,0.074]$
& $-$ 
& $0.082 \pm 0.004$ 
\\
&&&&&&&\\[-3mm]
$ R(A_{\lambda_\ell}^D)$ 
& $ - $ 
& $-0.332 \pm 0.003\phantom{-}$ 
& $-0.331 \pm 0.003\phantom{-}$ 
& $-$ 
& $-0.48 \pm 0.05\phantom{-}$ 
& $-$ 
& $-0.25 \pm 0.05\phantom{-}$ 
\\
&&&&&&&\\[-3mm]

\hline
&&&&&&&\\[-3mm]
$ R(A_{\lambda_\ell}^{D^\ast})$ 
& $ - $ 
& $0.47 \pm 0.02$ 
& $0.48 \pm 0.04$ 
& $0.48 \pm 0.02$ 
& $-$ 
& $0.36 \pm 0.04$ 
& $0.18 \pm 0.14$ 
\\
&&&&&&&\\[-3mm]
$ R(R_{L,T})$ 
& $ - $ 
& $0.79 \pm 0.02$ 
& $0.78 \pm 0.02$ 
& $0.80 \pm 0.02$ 
& $-$ 
& $0.95 \pm 0.05$ 
& $0.42 \pm 0.14$ 
\\
&&&&&&&\\[-3mm]
$ R(R_{A,B})$ 
& $ - $ 
& $0.520 \pm 0.004$ 
& $0.514\pm 0.005$ 
& $0.524 \pm 0.004$ 
& $-$ 
& $0.516 \pm 0.004$ 
& $0.64 \pm 0.07$ 
\\
&&&&&&&\\[-3mm]
$ R(A_{FB}^{D^\ast})$ 
& $ - $ 
& $0.23 \pm 0.04$ 
& $[-1.52,0.40]$ 
& $[-1.38,0.20]$
& $-$ 
& $0.00 \pm 0.06$ 
& $ -0.02 \pm 0.06\phantom{-}$ 
\\
&&&&&&&\\[-3mm]
$ R(A_3)$ 
& $ - $ 
& $0.62 \pm 0.01$ 
& $0.58 \pm 0.01$ 
& $0.63 \pm 0.02$ 
& $-$ 
& $0.56 \pm 0.02$ 
& $0.11 \pm 0.23$ 
\\
&&&&&&&\\[-3mm]
$ R(A_4)$ 
& $ - $ 
& $0.46 \pm 0.01$ 
& $0.45 \pm 0.01$ 
& $0.46 \pm 0.01$ 
& $-$ 
& $0.42 \pm 0.01$ 
& $0.06 \pm 0.18$ 
\\
&&&&&&&\\[-3mm]
$ R(A_5)$ 
& $ - $ 
& $1.15 \pm 0.02$ 
& $[-0.26,1.28]$
& $[-0.09,1.12]$
& $-$ 
& $1.24 \pm 0.05$
& $ 0.42 \pm 0.30$ 
\\
&&&&&&&\\[-3mm]
$ R(A_6)$ 
& $ - $ 
& $0.79 \pm 0.01$ 
& $[-0.96,0.96]$
& $[-0.76,0.76]$ 
& $-$ 
& $0.72 \pm 0.02$ 
& $0.15 \pm 0.25$ 
\\
&&&&&&&\\[-3mm]
$ D(A_7)$ 
& $ - $ 
& $0$ 
& $[-0.05,0.05]$
& $[-0.04,0.04]$
& $-$ 
& $0.00 \pm 0.01$
& $0.00 \pm 0.02$ 
\\
&&&&&&&\\[-3mm]
$ D(A_8)$ 
& $ - $ 
& $0$ 
& $[-0.03,0.03]$
& $[-0.03,0.03]$ 
& $-$ 
& $0$ 
& $0$ 
\\
&&&&&&&\\[-3mm]
$ D(A_9)$ 
& $ - $ 
& $0$ 
& $[-0.09,0.09]$
& $[-0.07,0.07]$ 
& $-$ 
& $0$ 
& $0$
\\
&&&&&&&\\
\hline
&&&&&&&\\[-3mm]
$ F_L^{D^*}$ 
& $ - $ 
& $0.47 \pm 0.02$ 
& $0.46 \pm 0.02$ 
& $0.46 \pm 0.02$ 
& $-$ 
& $0.51 \pm 0.02$ 
& $0.32 \pm 0.13$ \\[-3mm]
&&&&&&&\\
\hline
\end{tabular}
\caption{\label{tab:obsCLN}\small\sl The values of LFUV ratios/differences. Besides the SM results in each following column we show the results obtained by switching on the NP coupling $g_i$ [$i\in\{V,A,S,P,T\}$] corresponding to that column in such a way that we sweep through the values allowed by $R(D^{(\ast)})^\mathrm{exp}$ as shown in Fig.~\ref{fig:g_constraints}. For the quantities which barely exhibit a gaussian behavior (for the reasons discussed in the text) we provide the $2\sigma$ ranges. In the last line we also give the corresponding value of the fraction of longitudinally polarized $D^\ast$ in the sample of $B\to D^{(\ast)} \tau\bar{\nu}$.}
\end{sidewaystable}
\begin{sidewaystable}[!t!]
\fontsize{10.45}{13}\selectfont
\centering
\begin{tabular}{|c|c|c|c|c|c|c|}
\hline
&&&&&&\\[-1mm]
\textbf{Obs.} & Exp. & SM & $ g_{V_L} $ & $ g_{V_R} $ & $ g_{S_L} $ & $ {g_{S_R}} $    \\[1mm]
&&&&&&\\[-3mm]
\hline
&&&&&&\\[-3mm]
 $ R_D $  
& $0.334 \pm 0.029$ 
& $0.30 \pm 0.02$ 
& $0.33 \pm 0.02$ 
& $0.33 \pm 0.03$ 
& $0.36 \pm 0.03$ 
& $0.36 \pm 0.03$ 
\\
&&&&&&\\[-3mm]
 $ R_{D^*} $ 
& $0.297 \pm 0.014$ 
& $0.258 \pm 0.003$ 
& $0.30 \pm 0.01$ 
& $0.30 \pm 0.01$ 
& $0.26 \pm 0.01$ 
& $0.259 \pm 0.005$ 
\\
&&&&&&\\[-3mm]
\hline
&&&&&&\\[-3mm]
 $ R(A_{FB}^{D})$ 
 & $ - $ 
& $0.077 \pm 0.004$ 
& $0.074 \pm 0.003$ 
& $0.074 \pm 0.003$ 
& $0.06 \pm 0.01$ 
& $0.06 \pm 0.01$ 
\\
&&&&&&\\[-3mm]
 $ R(A_{\lambda_\ell}^D)$ 
 & $ - $ 
& $-0.332 \pm 0.003\phantom{-}$ 
& $-0.331 \pm 0.003\phantom{-}$ 
& $-0.331 \pm 0.003\phantom{-}$ 
& $-0.47 \pm 0.05\phantom{-}$ 
& $-0.47 \pm 0.05\phantom{-}$ 
\\
&&&&&&\\[-3mm]
\hline
&&&&&&\\[-3mm]
  $ R(A_{\lambda_\ell}^{D^\ast})$ 
 & $ - $ 
& $0.47 \pm 0.02$ 
& $0.48 \pm 0.02$ 
& $0.49 \pm 0.02$ 
& $0.47 \pm 0.05$ 
& $0.46 \pm 0.03$ 
\\
&&&&&&\\[-3mm]
 $ R(R_{L,T})$ 
 & $ - $ 
& $0.79 \pm 0.02$ 
& $0.79 \pm 0.02$ 
& $0.79 \pm 0.02$ 
& $0.81 \pm 0.05$ 
& $0.82 \pm 0.03$ 
\\
&&&&&&\\[-3mm]
 $ R(R_{A,B})$ 
 & $ - $ 
& $0.520 \pm 0.004$ 
& $0.520 \pm 0.004$ 
& $0.520 \pm 0.005$ 
& $0.520 \pm 0.004$ 
& $0.520 \pm 0.004$ 
\\
&&&&&&\\[-3mm]
 $ R(A_{FB}^{D^\ast})$ 
 & $ - $ 
& $0.23 \pm 0.04$ 
& $0.23 \pm 0.04$ 
& $-0.01 \pm 0.05\phantom{-}$ 
& $0.24 \pm 0.08$ 
& $0.23 \pm 0.08$ 
\\
&&&&&&\\[-3mm]
 $ R(A_3)$ 
 & $ - $ 
& $0.62 \pm 0.01$ 
& $0.62 \pm 0.02$ 
& $0.61 \pm 0.03$ 
& $0.61 \pm 0.02$ 
& $0.61 \pm 0.01$ 
\\
&&&&&&\\[-3mm]
 $ R(A_4)$ 
 & $ - $ 
& $0.46 \pm 0.01$ 
& $0.46 \pm 0.01$ 
& $0.46 \pm 0.01$ 
& $0.46 \pm 0.01$ 
& $0.45 \pm 0.01$ 
\\
&&&&&&\\[-3mm]
 $ R(A_5)$ 
 & $ - $ 
& $1.15 \pm 0.02$ 
& $1.15 \pm 0.03$ 
& $0.97 \pm 0.05$ 
& $1.04 \pm 0.03$ 
& $1.12 \pm 0.06$ 
\\
&&&&&&\\[-3mm]
 $ R(A_6)$ 
 & $ - $ 
& $0.79 \pm 0.01$ 
& $0.80 \pm 0.02$ 
& $0.55 \pm 0.05$ 
& $0.79 \pm 0.02$ 
& $0.78 \pm 0.02$ 
\\
&&&&&&\\[-3mm]
 $ D(A_7)$ 
 & $ - $ 
& $0$ 
& $0$ 
& \begin{tabular}{@{}c@{}}$-0.028 \pm 0.005\phantom{-}$ \\ $0.028 \pm 0.005$\end{tabular}  
& $[-0.02,0.02]$
&  $[-0.02,0.02]$ 
\\
&&&&&&\\[-3mm]
 $ D(A_8)$ 
 & $ - $ 
& $0$ 
& $0$ 
& \begin{tabular}{@{}c@{}}$-0.019 \pm 0.003\phantom{-}$ \\ $0.019 \pm 0.003$\end{tabular}  
& $0$ 
& $0$ 
\\
&&&&&&\\[-3mm]
 $ D(A_9)$ 
 & $ - $ 
& $0$ 
& $0$ 
& \begin{tabular}{@{}c@{}}$-0.05 \pm 0.01\phantom{-}$ \\ $0.05 \pm 0.01$\end{tabular} 
& $0$ 
& $0$ 
\\
&&&&&&\\
\hline
&&&&&&\\[-3mm]
$ F_{D^*}$ 
& $ - $ 
& $0.47 \pm 0.02$ 
& $0.46 \pm 0.02$ 
& $0.46 \pm 0.02$ 
& $0.47 \pm 0.05$ 
& $0.47 \pm 0.04$ \\[-3mm]
&&&&&&\\
\hline\end{tabular}
  \caption{\label{tab:obsCLN_LR} \sl Same as in Tab.~\ref{tab:obsCLN} but in the basis defined in the effective Hamiltonian~\eqref{eq:Heff_Jp}. We stress that for the form factors we use  CLN+HQET+LATT, as discussed in the text.}
\end{sidewaystable}

\subsection{Comment of $F_L^{D^\ast}$}

Very recently the Belle Collaboration presented the results of their first study of the fraction of the longitudinally polarized $D^\ast$'s in their full sample of $B\to D^{(\ast)} \tau\bar{\nu}$ and found $F_L^{D^\ast} = 0.60(9)$~\cite{Abdesselam:2019wbt}, where we combined the experimental errors in quadrature. That value turns out to be less than $2\sigma$ larger than predicted in the SM, which we find to be $(F_L^{D^\ast})^\mathrm{SM} = 0.47(2)$. As it can be seen in the last line of Tabs.~\ref{tab:obsCLN} and \ref{tab:obsCLN_LR}, it is very difficult to make the value of $F_L^{D^\ast} > (F_L^{D^\ast})^\mathrm{SM}$. Only a marginal enhancement is allowed by switching on $g_{S_L,S_R,P}$ while all the other non-zero NP coefficients would make  $F_L^{D^\ast} < (F_L^{D^\ast})^\mathrm{SM}$. We hope this quantity will be further scrutinized by the other experimental groups and its precision will be improved to the point that it can play an important role 
in discriminating among various models.

Since we consistently assume that NP arises from coupling to $\tau$, measuring this quantity in the case of $\mu$ or $e$ in the final state would be a very helpful check of our  assumption. In other words, if our assumption is right then the measured $F_L^{D^\ast}(\mu,e)$ should be equal to $\bigl( F_L^{D^\ast}(\mu,e)\bigr)^\mathrm{SM}=0.52(1)$.

\section{Summary\label{sec4}}

In this paper we discussed a possibility of using a set of observables that can be extracted from the angular distribution of the $B\to D^{(\ast)} \ell\bar{\nu}$ decays in order to study the effects of LFUV. In particular, we define $3$ such observables that can be obtained from the angular distribution of the $B\to D \ell\bar{\nu}$ decay, and $12$ from $B\to D^{\ast} \ell\bar{\nu}$. 

NP contribution to $B\to D^{(\ast)} \tau\bar{\nu}$ can be parametrized by the couplings $g_{V_L,V_R,S_L,S_R,T_L}$ (defined in the text), the values of which can be constrained by the experimentally measured $R{(D^{(\ast)})}$, found to be larger than its SM prediction. We explored the possibility of feeding  $R{(D^{(\ast)})}^\mathrm{exp}-R{(D^{(\ast)})}^\mathrm{SM}$ by turning on one coupling at the time and found that the measurement even of a subset of observables can indeed help disentangling among various possibilities. In other words, their measurement can considerably help in constraining the complex valued couplings $g_i$'s and thereby help us understanding the Lorentz structure of the NP contribution(s). For that purpose, like in the case of 
$R{(D^{(\ast)})}$ we point out that for the observables for which the SM prediction is non-zero it is convenient to consider the ratios between the value extracted from the $B\to D^{(\ast)} \tau\bar{\nu}$ mode and the one extracted from $B\to D^{(\ast)} l\bar{\nu}$ with $l\in (e,\mu)$. Instead, for the quantities which are zero in the SM we consider the differences. 

Since many of the proposed observables provide information of the NP couplings that cannot be accessed through the measurement of the branching fraction, we show that even in the case in which 
$R{(D^{(\ast)})}^\mathrm{exp}=R{(D^{(\ast)})}^\mathrm{SM}$ one can still have non-zero NP couplings which can be checked by measuring the ratios/differences of the observables deduced from the angular distributions of the considered decay modes. 

Even though the observables discussed in this paper are most interesting in the case of $\tau$-lepton in the final state, their experimental measurements in the case of $e$ and/or $\mu$ in the final state are very important too. They would help us checking on the assumption that is mostly made in the literature, namely that he NP affects only the couplings to $\tau$ and not to $e$ or $\mu$. In that case the measurements would coincide with the SM predictions. 

Furthermore, in some models of NP the coupling to $\mu$ can be large whereas the one to the electron negligibly small~\cite{Becirevic:2016oho}. In that situation it is very important to check also on the ratios of the observables discussed here in the case of $B\to D^{(\ast)} \mu\bar{\nu}$ with respect to those measured in the case of $B\to D^{(\ast)} e\bar{\nu}$.

Finally, in order to make this study quantitatively sound, beside the experimental input, one also needs the lattice QCD information concerning the shapes of hadronic form factors relevant to the $B\to D^{\ast} \ell\bar \nu$ decay, which are still missing.

\section*{Acknowledgements}
{\small We wish to thank Gudrun Hiller and Donal Hill for useful discussions during the completion of the manuscript. M.F. thanks the Fondazione Della Riccia for partial financial support during the completion of this work, and acknowledges the financial support from MINECO grant FPA2016- 76005-C2-1-P, Maria de Maetzu program grant MDM-2014-0367 of ICCUB and 2017 SGR 929. The work of I.N. is supported by BMBF under grant no. 05H18VKKB1. The work of A.T. has been supported  by the DFG Research Unit FOR 1873 ``Quark Flavour Physics and Effective Field Theories''. 
This  project  has also  received  support  from  the  European  Union's  Horizon  2020  research  and  innovation
programme under the Marie Sklodowska-Curie grant agreement N.~690575 and 674896.}

\FloatBarrier
\newpage
\appendix

\section{Form factors\label{app:FF}}

The hadronic matrix element are parametrized as follows:
\begin{equation}
\langle D(k)|\cbar\gamma_\mu b|\Bbar(p)\rangle = \left[(p+k)_\mu-{m_B^2-m_D^2 \over q^2}q_\mu\right]f_+(q^2)+q_\mu{m_B^2-m_D^2 \over q^2}f_0(q^2) \,,
\label{eq:vector_FF_D}
\end{equation}

\begin{equation}
\begin{split}
\langle D(k)|\cbar b|\Bbar(p)\rangle =& {1 \over m_b-m_c}q^\mu\langle D(k)|\cbar\gamma_\mu b|\Bbar(p)\rangle = {m_B^2-m_D^2 \over m_b-m_c}f_0(q^2) \,, \\
\langle D(k)|\cbar \gamma_5 b|\Bbar(p)\rangle =& 0 \,,
\end{split}
\label{eq:scalar_FF_D}
\end{equation}

\begin{equation}
\begin{split}
\langle D(k)|\cbar\sigma_{\mu\nu} b|\Bbar(p)\rangle =& -i\left(p_\mu k_\nu-k_\mu p_\nu\right) {2f_T(q^2) \over m_B+m_D} \,, \\
\langle D(k)|\cbar\sigma_{\mu\nu} \gamma_5 b|\Bbar(p)\rangle =& -{i \over 2} \epsilon_{\mu\nu\alpha\beta} \langle D(k)|\cbar\sigma^{\alpha\beta} b|\Bbar(p)\rangle = -\epsilon_{\mu\nu\alpha\beta} p^\alpha k^\beta {2f_T(q^2) \over m_B+m_D} \,.
\end{split}
\label{eq:tensor_FF_D}
\end{equation}

\begin{equation}
\begin{split}
\langle\Dst(k,\varepsilon)|\cbar\gamma_\mu b|\Bbar(p)\rangle =& -i\epsilon_{\mu\nu\alpha\beta}\varepsilon^{*\nu}p^\alpha k^\beta{2V(q^2) \over m_B+m_\Dst} \,, \\
\langle\Dst(k,\varepsilon)|\cbar\gamma_\mu\gamma_5 b|\Bbar(p)\rangle =& \varepsilon_{\mu}^* (m_B+m_\Dst)A_1(q^2) - (p+k)_\mu(\varepsilon^*q){A_2(q^2) \over m_B+m_\Dst} \\
& -q_\mu(\varepsilon^*q){2m_\Dst \over q^2}\Bigl[ A_3(q^2)-A_0(q^2)\Bigr] \,,
\end{split}
\label{eq:vector_FF_Dst}
\end{equation}
with
\begin{equation}
A_3(q^2) = {m_B+m_\Dst \over 2m_\Dst} A_1(q^2) - {m_B-m_\Dst \over 2m_\Dst} A_2(q^2) \,,
\end{equation}

\begin{equation}
\begin{split}
\langle\Dst(k,\varepsilon)|\cbar b|\Bbar(p)\rangle =& 0 \,, \\
\langle\Dst(k,\varepsilon)|\cbar\gamma_5 b|\Bbar(p)\rangle =& -{1 \over m_b+m_c}q^\mu\langle\Dst(k,\varepsilon)|\cbar\gamma_\mu\gamma_5 b|\Bbar(p)\rangle \\
=& -(\varepsilon^* q){2m_\Dst \over m_b+m_c}A_0(q^2) \,,
\end{split}
\label{eq:scalar_FF_Dst}
\end{equation}

The tensor contribution can be parametrized as
\begin{equation}
\begin{split}
\langle\Dst(k,\varepsilon)|\cbar\sigma_{\mu\nu} b|\Bbar(p)\rangle =& \epsilon_{\mu\nu\alpha\beta} \left[ \varepsilon^{*\alpha}(p+k)^\beta g_+(q^2) + \varepsilon^{*\alpha}q^\beta g_-(q^2) \right. \\
& \left. \qquad + (\varepsilon^* q)p^\alpha k^\beta g_0(q^2) \right] \,, \\
\langle\Dst(k,\varepsilon)|\cbar\sigma_{\mu\nu} \gamma_5 b|\Bbar(p)\rangle =& -{i \over 2} \epsilon_{\mu\nu\alpha\beta} \langle\Dst(k,\varepsilon)|\cbar\sigma^{\alpha\beta} b|\Bbar(p)\rangle \\
=& i \left\{ \left[ \varepsilon_\mu^*(p+k)_\nu - (p+k)_\mu\varepsilon_\nu^* \right] g_+(q^2)\right. \\
& \left. + \left[ \varepsilon_\mu^*q_\nu - q_\mu\varepsilon_\nu^* \right] g_-(q^2) + \left( \varepsilon^*q \right) \left[ p_\mu k_\nu - k_\mu p_\nu \right] g_0(q^2) \right\} \,,
\end{split}
\label{eq:tensor_FF_Dst}
\end{equation}
where $g_{\pm,0}$ can be related to the ``standard'' $T_{1-3}$ form factors as
\begin{equation}
\begin{split}
g_+(q^2) =& -T_1(q^2) \,, \\
g_-(q^2) =& {m_B^2-m_\Dst^2 \over q^2}[T_1(q^2)-T_2(q^2)] \,, \\
g_0(q^2) =& {2 \over q^2} \left[ T_1(q^2)-T_2(q^2) - {q^2 \over m_B^2 - m_\Dst^2} T_3(q^2) \right] \,.
\end{split}
\end{equation}

The additional form factors commonly used in the literature are defined as
\begin{equation}
\begin{split}
A_{12}(q^2) &= {1 \over 16m_B m_\Dst^2} \biggl[ (m_B^2-m_\Dst^2-q^2)(m_B+m_\Dst) A_1(q^2) - {\lambda_{B\Dst}(q^2) \over m_B+m_\Dst} A_2(q^2) \biggr] \,, \\
T_{23}(q^2) &= {1 \over 8m_B m_\Dst^2} \biggl[ (m_B^2+3m_\Dst^2-q^2)(m_B+m_\Dst) T_2(q^2) - {\lambda_{B\Dst}(q^2) \over m_B-m_\Dst} T_3(q^2) \biggr] \,.
\end{split}
\end{equation}

In order to cancel the divergence at $q^2=0$, the following conditions must be imposed
\begin{equation}
f_+(0)=f_0(0) \,, \quad A_0(0) = A_3(0) \,, \quad T_1(0)=T_2(0) \,, \quad A_{12}(0) = {m_B^2-m_\Dst^2 \over 8m_B m_\Dst} A_3(0) \,.
\end{equation}

In this work we use the convention $\epsilon_{0123}=1$ (or equivalently $\epsilon^{0123}=-1$). 
Note that for the alternative convention, $\epsilon_{0123}=-1$, the pseudo-tensor matrix elements in Eqs.\eqref{eq:tensor_FF_D},\eqref{eq:tensor_FF_Dst} change the sign since $\sigma_{\mu\nu}\gamma_5=-{\rm sgn}[\epsilon_{0123}](i/2)\epsilon_{\mu\nu\alpha\beta}\sigma^{\alpha\beta}$.

\section{Helicity amplitude formalism \label{app:helicity_amp}}

Using the property of the off-shell vector boson $V^*$ polarization vectors,
\begin{equation}
\sum_\lambda \eta_\mu^*(\lambda) \eta_\nu(\lambda) \delta_\lambda = g_{\mu\nu} \,, \quad\quad \delta_{0,\pm}=-\delta_t=-1\,,
\label{eq:delta_factor}
\end{equation}
one can write the $\Bbar \to MV^*\to M\ell\nubar$ ($M=D,\Dst$) amplitudes of general vector and tensor currents as
\begin{equation}
\begin{split}
\M_{V(A)}^{\lambda_M,\,\lambda_\ell} &\propto \langle M(\lambda_M) |J_{\rm had}^\mu|\Bbar\rangle \langle\ell(\lambda_\ell)\nubar|J_{\rm lep,\,\mu}|0\rangle \\
&= \sum_\lambda \eta_\mu^*(\lambda) \langle M(\lambda_M) |J_{\rm had}^\mu|\Bbar\rangle  \, \eta_\nu(\lambda) \langle\ell(\lambda_\ell)\nubar|J_{\rm lep}^\nu|0\rangle \delta_\lambda \\
&= \sum_\lambda \delta_\lambda H_{V(A),\,\lambda}^{\lambda_M} L_{V-A,\,\lambda}^{\lambda_\ell} \,, \\
\M_{T(T5)}^{\lambda_M,\,\lambda_\ell} \propto& \, \langle M(\lambda_M) |J_{\rm had}^{\mu\nu}|\Bbar\rangle \langle\ell(\lambda_\ell)\nubar|J_{\rm lep,\,\mu\nu}|0\rangle \\
&= \sum_{\lambda,\lambda^\prime} i\eta_\mu^*(\lambda)\eta_\nu^*(\lambda^\prime) \langle M(\lambda_M) |J_{\rm had}^{\mu\nu}|\Bbar\rangle  \, (-i)\eta_\alpha(\lambda) \eta_\beta(\lambda^\prime) \langle\ell(\lambda_\ell)\nubar|J_{\rm lep}^{\alpha\beta}|0\rangle \delta_\lambda \delta_{\lambda^\prime} \\
&= \sum_{\lambda,\lambda^\prime} \delta_\lambda \delta_{\lambda^\prime} H_{T(T5),\,\lambda\lambda^\prime}^{\lambda_M} L_{T-T5,\,\lambda\lambda^\prime}^{\lambda_\ell} \,,
\end{split}
\label{eq:VT_amp_formalism}
\end{equation}
where $H$ and $L$ denote the leptonic and hadronic helicity amplitudes defined in Eqs.~\eqref{eq:leptonic_amp},\eqref{eq:hadronic_amp}. The $\pm i$ factors in $\M_{T(T5)}$ are introduced for convenience, in order to make all hadronic $\Bbar\to M V^*$ and leptonic amplitudes real if $g_i\in\mathbb{R}$ and when $\chi\to0$ (cf. Fig.~\ref{fig:kinematics}).  The scalar amplitude is simply defined as
\begin{equation}
\M_{S(P)}^{\lambda_\ell} \propto \langle M(\lambda_M=0) |J_{\rm had}|\Bbar\rangle \langle\ell(\lambda_\ell)\nubar|J_{\rm lep}|0\rangle = H_{S(P)} L_{S-P}^{\lambda_\ell} \,.
\end{equation}
Here $\lambda_M$ and $\lambda^{(\prime)}$ denote the meson and the virtual boson helicities in the $\Bbar$ reference frame. The lepton helicity $\lambda_\ell$ is defined in the $\ell\nubar$ rest frame.

For the four-body final state $\Bbar\to\Dst V^*\to D\pi\ell\nubar$ decay the total amplitude has the form
\begin{equation}
\M_X^{\lambda_M,\,\lambda_\ell} \propto \langle D\pi|\Dst(\lambda_\Dst)\rangle \langle \Dst(\lambda_\Dst) |J_{\rm had}^X|\Bbar\rangle \langle\ell(\lambda_\ell)\nubar|J_{\rm lep,\,X}|0\rangle \, BW_\Dst \,,
\label{eq:BtoDpilnu_amp}
\end{equation}
where the propagation of the intermediate resonant state is parametrized by the Breit-Wigner function,
\begin{equation}
BW_\Dst(m_{D\pi}^2) = {1 \over m_{D\pi}^2-m_\Dst^2+im_\Dst\Gamma_\Dst} \,.
\end{equation}
Since the width of $\Dst$ is very small, one can use the narrow width approximation,
\begin{equation}
{1 \over (m_{D\pi}^2-m_\Dst^2)^2+m_\Dst^2\Gamma_\Dst^2} \xrightarrow{\Gamma_\Dst\ll m_\Dst} {\pi \over m_\Dst\Gamma_\Dst}\delta(m_{D\pi}^2-m_\Dst^2) \,,
\label{eq:NWA}
\end{equation}
and integrate out the $m_{D\pi}$ dependence in the phase space.

\subsection{Leptonic amplitudes \label{app:lep_amp}}

The leptonic amplitudes are defined as
\begin{equation}
\begin{split}
L_{V-A,\,\lambda}^{\lambda_\ell} (q^2,\chi,\theta_\ell) &= \eta_\mu(\lambda) \langle\ell(\lambda_\ell)\nubar|\lbar\gamma^\mu(1-\gamma_5)\nu|0\rangle \,, \\
L_{S-P}^{\lambda_\ell} (q^2,\chi,\theta_\ell) &= \langle\ell(\lambda_\ell)\nubar|\lbar(1-\gamma_5)\nu|0\rangle \,, \\
L_{T-T5,\,\lambda\lambda^\prime}^{\lambda_\ell} (q^2,\chi,\theta_\ell) &= -L_{T-T5,\,\lambda^\prime\lambda}^{\lambda_\ell} = -i\eta_\mu(\lambda)\eta_\nu(\lambda^\prime) \langle\ell(\lambda_\ell)\nubar|\lbar\sigma^{\mu\nu}(1-\gamma_5)\nu|0\rangle \,.
\end{split}
\label{eq:leptonic_amp}
\end{equation}
Using the polarization vectors and Dirac spinors, given in the Appendix~\ref{app:pol_spin}, and gamma-matrices in the Weyl (chiral) representation, one can obtain the explicit formulas for the vector type amplitudes:
\begin{equation}
\begin{split}
L_{V-A,\,+}^+ (q^2,\chi,\theta_\ell) &= \pm\sqrt2 m_\ell \beta_\ell \sin\thl \, e^{-2i\chi} \,, \\
L_{V-A,\,-}^+ (q^2,\chi,\theta_\ell) &= \pm\sqrt2 m_\ell \beta_\ell \sin\thl \,, \\
L_{V-A,\,0}^+ (q^2,\chi,\theta_\ell) &= 2 m_\ell \beta_\ell \cos\thl \, e^{-i\chi} \,, \\
L_{V-A,\,t}^+ (q^2,\chi,\theta_\ell) &= -2m_\ell \beta_\ell \, e^{-i\chi} \,, \\
L_{V-A,\,\pm}^- (q^2,\chi,\theta_\ell) &= \sqrt{2q^2} \beta_\ell (1\pm\cos\thl) \, e^{\mp i\chi} \,, \\
L_{V-A,\,0}^- (q^2,\chi,\theta_\ell) &= -2\sqrt{q^2} \beta_\ell \sin\thl \,, \\
L_{V-A,\,t}^- (q^2,\chi,\theta_\ell) &= 0 \,,
\end{split}
\label{eq:lepton_V_amp}
\end{equation}
with $\beta_\ell=\sqrt{1-m_\ell^2/q^2}$. Expressions for the scalar leptonic amplitudes read
\begin{equation}
\begin{split}
L_{S-P}^+ (q^2,\chi,\theta_\ell) &= -2\sqrt{q^2} \beta_\ell \, e^{-i\chi}  \,, \\
L_{S-P}^- (q^2,\chi,\theta_\ell) &= 0 \,.
\end{split}
\label{eq:lepton_S_amp}
\end{equation}
Finally, the tensor type amplitudes are given by
\begin{equation}
\begin{split}
L_{T-T5,\,+0}^+ (q^2,\chi,\theta_\ell) &= \sqrt{2q^2} \beta_\ell \sin\thl \, e^{-2i\chi} \,, \\
L_{T-T5,\,-0}^+ (q^2,\chi,\theta_\ell) &= \sqrt{2q^2} \beta_\ell \sin\thl \,, \\
L_{T-T5,\,+-}^+ (q^2,\chi,\theta_\ell) &= -L_{T-T5,\,0t}^+ = 2\sqrt{q^2} \beta_\ell \cos\thl e^{-i\chi} \,, \\
L_{T-T5,\,+t}^+ (q^2,\chi,\theta_\ell) &= \mp\sqrt{2q^2} \beta_\ell \sin\thl \, e^{-2i\chi} \,, \\
L_{T-T5,\,-t}^+ (q^2,\chi,\theta_\ell) &= \mp\sqrt{2q^2} \beta_\ell \sin\thl \,, \\
L_{T-T5,\,\pm0}^- (q^2,\chi,\theta_\ell) &= \pm\sqrt2 m_\ell \beta_\ell (1\pm\cos\thl) \, e^{\mp i\chi} \,, \\
L_{T-T5,\,+-}^- (q^2,\chi,\theta_\ell) &= -L_{T-T5,\,0t}^- = -2m_\ell \beta_\ell \sin\thl \,, \\
L_{T-T5,\,\pm t}^- (q^2,\chi,\theta_\ell) &= -\sqrt2 m_\ell \beta_\ell (1\pm\cos\thl) \, e^{\mp i\chi} \,.
\end{split}
\label{eq:lepton_T_amp}
\end{equation}
By setting $\chi\to0$ (i.e. the $x-z$ plane is defined by the lepton momentum) one ends up with the expressions coinciding with those given in Refs.~\cite{Tanaka:2012nw,Hagiwara:1989cu}.

\subsection{Hadronic amplitudes \label{app:had_amp}}

The general helicity amplitudes, in the operator basis~\eqref{eq:Heff}, read:
\begin{equation}
\begin{split}
H_{V(A),\,\lambda}^{\lambda_M}(q^2) &= (g_{V(A)}\pm1) \, \eta_\mu^*(\lambda) \, \langle M(\lambda_M) | \cbar \gamma^\mu (\gamma_5) b | \Bbar \rangle \,, \\
H_{S(P),\,\lambda}(q^2) &= g_{S(P)} \, \langle M(\lambda_M=0) | \cbar (\gamma_5) b | \Bbar \rangle \,, \\
H_{T(T5),\,\lambda\lambda^\prime}^{\lambda_M}(q^2) &= -H_{T,\,\lambda^\prime\lambda}^{\lambda_M}(q^2) = i \,g_{T(T5)}\,\eta_\mu^*(\lambda)\eta_\nu^*(\lambda^\prime) \, \langle M(\lambda_M) | \cbar \sigma^{\mu\nu} (\gamma_5) b | \Bbar \rangle \,, \\
\end{split}
\label{eq:hadronic_amp}
\end{equation} 
which can be written explicitly for the $\Bbar\to D$ case as~\footnote{To avoid confusion and simplify notation, we denote the $\Bbar\to D$ amplitudes by $h$ and omit the super-index $\lambda_D=0$.} 
\begin{subequations}
\begin{align}
h_0(q^2) &\equiv h_{V,\,0}(q^2) = (g_V+1) \sqrt{\lambda_{BD}(q^2) \over q^2} f_+(q^2) \,, \\
& \nonumber \\
h_t(q^2) &\equiv h_{V,\,t}(q^2) = (g_V+1) {m_B^2-m_D^2 \over \sqrt{q^2}} f_0(q^2) \,, \\
& \nonumber \\
h_S(q^2) &\simeq g_S {m_B^2-m_D^2 \over m_b-m_c} f_0(q^2) \,, \\
& \nonumber \\
h_{T,\,0t}(q^2) &= -{g_T \over g_{T5}} h_{T5,\,+-}(q^2) = \, -g_T {\sqrt{\lambda_{BD}(q^2)} \over m_B+m_D} f_T(q^2) \,, \\
\label{eq:hT}
& \nonumber \\
h_{A,\,\lambda}(q^2) &= h_P(q^2) = h_{T,\,+-}(q^2) = h_{T5,\,0t}(q^2) = 0 \,,
\end{align} 
\label{eq:BtoD_amp}
\end{subequations}
and for the $\Bbar\to\Dst$ case as
\begin{subequations}
\begin{align}
H_{V,\,\pm}^\pm(q^2) &= \mp (g_V+1) { \sqrt{\lambda_{B\Dst}(q^2)} \over m_B+m_\Dst } V(q^2) \,, \\
& \nonumber \\
H_{A,\,\pm}^\pm(q^2) &= -(g_A-1) (m_B+m_\Dst) A_1(q^2) \,, \\
& \nonumber \\
H_0(q^2) &\equiv H_{A,\,0}^0(q^2) = (g_A-1){8m_B m_\Dst \over \sqrt{q^2} }  A_{12}(q^2) \,, \\
& \nonumber \\
H_t(q^2) &\equiv H_{A,\,t}^0(q^2) = (g_A-1)\sqrt{ \lambda_{B\Dst}(q^2) \over q^2 } A_0(q^2) \,, \\
& \nonumber \\
H_P(q^2) &\simeq -g_P {\sqrt{\lambda_{B\Dst}(q^2)} \over m_b+m_c} A_0(q^2) \,, \\
& \nonumber \\
H_{T,\,\pm0}^\pm(q^2) &= \mp {g_T \over g_{T5}} H_{T5,\,\pm t}^\pm(q^2) = \pm g_T {m_B^2-m_\Dst^2 \over \sqrt{q^2}} T_2(q^2) \,, \\
& \nonumber \\
H_{T,\,\pm t}^\pm(q^2) &= \mp {g_T \over g_{T5}} H_{T5,\,\pm 0}^\pm(q^2) = \, \pm g_T \sqrt{\lambda_{B\Dst}(q^2) \over q^2} T_1(q^2) \,, \\
& \nonumber \\
H_{T,\,+-}^0(q^2) &= -{g_T \over g_{T5}} H_{T5,\,0t}^0(q^2) = -g_T {4m_B m_\Dst \over m_B+m_\Dst} T_{23}(q^2) \,, \\
& \nonumber \\
H_{V,\,0}^0(q^2) &= H_{V,\,t}^0(q^2) = H_S(q^2) = H_{T,\,0t}^0(q^2) = H_{T5,\,+-}^0(q^2) = 0 \,,
\end{align}
\label{eq:BtoDst_amp}
\end{subequations}
where again $\lambda_{BM}(q^2) = m_B^4 + m_M^4 + q^4 - 2(m_B^2 m_M^2 + m_B^2 q^2 + m_M^2 q^2)$. 
In deriving the expressions for the above amplitudes we used the decomposition of the hadronic matrix elements in terms of form factors listed in Appendix~\ref{app:FF}.

\subsection{$\bm{\Dst\to D\pi}$ amplitude}

The $\Dst\to D\pi$ amplitude can be parametrized as
\begin{equation}
\langle D\pi|\Dst(\lambda_\Dst)\rangle = g_{\Dst D\pi} \, \varepsilon_\mu(\lambda_\Dst) p_D^\mu \,,
\end{equation}
where the coupling $g_{\Dst D\pi}$ parameterizes the physical $\Dst\to D\pi$ decay and can be determined from the numerical simulations of QCD on the lattice or extracted from the measured width of $D^{*+}$,
\begin{equation}
\Gamma(\Dst\to D\pi) = {C \over 24 \pi m_\Dst^2}\  g_{\Dst D\pi}^2  |\p_D|^3 \,,
\label{eq:G_DstDpi}
\end{equation}
where $C=1$ if the outgoing pion is charged, and $C=1/2$ if it is neutral, and $\p_D$ is the $D$ three-momentum in the $\Dst$ rest frame. It must be stressed that $g_{\Dst D\pi}$ is $m_{D\pi}^2$-independent, and the entire dependence of the amplitude \eqref{eq:BtoDpilnu_amp} on $m_{D\pi}^2$ is assumed to be described by the Breit-Wigner function.

The amplitudes are computed in the $\Dst$ reference frame, where the $D,\pi$ momenta are in the $x-z$ plane, and are given by
\begin{equation}
\begin{split}
\langle D\pi|\Dst(\pm1)\rangle =& \pm{1\over\sqrt2} g_{\Dst D\pi} |\p_D| \sin\thD \,, \\
\langle D\pi|\Dst(0)\rangle =& -g_{\Dst D\pi} |\p_D| \cos\thD \,.
\end{split}
\end{equation}

\subsection{Relations between amplitudes \label{app:amp_relations}}

One can notice from Eqs.~\eqref{eq:lepton_V_amp} and \eqref{eq:lepton_S_amp} that
\begin{equation}
L_{S-P}^{\lambda_\ell}={\sqrt{q^2}\over m_\ell}L_{V-A,\,t}^{\lambda_\ell} \,.
\end{equation}
Therefore, in order to further simplify the expressions, one can absorb the hadronic $S/P$ amplitudes into the $V/A$ time-like ones and redefine,
\begin{equation}
\Htil_{V/A,\,t}^{\lambda_M(=0)} \equiv H_{V/A,\,t}^0 + {\sqrt{q^2}\over m_\ell} H_{S/P} \,.
\end{equation}
Moreover, from Eqs.~\eqref{eq:lepton_V_amp} and \eqref{eq:lepton_T_amp} it can be seen that
\begin{equation}
\begin{split}
L_{T-T5,\,\pm0}^+ &= \pm {\sqrt{q^2} \over m_\ell} L_{V-A,\,\pm}^+ \,, \\
L_{T-T5,\,+-}^+ &= \, -L_{T-T5,\,0t}^+ = \, {\sqrt{q^2} \over m_\ell} L_{V-A,\,0}^+ \,, \\
L_{T-T5,\,\pm t}^+ &= -{\sqrt{q^2} \over m_\ell} L_{V-A,\,\pm}^+ \,, \\
L_{T-T5,\,\pm0}^- &= \pm {m_\ell \over \sqrt{q^2}} L_{V-A,\,\pm}^- \,, \\
L_{T-T5,\,+-}^- &= \, -L_{T-T5,\,0t}^- = \, {m_\ell \over \sqrt{q^2}} L_{V-A,\,0}^- \,, \\
L_{T-T5,\,\pm t}^- &= -{m_\ell \over \sqrt{q^2}} L_{V-A,\,\pm}^- \,.
\end{split}
\label{eq:LV-LT_relations}
\end{equation}
In this way, summing over the off-shell boson polarizations $\lambda,\lambda^\prime$ and taking into account the $\delta_\lambda$ factors in Eq.~\eqref{eq:VT_amp_formalism}, one can write the general total amplitude as
\begin{equation}
\begin{split}
\M^{\lambda_M,\,\lambda_\ell} \propto& -\sum_{\lambda=\pm,0} (H_{V,\,\lambda}^{\lambda_M} + H_{A,\,\lambda}^{\lambda_M}) L_{V-A,\,\lambda}^{\lambda_\ell} + (\Htil_{V,\,t}^{\lambda_M} + \Htil_{A,\,t}^{\lambda_M}) L_{V-A,\,t}^{\lambda_\ell} \\
& + 2\sum_{\lambda=\pm} \biggl[ (H_{T,\,\lambda0}^{\lambda_M} + H_{T5,\,\lambda0}^{\lambda_M}) L_{T-T5,\,\lambda0}^{\lambda_\ell} - (H_{T,\,\lambda t}^{\lambda_M} + H_{T5,\,\lambda t}^{\lambda_M}) L_{T-T5,\,\lambda t}^{\lambda_\ell} \biggr] \\
& + 2(H_{T,\,+-}^{\lambda_M} + H_{T5,\,+-}^{\lambda_M}) L_{T-T5,\,+-}^{\lambda_\ell} - 2(H_{T,\,0t}^{\lambda_M} + H_{T5,\,0t}^{\lambda_M}) L_{T-T5,\,0t}^{\lambda_\ell} \,.
\end{split}
\end{equation}
Using the relations \eqref{eq:LV-LT_relations} one can write it in a more compact form :
\begin{equation}
\M^{\lambda_M,\,\lambda_\ell} \propto -\sum_{\lambda=\pm,0} \Htil_\lambda^{\lambda_M,\,\lambda_\ell} L_{V-A,\,\lambda}^{\lambda_\ell} + \Htil_t^{\lambda_M} L_{V-A,\,t}^{\lambda_\ell}\,, \\
\end{equation}
where the redefined amplitudes are given as,
\begin{equation}
\begin{split}
\Htil_\pm^{\lambda_M,\,+} &\equiv H_{V,\,\pm}^{\lambda_M} + H_{A,\,\pm}^{\lambda_M} - 2{\sqrt{q^2} \over m_\ell} \bigl( H_{T,\,\pm t}^{\lambda_M} \pm H_{T,\,\pm0}^{\lambda_M} + H_{T5,\,\pm t}^{\lambda_M} \pm H_{T5,\,\pm0}^{\lambda_M} \bigr) \,, \\
\Htil_0^{\lambda_M,\,+} &\equiv H_{V,\,0}^{\lambda_M} + H_{A,\,0}^{\lambda_M} - 2{\sqrt{q^2} \over m_\ell} \bigl( H_{T,\,+-}^{\lambda_M} + H_{T,\,0t}^{\lambda_M} + H_{T5,\,+-}^{\lambda_M} + H_{T5,\,0t}^{\lambda_M} \bigr) \,, \\
\Htil_\pm^{\lambda_M,\,-} &\equiv H_{V,\,\pm}^{\lambda_M} + H_{A,\,\pm}^{\lambda_M} - 2{m_\ell \over \sqrt{q^2}} \bigl( H_{T,\,\pm t}^{\lambda_M} \pm H_{T,\,\pm0}^{\lambda_M} + H_{T5,\,\pm t}^{\lambda_M} \pm H_{T5,\,\pm0}^{\lambda_M} \bigr) \,, \\
\Htil_0^{\lambda_M,\,-} &\equiv H_{V,\,0}^{\lambda_M} + H_{A,\,0}^{\lambda_M} - 2{m_\ell \over \sqrt{q^2}} \bigl( H_{T,\,+-}^{\lambda_M} + H_{T,\,0t}^{\lambda_M} + H_{T5,\,+-}^{\lambda_M} + H_{T5,\,0t}^{\lambda_M} \bigr) \,, \\
\Htil_t^{\lambda_M} &\equiv \Htil_{V,\,t}^{\lambda_M} + \Htil_{A,\,t}^{\lambda_M} \,.
\label{eq:Htil_general}
\end{split}
\end{equation}

Absorbing the scalar and tensor amplitudes into $\htil$ and $\Htil$ allows to significantly simplify the calculations and to write the expressions for the angular coefficients of the differential decay rate in a much more compact way. To be more specific, we introduce the following linear combinations :
\begin{equation}
\begin{split}
\htil_0^+(q^2) &\equiv h_0(q^2)- 2{\sqrt{q^2} \over m_\ell} h_T(q^2) \,, \\
\htil_0^-(q^2) &\equiv h_0(q^2) - 2{m_\ell \over \sqrt{q^2}} h_T(q^2) \,, \\
\htil_t(q^2) &\equiv h_t(q^2) +{\sqrt{q^2} \over m_\ell^2} h_S(q^2) \,,
\end{split}
\label{eq:htil}
\end{equation}
and
\begin{equation}
\begin{split}
\Htil_\pm^+(q^2) &\equiv H_\pm(q^2) - 2{\sqrt{q^2} \over m_\ell} H_{T,\,\pm}(q^2) \,, \\
\Htil_0^+(q^2) &\equiv H_0(q^2) - 2{\sqrt{q^2} \over m_\ell} H_{T,\,0}(q^2) \,, \\
\Htil_\pm^-(q^2) &\equiv H_\pm(q^2) - 2{m_\ell \over \sqrt{q^2}} H_{T,\,\pm}(q^2) \,, \\
\Htil_0^-(q^2) &\equiv H_0(q^2) - 2{m_\ell \over \sqrt{q^2}} H_{T,\,0}(q^2) \,, \\
\Htil_t(q^2) &\equiv H_t(q^2) +{\sqrt{q^2} \over m_\ell^2} H_P(q^2) \,,
\end{split}
\label{eq:Htil}
\end{equation}
with
\begin{equation}
\begin{split}
h_T(q^2) &\equiv h_{T,\,0t}(q^2) + h_{T5,\,+-}(q^2) \,, \\
H_\pm(q^2) &\equiv H_{V,\,\pm}^\pm(q^2) + H_{A,\,\pm}^\pm(q^2) \,, \\
H_{T,\,\pm}(q^2) &\equiv H_{T,\,\pm t}^\pm(q^2) \pm H_{T,\,\pm0}^\pm(q^2) + H_{T5,\,\pm t}^\pm(q^2) \pm H_{T5,\,\pm0}^\pm(q^2) \,, \\
H_{T,\,0}(q^2) &\equiv H_{T,\,+-}^0(q^2) +H_{T5,\,0t}^0(q^2) \,.
\end{split}
\end{equation}
To simplify and shorten the final expressions of angular observables, we omit the super-index $\lambda_M$ in $\htil$ and $\Htil$.

Note that using the relations between various hadronic tensor amplitudes [Eqs.~(\ref{eq:BtoD_amp},\ref{eq:BtoDst_amp})], the tensor contribution to $\htil$ and $\Htil$ in Eqs.~(\ref{eq:htil},\ref{eq:Htil}) vanishes if $g_T=g_{T5}$. This is reasonable since the operator $\cbar \sigma_{\mu\nu} (1+\gamma_5) b \, \lbar \sigma^{\mu\nu}(1-\gamma_5) \nu$ identically vanishes due to the Fierz transformations.

\section{Polarization vectors and spinors\label{app:pol_spin}}

In the $B$ rest frame the four-momenta of $B$ ($p$), $\DDst$ ($k$) and $q$ are
\begin{equation}
p^\mu = \left(
\begin{array}{c}
m_B \\
0 \\
0 \\
0
\end{array}\right) \,, \quad\quad
k^\mu = \left(
\begin{array}{c}
E_\DDst \\
0 \\
0 \\
|\q|
\end{array}\right) \,, \quad\quad
q^\mu = \left(
\begin{array}{c}
q_0 \\
0 \\
0 \\
-|\q|
\end{array}\right) \,.
\label{eq:momenta}
\end{equation}

The polarization vectors of $D^*$ ($\varepsilon$) and the virtual vector boson ($\eta$) are defined in the $B$ rest frame as in Ref.~\cite{Hagiwara:1989cu}:
\begin{equation}
\varepsilon^\mu(\pm) = {1 \over \sqrt2}\left(
\begin{array}{c}
0 \\
\mp1 \\
-i \\
0
\end{array}\right)\,,\quad\quad
\varepsilon^\mu(0) = {1 \over m_\Dst}\left(
\begin{array}{c}
|\q| \\
0 \\
0 \\
E_\Dst
\end{array}\right) \,,
\label{eq:pol_Dst}
\end{equation}
and
\begin{equation}
\eta^\mu(\pm) = {1 \over \sqrt2}\left(
\begin{array}{c}
0 \\
\mp 1 \\
i \\
0
\end{array}\right) \,,\quad
\eta^\mu(0) = {1 \over \sqrt{q^2}}\left(
\begin{array}{c}
|\q| \\
0 \\
0 \\
-q_0
\end{array}\right) \,, \quad
\eta^\mu(t) = {1 \over \sqrt{q^2}}\left(
\begin{array}{c}
q_0 \\
0 \\
0 \\
-|\q|
\end{array}\right) \,,
\label{eq:pol_boson}
\end{equation}
with
\begin{equation}
|\q| = {\sqrt{\lambda_{B\DDst}(q^2)} \over 2m_B} \,, \quad\quad q_0={m_B^2-m_\DDst^2+q^2 \over 2m_B}\,, \quad\quad E_\DDst={m_B^2+m_\DDst^2-q^2 \over 2m_B} \,.
\end{equation}
Other alternative parametrization can be found in the seminal paper~\cite{Korner:1989qb}, in which the $z$-axis is also along $\DDst$ momentum. Note that in Ref.~\cite{Korner:1989qb} all four-vectors are defined as covariant, while in this work we define all vectors as contravariant.

The Dirac spinors, used in the leptonic helicity amplitudes \eqref{eq:leptonic_amp} calculation, are defined as \cite{Hagiwara:1989cu}
\begin{equation}
u(\lambda=\pm1/2) = \left(
\begin{array}{c}
\sqrt{E \mp |{\bm p}|} \, \xi_\pm \\
\sqrt{E \pm |{\bm p}|} \, \xi_\pm
\end{array}\right)\,,\quad
v(\lambda=\pm1/2) = \left(
\begin{array}{c}
-\sqrt{E \pm |{\bm p}|} \, \xi_\mp \\
\sqrt{E \mp |{\bm p}|} \, \xi_\mp
\end{array}\right)\,,
\label{eq:spinors}
\end{equation}
where the helicity eigenspinors
\begin{equation}
\xi_+ = \left(
\begin{array}{c}
\cos{\theta\over2} \\
\sin{\theta\over2} e^{i\phi}
\end{array}\right)\,,\quad\quad
\xi_- = \left(
\begin{array}{c}
-\sin{\theta\over2} e^{-i\phi} \\
\cos{\theta\over2}
\end{array}\right)\,,
\label{eq:xi_spinors}
\end{equation}
describe either particles of helicity $\pm1/2$ respectively or antiparticles of helicity $\mp1/2$. Since in this work we assume that neutrino is only left-handed, $\lambda_{\nubar}=1/2$. The $\nubar$ spinor, $v(\lambda_{\nubar}=1/2)$, is defined by Eqs.~\eqref{eq:spinors},\eqref{eq:xi_spinors} with $\theta_{\nubar}=\pi-\thl$ and $\phi_{\nubar}=\phi_\ell+\pi$. In our chosen system of coordinates the leptonic azimuthal angle $\phi_\ell\equiv\chi$.

\section{Four-body phase space\label{app:PS}}

The four-body phase space can be reduced to the product of the two-body phase spaces:
\begin{equation}
\begin{split}
d\Phi_4 =& (2\pi)^4\int\prod_{i=1}^4{d^3p_i \over (2\pi)^32E_i}\delta\biggl(P-\sum_{j=1}^4 p_j\biggr) \\
=& {dm_{12}^2 \over 2\pi}{dm_{34}^2 \over 2\pi}\,d\Phi_2(p_{12},\,p_{34})\,d\Phi_2(\hat p_1,\,\hat p_2)\,d\Phi_2(\hat p_3,\,\hat p_4) \,,
\end{split}
\label{eq:PS4}
\end{equation}
where $m_{ij}^2=p_{ij}^2=(p_i+p_j)^2$. The two-body phase space is given by the standard expression
\begin{equation}
d\Phi_2(\hat p_i,\,\hat p_j)={1 \over 16\pi^2}{|\p_i| \over m_{ij}}\,d\cos\theta_i\,d\phi_i \,,
\end{equation}
with three-momentum $\p_i$ defined in the $ij$ rest frame.

Using Eq.~\eqref{eq:PS4} one can write the phase space for the $\Bbar\to \Dst(\to D\pi)\ell\nubar_\ell$ as,
\begin{equation}
d\Phi_4 = {1 \over 64(2\pi)^8}dm_{D\pi}^2dq^2\,{|\p_{D\pi}| \over m_B}\,d\cos\theta_{D\pi}\,d\phi_{D\pi}\,{|\p_D| \over m_{D\pi}}\,d\cos\thD\,d\phi_D\,{|\p_\ell| \over \sqrt{q^2}}\,d\cos\thl\,d\phi_\ell \,,
\label{eq:PS4_2}
\end{equation}
where $\p_{D\pi}(=-\q)$, $\p_D$, $\p_\ell$ and the corresponding angles are defined in the $B$, $D\pi$ and $\ell\nubar$ rest frames respectively, namely,
\begin{equation}
|\p_{D\pi}| = {\sqrt{\lambda(m_B^2,m_{D\pi}^2,q^2)} \over 2m_B},\quad |\p_D| = {\sqrt{\lambda(m_{D\pi}^2,m_D^2,m_\pi^2)} \over 2m_{D\pi}},\quad |\p_\ell| = {q^2-m_\ell^2 \over 2\sqrt{q^2}} \,,
\end{equation}
where, as before, $\lambda(a,b,c)=a^2+b^2+c^2-2(ab+ac+bc)$.

Integrating over the polar and azimuthal angles of the $\Dst$ momentum ($\theta_{D\pi},\,\phi_{D\pi}$) and over the azimuthal angle of the $D$ momentum ($\phi_D$), one obtains
\begin{equation}
d\Phi_4 = {1 \over 64(2\pi)^6}{|\q| \over m_B}{|\p_D| \over m_{D\pi}}\left(1-{m_\ell^2\over q^2}\right) dq^2dm_{D\pi}^2d\cos\thD d\cos\thl d\chi \,.
\label{eq:PS4_3}
\end{equation}
Here we defined the angle $\phi_\ell=\chi$ with respect to the $D\pi$ rest frame.

Similarly, one can obtain the three-body phase space for the $\Bbar \to D\ell\nubar$ decay :
\begin{equation}
d\Phi_3 = {dq^2 \over 2\pi} \, d\Phi_2(p_D,\,q) \, d\Phi_2(\hat p_\ell, \,\hat p_\nu) \to {1\over16(2\pi)^3} \, {|\q| \over m_B} \, \left(1-{m_\ell^2\over q^2}\right) \, dq^2 \, d\cos\thl \,.
\end{equation}

\section{Fit results with NP couplings at best fit values\label{app:bestfit_tab}}
In here we show the fit results obtained following the same procedure as the one explained in 
Sec.~\ref{sec:fitres}, but fixing for each scenario the NP coupling at the best fit value, which can be found 
at Eqs.~(\ref{eq:bestfits},\ref{eq:bestfits_LR}). Similarly to Tab.~\ref{tab:obsCLN}, these results have been 
obtained enforcing the experimental results for $R(D)$ and $R(D^*)$ and allowing for NP effects in one coefficient at a time. In particular, we show in Tab.~\ref{tab:obsCLNfixed} the results for the NP coefficients employed in 
Eq.~\eqref{eq:Heff}, while we report in Tab.~\ref{tab:obsCLNfixed_LR} the results for the ones introduced in
Eq.~\eqref{eq:Heff_Jp}. Since all the predicted observables behave in a Gaussian manner, we write for
all of them the mean values and the standard deviations.

\begin{sidewaystable}[!t]
\fontsize{10.45}{13}\selectfont
\centering
\begin{tabular}{|c|c|c|c|c|c|c|c|}
\hline
&&&&&&&\\[-1mm]
\textbf{Obs.} & Exp. & SM & $ g_V $ & $ g_A $ & $ g_S $ & $ g_P $ & $ g_T $ \\[1mm]
&&&&&&&\\[-3mm]
\hline
&&&&&&&\\[-3mm]
$ R_D $
& $0.334 \pm 0.029$ 
& $0.30 \pm 0.02$ 
& $0.40 \pm 0.01$ 
& $-$ 
& $0.37 \pm 0.02$ 
& $-$ 
& $0.33 \pm 0.01$ 
\\
&&&&&&&\\[-3mm]
$ R_{D^*} $
& $0.297 \pm 0.015$ 
& $0.258 \pm 0.003$ 
& $0.262 \pm 0.003$ 
& $0.299 \pm 0.004$ 
& $-$ 
& $0.284 \pm 0.006$ 
& $0.30 \pm 0.01$ 
\\
&&&&&&&\\[-3mm]
\hline
&&&&&&&\\[-3mm]
$ R(A_{FB}^{D})$ 
& $ - $ 
& $0.077 \pm 0.004$ 
& $0.071 \pm 0.002$ 
& $-$ 
& $0.067 \pm 0.002$
& $-$ 
& $0.081 \pm 0.003$ 
\\
&&&&&&&\\[-3mm]
$ R(A_{\lambda_\ell}^D)$ 
& $ - $ 
& $-0.332 \pm 0.003\phantom{-}$ 
& $-0.331 \pm 0.003\phantom{-}$ 
& $-$ 
& $-0.495 \pm 0.003\phantom{-}$ 
& $-$ 
& $-0.25 \pm 0.05\phantom{-}$ 
\\
&&&&&&&\\[-3mm]

\hline
&&&&&&&\\[-3mm]
$ R(A_{\lambda_\ell}^{D^\ast})$ 
& $ - $ 
& $0.47 \pm 0.02$ 
& $0.49 \pm 0.02$ 
& $0.48 \pm 0.02$ 
& $-$ 
& $0.34 \pm 0.03$ 
& $0.22 \pm 0.02$ 
\\
&&&&&&&\\[-3mm]
$ R(R_{L,T})$ 
& $ - $ 
& $0.79 \pm 0.02$ 
& $0.76 \pm 0.02$ 
& $0.80 \pm 0.02$ 
& $-$ 
& $0.99 \pm 0.04$ 
& $0.46 \pm 0.02$ 
\\
&&&&&&&\\[-3mm]
$ R(R_{A,B})$ 
& $ - $ 
& $0.520 \pm 0.004$ 
& $0.510\pm 0.004$ 
& $0.524 \pm 0.004$ 
& $-$ 
& $0.515 \pm 0.004$ 
& $0.613 \pm 0.007$ 
\\
&&&&&&&\\[-3mm]
$ R(A_{FB}^{D^\ast})$ 
& $ - $ 
& $0.23 \pm 0.04$ 
& $0.37 \pm 0.03$ 
& $-0.37 \pm 0.04\phantom{-}$
& $-$ 
& $-0.04 \pm 0.04\phantom{-}$ 
& $ -0.03 \pm 0.03\phantom{-}$ 
\\
&&&&&&&\\[-3mm]
$ R(A_3)$ 
& $ - $ 
& $0.62 \pm 0.01$ 
& $0.58 \pm 0.01$ 
& $0.63 \pm 0.01$ 
& $-$ 
& $0.55 \pm 0.01$ 
& $0.16 \pm 0.07$ 
\\
&&&&&&&\\[-3mm]
$ R(A_4)$ 
& $ - $ 
& $0.46 \pm 0.01$ 
& $0.45 \pm 0.01$ 
& $0.46 \pm 0.01$ 
& $-$ 
& $0.41 \pm 0.01$ 
& $0.12 \pm 0.04$ 
\\
&&&&&&&\\[-3mm]
$ R(A_5)$ 
& $ - $ 
& $1.15 \pm 0.02$ 
& $1.24 \pm 0.02$
& $0.71 \pm 0.03$
& $-$ 
& $1.27 \pm 0.02$
& $ 0.55 \pm 0.05$ 
\\
&&&&&&&\\[-3mm]
$ R(A_6)$ 
& $ - $ 
& $0.79 \pm 0.01$ 
& $0.92 \pm 0.02$
& $0.21 \pm 0.01$ 
& $-$ 
& $0.71 \pm 0.02$ 
& $0.25 \pm 0.05$ 
\\
&&&&&&&\\[-3mm]
$ D(A_7)$ 
& $ - $ 
& $0$ 
& $(7.3 \pm 0.4)\cdot10^{-3}$
& $0.035 \pm 0.002$
& $-$ 
& $(-5.6 \pm 0.2)\cdot10^{-3}$
& $-0.02 \pm 0.01\phantom{-}$ 
\\
&&&&&&&\\[-3mm]
$ D(A_8)$ 
& $ - $ 
& $0$ 
& $(5.2 \pm 0.2)\cdot10^{-3}$
& $0.025 \pm 0.001$ 
& $-$ 
& $0$ 
& $0$ 
\\
&&&&&&&\\[-3mm]
$ D(A_9)$ 
& $ - $ 
& $0$ 
& $(13.2 \pm 0.4)\cdot10^{-3}$
& $0.063 \pm 0.002$ 
& $-$ 
& $0$ 
& $0$
\\
&&&&&&&\\
\hline
\end{tabular}
\caption{\label{tab:obsCLNfixed} All the observables obtained by using the values of the NP couplings given in Eq.~\ref{eq:bestfits}.}
\end{sidewaystable}

\begin{sidewaystable}[!t!]
\fontsize{10.45}{13}\selectfont
\centering
\begin{tabular}{|c|c|c|c|c|c|c|}
\hline
&&&&&&\\[-1mm]
\textbf{Obs.} & Exp. & SM & $ g_{V_L} $ & $ g_{V_R} $ & $ g_{S_L} $ & $ {g_{S_R}} $ \\[1mm]
&&&&&&\\[-3mm]
\hline
&&&&&&\\[-3mm]
 $ R_D $  
& $0.334 \pm 0.029$ 
& $0.30 \pm 0.02$ 
& $0.33 \pm 0.01$ 
& $0.33 \pm 0.01$ 
& $0.33 \pm 0.01$ 
& $0.37 \pm 0.01$ 
\\
&&&&&&\\[-3mm]
 $ R_{D^*} $ 
& $0.297 \pm 0.014$ 
& $0.258 \pm 0.003$ 
& $0.300 \pm 0.004$ 
& $0.300 \pm 0.004$ 
& $0.275 \pm 0.005$ 
& $0.263 \pm 0.004$ 
\\
&&&&&&\\[-3mm]
\hline
&&&&&&\\[-3mm]
 $ R(A_{FB}^{D})$ 
 & $ - $ 
& $0.077 \pm 0.004$ 
& $0.074 \pm 0.003$ 
& $0.074 \pm 0.003$ 
& $0.044 \pm 0.002$ 
& $0.068 \pm 0.002$ 
\\
&&&&&&\\[-3mm]
 $ R(A_{\lambda_\ell}^D)$ 
 & $ - $ 
& $-0.332 \pm 0.003\phantom{-}$ 
& $-0.331 \pm 0.003\phantom{-}$ 
& $-0.331 \pm 0.003\phantom{-}$ 
& $-0.403 \pm 0.003\phantom{-}$ 
& $-0.501 \pm 0.003\phantom{-}$ 
\\
&&&&&&\\[-3mm]
\hline
&&&&&&\\[-3mm]
  $ R(A_{\lambda_\ell}^{D^\ast})$ 
 & $ - $ 
& $0.47 \pm 0.02$ 
& $0.48 \pm 0.02$ 
& $0.48 \pm 0.02$ 
& $0.38 \pm 0.02$ 
& $0.44 \pm 0.02$ 
\\
&&&&&&\\[-3mm]
 $ R(R_{L,T})$ 
 & $ - $ 
& $0.79 \pm 0.02$ 
& $0.79 \pm 0.02$ 
& $0.80 \pm 0.02$ 
& $0.92 \pm 0.03$ 
& $0.85 \pm 0.02$ 
\\
&&&&&&\\[-3mm]
 $ R(R_{A,B})$ 
 & $ - $ 
& $0.520 \pm 0.004$ 
& $0.520 \pm 0.004$ 
& $0.520 \pm 0.004$ 
& $0.516 \pm 0.004$ 
& $0.519 \pm 0.004$ 
\\
&&&&&&\\[-3mm]
 $ R(A_{FB}^{D^\ast})$ 
 & $ - $ 
& $0.23 \pm 0.04$ 
& $0.23 \pm 0.04$ 
& $-0.02 \pm 0.03\phantom{-}$ 
& $0.07 \pm 0.04$ 
& $0.13 \pm 0.04$ 
\\
&&&&&&\\[-3mm]
 $ R(A_3)$ 
 & $ - $ 
& $0.62 \pm 0.01$ 
& $0.62 \pm 0.01$ 
& $0.62 \pm 0.01$ 
& $0.57 \pm 0.01$ 
& $0.60 \pm 0.01$ 
\\
&&&&&&\\[-3mm]
 $ R(A_4)$ 
 & $ - $ 
& $0.46 \pm 0.01$ 
& $0.455 \pm 0.005$ 
& $0.455 \pm 0.005$ 
& $0.41 \pm 0.01$ 
& $0.45 \pm 0.01$ 
\\
&&&&&&\\[-3mm]
 $ R(A_5)$ 
 & $ - $ 
& $1.15 \pm 0.02$ 
& $1.16 \pm 0.02$ 
& $1.00 \pm 0.02$ 
& $1.19 \pm 0.02$ 
& $1.02 \pm 0.02$ 
\\
&&&&&&\\[-3mm]
 $ R(A_6)$ 
 & $ - $ 
& $0.79 \pm 0.01$ 
& $0.79 \pm 0.01$ 
& $0.57 \pm 0.02$ 
& $0.73 \pm 0.01$ 
& $0.77 \pm 0.01$ 
\\
&&&&&&\\[-3mm]
 $ D(A_7)$ 
 & $ - $ 
& $0$ 
& $0$ 
& $-0.027 \pm 0.001\phantom{-}$
& $0.019 \pm 0.001$ 
& $(-2.2 \pm 0.1)\cdot10^{-3}$
\\
&&&&&&\\[-3mm]
 $ D(A_8)$ 
 & $ - $ 
& $0$ 
& $0$ 
& $-0.019 \pm 0.001\phantom{-}$
& $0$ 
& $0$ 
\\
&&&&&&\\[-3mm]
 $ D(A_9)$ 
 & $ - $ 
& $0$ 
& $0$ 
& $-0.047 \pm 0.001\phantom{-}$
& $0$ 
& $0$ 
\\
&&&&&&\\
\hline
\end{tabular}
\caption{\label{tab:obsCLNfixed_LR}Same as in Tab.~\ref{tab:obsCLNfixed} but in different basis of operators and for values of NP couplings given in Eq.~\ref{eq:bestfits_LR}.}
\end{sidewaystable}
\FloatBarrier

\bibliographystyle{utphys}
\bibliography{bibliography}

\end{document}